\documentclass[pre,twocolumn,showpacs,preprintnumbers,amsmath,amssy mb,floatfix]{revtex4}
\usepackage{graphics}

\usepackage{graphicx}% Include figure files
\usepackage{dcolumn}% Align table columns on decimal point
\usepackage{bm}% bold math

\newcommand{\beq}{\begin{equation}}
\newcommand{\eeq}{\end{equation}}    

\newcommand{\bea}{\begin{eqnarray}}
\newcommand{\eea}{\end{eqnarray}}    

\begin{document}

\font\small=cmr8           % kleine Schrift fuer Zitate
\font\petit=cmcsc10        % Grossbuchstaben verschiedener Groesse !
                           % usage : {\petit .....}

\font\bbf=cmbx10 scaled\magstep1 % for titles
\font\bbbf=cmbx10 scaled\magstep2 % for big titles
\font\bbbbf=cmbx10 scaled\magstep3 % for big titles

\def\subti#1{\par\vskip0.8cm{\bf\noindent #1}\par\vskip0.4cm}
\def\ti#1{\par\vskip1.6cm{\bbf\noindent #1}\par\vskip0.8cm}
\def\bigti#1{\par\vfil\eject{\bbbf\noindent #1}\par\vskip1.6cm}

% Centered titles
% ---------------
\def\cbigti#1{\par\vskip2.0cm{\bbbf\noindent\centerline {#1}\par\vskip0.5cm}}
\def\cti#1{\par\vskip1.0cm{\bbf\noindent\centerline {#1}\par\vskip0.4cm}}
\def\csubti#1{\par\vskip0.5cm{\bf\noindent\centerline {#1}\par\vskip0.3cm}}

\def\doublespace {\baselineskip 22pt}        % doppelter Zeilenabstand 

\def\eqd{\buildrel \rm d \over =}    % equal in distribution (statistics)
\newcommand{\p}{\partial}           % for partial derivatives
\def\px{\partial _x}           % for partial derivatives
\def\py{\partial _y}           % for partial derivatives
\def\pz{\partial _z}           % for partial derivatives
\def\pt{\partial _t}           % for partial derivatives
\def\ssum{\textstyle\sum}
\def\arr{\rightarrow}
\def\id{\equiv}
\def\eqv{\leftrightarrow}
\def\fol{\rightarrow}
\let\prop=\sim
\def\gapprox{\;\rlap{\lower 2.5pt            % > ungefaehr =
 \hbox{$\sim$}}\raise 1.5pt\hbox{$>$}\;}       
\def\lapprox{\;\rlap{\lower 2.5pt            % < ungefaehr =
 \hbox{$\sim$}}\raise 1.5pt\hbox{$<$}\;} 

\def\ang{\,{\rm\AA}}
\def\cm{\,{\rm cm}}
\def\km{\,{\rm km}}
\def\kpc{\,{\rm kpc}}
\def\second{\,{\rm sec}}     % '\sec' used to number sections !
\def\erg{\,{\rm erg}}
\def\ev{\,{\rm e\kern-.1em V}}
\def\kev{\,{\rm ke\kern-.1em V}}
\def\k{\,{\rm K}}
\def\K{\,{\rm K}}
\def\gauss{\,{\rm gauss}}
\def\SFU{\,{\rm SFU}}

\newcommand{\R}{\,{\rm I\kern-.15em R}}
\def\N{\,{\rm I\kern-.15em N}}
\def\N{\,{\rm /\kern-.15em R}}

\def\mhz{\,{\rm MHz}}

% vereinfachende und redundante Abkuerzungen
\def\n{\noindent}
\def\lead{\leaders\hbox to 10pt{\hfill.\hfill}\hfill}
\def\a{\"a}
\def\o{\"o}
\def\u{\"u}
\def\infinit{\infty}
\def\upr#1{\rm#1}

% dimension and correlation integral:
\def\dd{D^{\left(2\right)}}
\def\cc{C_d ^{\left(2\right)} \left(\epsilon\right)}

%o.k., when printer o.k
%\vbox{}\vskip -1.5truecm
%\vbox{}\vskip -3.5truecm

% to number the equations:
\newcount\glno
\def\no{\global\advance\glno by 1 \the\glno}

% to number the sections and equations:
\newcount\secno
\def\newsec{\global\advance\secno by 1}
\def\sec{\the\secno}

% to number the chapters :
\newcount\chapno
\def\newchap{\global\advance\chapno by 1}
\def\chap{\the\chapno}

% \thesaurus{09 (06.06.3; 02.13.2;  02.03.1; 02.20.1)}

%\title{Random walk on fractal supports}
%\subtitle{An example of Levy flights}

\preprint{....}

%\title{Random walk through fractal environments II:\\
%A random walk approach to stochastic acceleration.\\
%Internal report, 2005}

\title{A random walk approach to anomalous particle and energy transport}
%Combined Continuous Time Random Walk in Position 
%and Momentum Space\\
%Internal report, 2007}

\author{H.\ Isliker}
%\email{isliker@helios.astro.auth.gr}

\affiliation{
Section of Astrophysics, Astronomy and Mechanics \\
Department of Physics, University of Thessaloniki \\
Association Euratom-Hellenic Republic,\\
GR 54006 Thessaloniki, GREECE}

\date{\today}

%\date{Received ...; accepted ...}

%\titlerunning{.......}

\begin{abstract}
The combined Continuous Time Random Walk (CTRW) in position
and momentum space is introduced, in the form of two coupled integral equations
that describe the evolution of the probability distribution for finding a particle
at a certain position and with a certain momentum as a function of time.
The integral equations are solved numerically with a pseudospectral method that is 
based on the expansion of the unknown functions in terms of Chebyshev
polynomials. In parallel, Monte-Carlo simulation are performed. 
Through the inclusion of momentum space, the combined CTRW is able to yield results on density and 
temperature profile evolution, on particle and heat fluxes and diffusivities, and on kinetic energy 
distributions. Depending on the choice of the probability distributions of the particle 
displacements in position and momentum space, the combined CTRW is non-local in position-space,
in momentum-space, and in time (non-Markovian), and it is able to model phenomena of anomalous
transport in position as well as in momentum (or energy or velocity) space.
An application is made to a toroidally confined plasma that
undergoes off-center injection of cold plasma (off-axis fueling), using two variants of the
model, the mixed model and the critical gradient model. 
The phenomenon of profile stiffness is addressed, and it is shown that it can be reproduced 
by the combined CTRW with varying success, both for the density and for the
temperature profile, respectively. The particle and energy confinement times are determined,
and their dependence on the applied intensity of plasma heating is discussed.
Finally, the analysis of the particle and heat fluxes shows that the dynamics realized in the combined
CTRW is incompatible with the classical approach of Fick's or 
Fourier's law for particle and heat transport, respectively, so that
particle and heat diffusivities determined through the latter are not an adequate 
characterization of the actual transport process in position and momentum space.
\end{abstract}



\maketitle

\section{Introduction}

Anomalous transport phenomena are conveniently characterized 
by the scaling of the mean square displacement $\langle r^2\rangle$ 
of an ensemble of particles with time $t$. Often, a power-law scaling is 
observed, 
\beq
\langle r^2\rangle \propto t^\gamma  ,
\label{msd}
\eeq
where the characteristic index $\gamma$ is used to discern 
normal or classical diffusion with $\gamma=1$ from anomalous
diffusion with $\gamma\ne 1$, and in particular sub-diffusion 
for $\gamma<1$ and super-diffusion for $\gamma>1$. 
Continuous time random walk (CTRW), introduced in \cite{{Montroll65}}, 
has successfully been applied to model various phenomena of anomalous 
transport, including sub- and super-diffusive phenomena, 
in the fields of physics, chemistry, astronomy, biology, and economics (see e.g.\ the
references in \cite{Metzler00}).

Most applications of the CTRW in physics model the random walk of particles in position space.
Two variants of CTRW are used, the waiting or trapping model, introduced by \cite{Montroll65}, 
and the velocity model, introduced by \cite{Shlesinger87}.
The two models differ in the timing of the random walk. In both models,
the random walker (particle) takes steps of random size and/or direction in position 
space. In the trapping model, the random walker waits for a 
random time at the 
location it was brought to by its last step before it moves
a new step away, in the step itself no time is consumed. In the 
velocity model, a particle is constantly 
moving, and the travel time, the time it takes a particle to complete 
its spatial step, determines the timing. To determine the travel time, a 
velocity has to be 
assumed, and since velocity space dynamics usually was not taken into account 
so-far, this velocity was usually assumed to be constant and for 
convenience set equal to one.

The main purpose of this article is to extend the position space CTRW in the 
variant of the velocity model, and to include also momentum space dynamics, 
so that the velocity does not play anymore just
a dummy role, but reflects the dynamics that take place in velocity
(or momentum) space, in parallel with the position
space evolution. This extension to the combined CTRW in position and momentum 
space is relevant in applications to turbulent fluids, and, most prominently, in
applications to  
turbulent laboratory and astrophysical
plasmas, where localized turbulent electric fields, magnetic field 
discontinuities (e.g.\ at magnetic reconnection sites), externally or 
internally generated waves or other forms of induced plasma heating, etc., are the cause
of a highly dynamic evolution in momentum space. 
The combined CTRW in such applications is able to yield information about 
kinetic energy distributions, temperature
profiles, heat fluxes, and heat diffusivities, together of course with information about
particle densities, particle fluxes, and particle diffusivities.

The CTRW formalism was first applied to 
plasmas by Ref.\ 
\cite{Balescu95}, and it was shown that in the form of the critical 
gradient model (\cite{vanMil04}, \cite{vanMil04b}), the CTRW can successfully 
model various observed 
phenomena of anomalous transport in confined turbulent plasmas, where though
so-far only the evolution in position space was taken into account.
In \cite{Vlahos04}, we performed a Monte Carlo simulation study of the combined
CTRW in position and momentum space, in application to the plasma in the solar corona, 
where the main physical interest was in modeling the appearance of non-thermal energy distributions
and the related electromagnetic emission spectra during solar flares. 
In this article, we introduce a set of equations
that describe the combined CTRW in position and momentum space, 
and we present a method for solving them numerically.

In order to illustrate the capabilities of the combined CTRW and to demonstrate the importance
of the inclusion of momentum space in the random walk approach, we
present an application 
to laboratory plasmas, 
confined in toroidal devices such as the tokamak.
Toroidally confined plasmas exhibit a variety of 
anomalous transport phenomena, both what particle and heat transport are concerned, respectively,
and which clearly contradict what classical diffusion would
predict. In view of the intended application, we just mention a few
manifestations of anomalous transport in confined plasmas:
(i) Measured diffusion coefficients (for particle and for heat diffusion)
are usually larger than the neoclassical values, i.e.\ those derived from collisional 
effects in toroidal geometry (e.g. \cite{Garbet04}).
(ii) A characteristic property in confined plasmas is profile stiffness (also termed 
resilience or consistency), which denotes the preference of a plasma to stay close to a certain
density and temperature profile (in direction of the minor radius of the toroidal 
confinement device), i.e.\ the profiles are usually
peaked at the center, and they are to a large degree unaffected by the way the plasma is distorted
externally, e.g.\ through the localized off-center injection of particles or heat. 
Profile stiffness is an anomalous transport phenomenon in the sense that 
particles or energy are transported 'up-hill', against the density or temperature 
gradient, respectively, which, in the case of normal diffusion, would drive diffusion 
'down-hill'.
Profile stiffness is discussed e.g.\ in 
Ref.\ \cite{Garbet04}, and examples of related experiments
include \cite{Luce92}, \cite{Petty94}, 
\cite{Ryter03} and are reviewed e.g.\ in \cite{Ryter01}. 
(iii) The experimental analysis in Ref.\ \cite{Lemoine05} finds that plasma diffusion is characterized
by particle displacements that can be characterized statistically with a probability 
distribution that exhibits 
a power-law tail with a power-law index 2. This implies that particles undergo
occasionally large displacements, so that the spatial diffusion process can be non-local
in nature.
(iv) Ref.\ \cite{Hoang01} and, similarly, Ref.\ \cite{Baker01} infer from experiments that 
the electron heat transport is threshold
dependent, in the sense that transport is activated only if the temperature 
gradient $\nabla T$ exceeds a certain threshold, $\vert \nabla T \vert > \nabla T_{crit}$.
We just note that the properties (ii) and (iv) are very reminiscent
of Self-Organized Criticality (SOC; \cite{Bak87}).

Based on these experimental results, the 
application we present is to an experiment where a colder plasma is
injected localized off the center (off-axis fueling), and we will focus
on phenomena of density and 
temperature profile stiffness, the quality of the particle and the energy 
confinement, and the possibility to characterize transport with particle and heat
diffusivities. 
In accordance with the mentioned experimental result (iii),
particle displacements in position space will partly be 
allowed to be large, i.e.\ of system size, so that position space transport will throughout be
of partly non-local nature.
To explore the influence of the momentum space evolution
on the over-all dynamics, we will in all applications consider two cases 
for displacements in momentum space, namely small displacements that correspond to a very low
level of energy injection (heating) into the system and to classical random walk in momentum space, 
and large displacements in momentum that
follow a power-law distribution, and which represent the case of
intense heating and non-local transport in momentum space.

In Sec.\ \ref{SecII}, the equations for the combined CTRW in position
and momentum space are introduced, first in a general three dimensional form
and then in a one dimensional version that will be used in the applications.
Also, short explanations on Monte-Carlo simulations are given, and the pseudospectral method
based on Chebyshev polynomials, with which the combined CTRW equations are solved numerically, 
is shortly presented. 
Sec.\ \ref{SecIII} contains the applications to the off-center
plasma injection experiment. In order to specify the combined CTRW to the 
set-up of a toroidally confined plasma, two models will be introduced,
the mixed model and the critical gradient model, which realizes the mentioned
experimental feature (iv).  
In Sec.\ \ref{SecIV}, the particle and heat confinement
times, fluxes and diffusivities are discussed. Finally, Sec.\ \ref{SecV} presents
a summary and the conclusions, and a more detailed presentation of the pseudospectral
numerical method is given in App.\ \ref{AppA}.

\section{Continuous Time Random Walk in position and momentum space \label{SecII}}

\subsection{The distribution of increments}

In the random walk approach, the different processes a particle
can undergo in its evolution are formally separated, and for
the random walk of a particle in a turbulent plasma they consist in 
(i) collision and heating or acceleration events, 
in which mainly the momentum and much less the position
of a particle changes, and which 
include particle-particle 
collisions, particle-wave collisions, absorption of electromagnetic waves, 
interaction with localized, turbulent electric fields; (ii) trapping events, 
mostly in inhomogeneous magnetic field structure, and in which neither 
position nor momentum of a particle is changed;
and (iii) free travel or drift events events, during which the energy
of a particle remains unchanged. In each of these events, a particle
spends a certain time. 

For simplicity, we will throughout the following omit
the trapping events, and we only consider acceleration and free 
travel events. Moreover, we take 
into account only the time spent in the free travels, the acceleration time 
is considered negligible and neglected. 
We note that the free flight times provide a basic
coupling between momentum and position space dynamics, since the acceleration 
events have a direct influence on the velocity of a particle, and the 
latter in turn determines, together with the travel distance, the free 
flight time and thus the overall timing of the random walk.
(If one alternatively would take into account only the 
trapping times, then the dynamics in position and momentum space would
be decoupled, unless the trapping time would depend in some way
on momentum and position.)

The basic quantity for the random walk is the probability density function 
(pdf) $\psi(\Delta\vec p,\Delta\vec r,\tau)$ of random walk increments or 
steps, which determines the probability for a
particle to perform a jump $\Delta\vec p$ in momentum space (corresponding
to heating, acceleration, or possibly also dissipation), to freely
travel a directed distance $\Delta\vec r$ in position space, and 
to spend on this free travel a time $\tau$. The free travel time is defined
as $\tau \equiv \Delta r/v$, with the 
jump-length $\Delta r\equiv \vert\Delta\vec r\vert$ 
and the instantaneous velocity 
$v\equiv \vert\vec v\vert$, which is a direct function of the 
instantaneous momentum $\vec p$. 
We assume $\Delta\vec p$ and $\Delta\vec r$ to be 
independent random variables, 
with pdf $q_{\Delta\vec p}(\Delta\vec p)$ and $q_{\Delta\vec r}(\Delta\vec r)$,
respectively, so that the
joint pdf $\psi$ decouples to
\bea
\psi(\Delta\vec p,\tau,\Delta\vec r) &=& q_{\Delta\vec
p}(\Delta\vec p)\, q_{\Delta\vec r}(\Delta\vec r) \nonumber \\
&&  \times        \phi(\tau \,\vert\, \Delta r;v)  ,
\label{rw1}
\eea
where the conditional probability for the free flight time, given the length 
of the step and the particle's velocity, can be written as  
$\phi(\tau \,\vert\, \Delta r;v)= \delta(\tau-\Delta r/ v)$,
so that the pdf of increments takes the form
\begin{eqnarray}
\psi(\Delta\vec p,\tau,\Delta\vec r,\tau) &=& q_{\Delta \vec
p}(\Delta\vec p)\, q_{\Delta\vec r}(\Delta\vec r)
                                                            \nonumber \\
&&  \times      \delta(\tau-\Delta r/ v) .
\label{distinc}
\end{eqnarray}
We just note that a different choice for the delta function seems to be
$\delta(\Delta r - v\tau)$, with this choice though, the joint
pdf $\delta(\Delta r - v\tau)q_{\Delta\vec r}(\Delta\vec r)$
would have wrong units, and also the marginal distributions
calculated from it would be inconsistent.

\subsection{The pdf of the turning-points}

In the derivation of the equations for the combined CTRW equations in position and
momentum space, we follow the formalism of e.g.\ \cite{Zumofen93} for the CTRW equations 
in position space alone, which we extend by adding momentum. The basic CTRW equations 
in \cite{Zumofen93} are a set of two integral equations that express the conservation of 
particles in integral form. Here, we use the variant of the velocity model, since we want 
to take the free flight times explicitly into account, 

Following \cite{Zumofen93}, we 
introduce the concept of turning points,
at which a particle takes a new step in its random walk.
More precisely, as turning points of the random walk we define the points 
in position 
($\vec r$) and momentum ($\vec p$) space where the particles arrive at and 
undergo an acceleration event. 
Two turning-points are thus separated by an acceleration event and
a free jump in position-space, and we are in principle free 
to choose in which order the two processes happen, for
practical reasons though we let the cycle start with an acceleration event. 
We define $Q(\vec r,\vec p,t)$ as the distribution of the turning
points, which describes the rate at which particles arrive at time $t$ 
at the turning point that is located at $(\vec r,\vec p)$ (as a rate 
$Q$ has units 
$\left(cm\, g\frac{cm}{sec}\right)^{-3}\, \left(sec\right)^{-1}$). 
Adding the momentum to the evolution equation for $Q$ in \cite{Zumofen93} 
and using the distribution of increments of Eq.\ (\ref{distinc})
yields
\begin{eqnarray}
Q(\vec r,\vec p,t) &=& 
      \int\limits_{-\infty}^{\infty} d^3\vec p^{\,\prime}   
\int\limits_{\vert \vec r- \vec r^{\,\prime}\vert \leq v(\vec p)t} 
                  \!\!\!\!\!\!\!\!\!\!\!\!\!\! d^3\vec r^{\,\prime} 
              \ \int\limits_0^t dt^\prime                  
                             \nonumber \\
&& \ \times Q(\vec r^{\,\prime},\vec p^{\,\prime},t^\prime) \nonumber \\
&& \ \times 
       \delta\left(t-t^\prime-\vert\vec r-\vec r^{\,\prime}\vert /v(\vec p)\right)\,
           q_{\Delta \vec r}(\vec r-\vec r^{\,\prime}) 
             \nonumber \\
&& \ \times   q_{\Delta \vec p}(\vec p-\vec p^{\,\prime})\,
               \nonumber \\
&+&\ \delta(t)\, 
     P(\vec r,\vec p,t=0) + S(\vec r,\vec p,t)   .
\label{rwQ}
\end{eqnarray}
The first term on the right hand side just describes a completed 
cycle of a CTRW step in position space, momentum space, and time:
in order to arrive at a turning point at $(\vec r,\vec p)$ at time $t$, a particle
must have arrived at a turning point $(\vec r^{\,\prime},\vec p^{\,\prime})$ 
at an earlier time $t^{\prime}$, where after 
it has performed a step $\vec p-\vec p^{\,\prime}$ in momentum space and then a step 
$\vec r-\vec r^{\,\prime}$ in position
space, for which the particle has spent a time $t-t^{\prime}$ that must 
equal the free flight time, 
$t-t^\prime = \vert\vec r-\vec r^{\,\prime}\vert /v(\vec p)$.  
Since we assume the free flight to take place after the 
acceleration event, the momentum during the free flight is $\vec p$,
so that $v=v(\vec p)$
(if we
would assume the free flight to take place before the acceleration event, 
then the free flight time 
would depend on $p_x^\prime$, and the integrals would become more
complicated in their formal structure).
The limits of the $\vec r^{\,\prime}$ integration
are imposed by causality, we cannot consider at time $t$ spatial increments
that have free flight times $\tau$ larger than $t$.
The second term on the right hand side of Eq.\ (\ref{rwQ}) takes the initial conditions
into account, $P(\vec r,\vec p,t=0)$ is the particle distribution
at time $t=0$. 
Finally, $S$ in Eq.\ (\ref{rwQ}) is a source term that represents
a continuous particle source 
(more precisely, $S$ is the source rate, with units
$\left(cm\, g\frac{cm}{sec}\right)^{-3} sec^{-1}$).
Writing the source in this form implies that 
particles are injected at turning points, which means  
that injected particles are immediately %either 
accelerated after their injection.

We just note that if also trapping and acceleration times were 
taken into account, then two more temporal integrals would have to be added to
the equation, which is formally 
possible, it increases though the numerical complexity, computing time would 
be increased in the numerical solution, and good numerical precision would 
be more difficult to be achieved.

\subsection{The propagator}

The propagator $P(\vec r,\vec p,t)$ is defined as the probability distribution
for a particle 
to be at time $t$ at position $(\vec r,\vec p)$ anywhere at a turning point
or in-between two turning points (the 
units of $P$ are 
$\left(cm\, g\frac{cm}{sec}\right)^{-3}$). 
The propagator evolution equation is again determined by generalizing the
corresponding equation in \cite{Zumofen93} to include also momentum, 
\begin{eqnarray}
P(\vec r,\vec p,t) &=& 
  \int\limits_{-\infty}^\infty \!\! d^3  \vec p^{\,\prime}
     \int\limits_{\vert \vec r- \vec r^{\,\prime}\vert \leq v(\vec p)t} 
                  \!\!\!\!\!\!\!\!\!\!\!\!\!\! d^3 \vec r^{\,\prime}
    \  \int\limits_0^t\!\! d t^\prime 
                                                   \nonumber \\
  &&\ \ \ \ \ \times Q( \vec r^{\,\prime},
                      \vec p^{\,\prime},t^\prime) 
                                                        \nonumber\\
  &&\ \ \ \ \ \times 
    \Phi_{\Delta\vec r}(\vec r- \vec r^{\,\prime},t-t^\prime;v(\vec p)) \,
                                                        \nonumber\\
  &&\ \ \ \ \ \times    q_{\Delta\vec p}(\vec p-\vec p^{\,\prime})  .
\label{rwP}
\end{eqnarray}
The limits of the spatial integral express again the fact that particles cannot
travel spatial distances at time $t$ that take flight times $\tau$
longer than $t$.
$\Phi_{\Delta\vec r}(\Delta \vec r,\tau;v)$ is the probability 
for a particle to be found at a certain spatial location 
on its free travel in-between two turning-points, which equals
the probability to make a spatial jump in the direction of $\Delta\vec r$,
with length at least $\vert \Delta\vec r\vert$ and of duration at least
$\tau$, being though at time $\tau$ exactly at position
$\Delta\vec r$. In spherical coordinates with 
$\Delta\vec r=(\Delta r,\theta,\phi)$, and where the
angles $\theta$ and $\phi$ determine the direction of a jump and 
$\Delta r=\vert \Delta\vec r\vert$ is its length, $\Phi_{\Delta\vec r}(\Delta \vec r,\tau;v)$
can be expressed as (see the explanations below)
\begin{eqnarray}
\Phi_{\Delta\vec r}(\Delta \vec r,\tau;v) &=& 
\delta(\Delta r - v\tau)
                                  \nonumber\\
&\times&\!\!\!\! 
\frac{1}{\Delta r^{2}} 
\int\limits_{\Delta r^{\prime} \geq \Delta r,\, 
             \theta^{\prime}=\theta,\,
             \varphi^{\prime}=\varphi}
                                    \!\!\!\!\!\!\!\!
                     d \Delta r^{\,\prime} 
                         \, \Delta r^{\prime 2} \nonumber \\
&&\times       \int\limits_{\tau^{\prime} \geq \tau}
                          \!\!\!\! d\tau^{\prime}
           \delta(\tau^{\prime}-\Delta r^{\,\prime}/v)
            \,q_{\Delta\vec r}(\Delta\vec r^{\,\prime})  ,
\label{rwPhi}
\end{eqnarray}
%where $\Delta r = \vert \Delta\vec r\vert$ and 
where $\Delta r^\prime = \vert \Delta\vec r^{\,\prime}\vert$
and $\Delta\vec r^{\,\prime}=(\Delta r^\prime,\theta^\prime,\varphi^\prime)$.
Note the form of the first delta function, in contrast to
the conditional pdf for the free flight times --- the seemingly 
alternative choice $\delta(\tau^\prime - \Delta r^\prime/v)$
would cause $\Phi_{\Delta\vec r}(\Delta \vec r^{\,\prime},\tau^\prime)$
to have wrong units, i.e.\ $1/(sec\,\,cm^2)$ instead of the needed
$1/cm^3$.

Eq.\ (\ref{rwP}) states that a particle is at position $(\vec r, \vec p)$
at time $t$ if it was at a turning point
$(\vec r^\prime,\vec p^\prime)$ at time $t^\prime$, it 
underwent an acceleration event which changed its momentum 
by $\vec p-\vec p^\prime$, and it is now on a free flight event 
whose duration is at least $t-t^\prime$, being though at time $t$  
exactly at position $(\vec r, \vec p)$.
Since no time is assumed to be consumed in the acceleration events, we cannot 
locate the particles during acceleration events, but only during the free
flights in position space.
Again, the assumption that the turning points are the points where a 
particle starts undergoing first an acceleration event,  
and the acceleration event is then followed by a free 
flight event, was used in the formulation of Eq.\ (\ref{rwP}). 

\subsubsection{Explanations on the form of $\Phi_{\Delta\vec r}$}

In spherical coordinates $(\Delta r,\theta,\phi)$, the probability (not the density) 
to make a jump of length $\Delta\vec r$ into the direction $(\theta,\phi)$ and 
in time $\tau$ is given
by multiplying Eq.\ (\ref{distinc}) by the differentials of the coordinates,
\beq
\delta(\tau-\Delta r/v) \,q_{\Delta\vec r}(\Delta r,\theta,\phi) 
\, \Delta r^2\,\sin(\theta)\, d\Delta r\,d\theta\,d\phi\,d\tau  ,
\eeq
and the probability to make a jump in the direction $(\theta,\phi)$
larger than $\Delta r$ and $\tau$ is 
\bea
\sin(\theta)\,d\theta\,d\phi\,  \nonumber \\
\times \int\limits_{\tau^\prime \geq \tau}  \!\!\!\!\! d\tau^\prime 
\int\limits_{\Delta r^\prime \geq \Delta r} \!\!\!\!\!\!\! d\Delta r^\prime 
\delta(\tau^\prime-\Delta r^\prime/v) \,q_{\Delta\vec r}(\Delta r^\prime,\theta,\phi) 
\, \Delta r^{\prime\,2} .
\label{expl1}
\eea
We furthermore demand that the walker has traveled a distance exactly
 $\Delta r$ at time $\tau$ into the fixed direction 
$(\theta,\phi)$, i.e.\ $\Delta r= v\tau$, which is conveniently 
enforced through a delta function, $\delta(\Delta r - v\tau)$
(the variant $\delta(\tau - \Delta r/v)$ has wrong units, see the remark above).
%has units $1/sec$, which will lead to wrong units
%of $\Phi_{\Delta\vec r}$ and therewith of the total expression for $P$, so the correct choice is
%, which has units $1/cm$. 
In spherical coordinates, the delta function is of the form
\beq
\delta(\Delta\vec r) = \frac{1}{\Delta r^2} \delta(\Delta r)\,
\delta(\cos\theta)\,\delta(\phi)
\eeq 
(e.g.\ \cite{Jackson62}), %Jackson or Arfken), 
of which we need the spatial part only, since the direction
is explicitly kept fixed, so that the condition $\Delta r= v\tau$
must be written as 
\beq
\frac{1}{\Delta r^2} \delta(\Delta r-v\tau)  .
\label{expl2}
\eeq 
Combining Eqs.\ (\ref{expl1}) and (\ref{expl2}) leads to Eq.\ (\ref{rwPhi}),
whereby the angular differential $\sin(\theta)\,d\theta\,d\phi$ in
Eqs.\ (\ref{expl1}) is understood as part of $d^3\vec r^{\,\prime}$ 
in Eq.\ (\ref{rwPhi}).

\subsubsection{Marginal distribution\label{margi}}

Once the propagator is determined, the particle density distribution $n(\vec r,t)$
and the momentum distribution function $f_p(\vec p,t)$ are given as marginal
distributions of $P(\vec r,\vec p,t)$,
\beq
n(\vec r,t) = \int d^3p \,P(\vec r,\vec p,t) ,
\eeq
and
\beq
f_p(\vec p,t) = \int d^3r \,P(\vec r,\vec p,t) ,
\eeq
respectively.
For the kinetic energy distribution $f_E(E_{kin},t)$, we first define
the distribution of $p\equiv \vert \vec p\vert$,
\begin{equation}
P_p(p) = \int\!\! p^2 \sin\theta_p \, d\theta_p \,d\phi_p \,
   P_{\vec p}(\vec p)  .
\label{rw7}
\end{equation}
The distribution of $E_{kin}$ is then found from the distribution of $p$ through
the relation $f_E(E_{kin})dE_{kin}=p(p_x)dp_x$, or 
$f_E(E_{kin})=p(p_x)dp_x/dE_{kin}$. From the expression for 
the total energy $E$ in terms of the momentum, $E^2=p^2c^2+m^2c^4$, and
the definition of the kinetic energy, $E_{kin}=E-mc^2$, it follows that 
$dp_x/dE_{kin}=(E_{kin}+mc^2)/c^2\sqrt{E_{kin}^2/c^2+2E_{kin}m}$,
where $m$ is the particle mass and $c$ the speed of light.  

Defining the temperature as the mean kinetic energy per particle,
$\frac{3}{2} k_B T = \langle E_{kin} \rangle$, 
with $k_B$ the Boltzmann constant,
we can determine the temperature profile as
\beq
T(\vec r,t) = \frac{1}{n(r,t)} \int\!\! d^3\vec p \,
   P(\vec r,\vec p,t) \, m(\gamma-1) c^2,
\label{rw8}
\eeq
with $\gamma=1/\sqrt{(1-v^2/c^2)}$ and $v=v(\vec p)$.

\subsection{Remarks on possible reformulating of the equations}

The explicit appearance of the time, position, and momentum in the integration
limits of Eqs.\ (\ref{rwQ}) and (\ref{rwP}) make the straightforward application
of Fourier and Laplace transforms with the corresponding convolution theorems
impossible. As a consequence, the equations
cannot trivially be reformulated into one equation for
the propagator $P(\vec x,\vec p,t)$ alone, and also a transformation to a different
type of equation (integro-differential or possibly fractional diffusion equation)
seems at least difficult. 

Also direct ways of reformulating into one equation do not work, e.g.\ when inserting $Q$ 
into $P$ and trying to change the order of the integrations to identify $P$ under the outer 
integrals, it turns out that the free flight delta function does not allow the change of the 
order of integrations because of appearance of the velocity $v$, which is a function
of $\vec p$.

\subsection{The 1-dimensional case}

We specify the general equations to the 1-dimensional form, using the variables
$x$ for position and $p_x$ for momentum, and, in view of the intended application,
we assume the system to be finite in $x$-direction, $x\in[-L,L]$, with $L$ half the 
system size.  
The turning point equation [Eq.\ (\ref{rwQ})] in one dimension takes the form
\begin{eqnarray}
Q(x,p_x,t) &=& \int\limits_{-\infty}^\infty d p_x^\prime    
\int\limits_{\max[x-\vert v_x\vert t,\,-L]}^{\min[x+\vert v_x\vert t,\,L]} 
                          d x^\prime 
\int\limits_0^t dt^\prime   
                             \nonumber \\
&& \ \times Q(x^\prime,p_x^\prime,t^\prime) \nonumber \\
&& \ \times \delta\left(t-t^\prime-\vert x- x^\prime\vert /\vert v_x(p_x)\vert\right)
               \,    q_{\Delta x}(x- x^\prime)
               \nonumber \\
	      && \ \times  q_{\Delta p_x}(p_x- p_x^\prime)\,
                            \nonumber \\
&+&\delta(t)\, P(x,p_x,t=0) + S(x,p_x,t) .
\label{rwQ1D}
\end{eqnarray}
The spatial integration limits 
are equivalent to the condition $\vert x -x^\prime\vert \leq \vert v_x\vert t$ 
from Eq.\ (\ref{rwQ}), and moreover they restrict $x^\prime$ to   
$x^\prime\in[-L,L]$ to account for the finiteness of the system.

In one dimension, the propagator
equation [Eq.\ (\ref{rwP})] becomes
\begin{eqnarray}
P(x,p_x,t) &=& 
    \int\limits_{-\infty}^\infty \!\! d p_x^{\prime}   
     \int\limits_{\max[x-\vert v_x\vert t,\,-L]}^{\min[x+\vert v_x\vert t,\,L]}
  \!\!\!\!\!\!\!\! dx^{\prime}
      \int\limits_0^t\!\! dt^\prime 
                                                   \nonumber \\
  &&\ \ \ \ \ \times Q(x^{\prime},
                       p_x^{\prime},t^\prime) 
                                                        \nonumber\\
  &&\ \ \ \ \ \times 
    \Phi_{\Delta x}(x-  x^{\prime},t-t^\prime;v(p_x)) \,
                                                        \nonumber\\
  &&\ \ \ \ \ \times    q_{\Delta p_x}(p_x-p_x^{\prime}) .
\label{rwP1D}
\end{eqnarray}
%(units 
%$\left(cm\, g\frac{cm}{sec}\right)^{-1}$).
The spatial integration limits are the same as for $Q$ in Eq.\ (\ref{rwQ1D}).
 
For $\Phi_{\Delta x}(\Delta  x,\tau;v_x)$, the probability for a particle
to make a spatial jump of 
length at least $\vert \Delta x\vert$ in the direction 
$\Delta x/\vert \Delta x\vert$ and of duration at least
$\tau$, being though at time $\tau$ exactly at position
$\Delta x$ [see Eq.\ (\ref{rwPhi})], there are two choices of interest. 
First, $\Delta x$
can be independent of the direction of $v_x$, and the jump direction 
is given by 
the pdf of increments $q_{\Delta x}(\Delta x)$, which
is two-sided and includes the sign of $\Delta x$, so that  
$\Phi_{\Delta x}$ can be written as 
\begin{eqnarray}
\Phi_{\Delta x}^{(mag)}(\Delta  x,\tau;v_x) &=& 
\frac{1}{2}\delta(\vert \Delta x\vert - \vert v_x\vert\tau)
                                  \nonumber\\
&\times&\!\!\!\! 
\int\limits_{\vert \Delta x^{\prime}\vert \geq 
             \vert \Delta x\vert}
                                    \!\!\!\!\!\!\!\!
                     d \Delta  x^{\,\prime} 
                         \nonumber \\
&\times&       \int\limits_{\tau^{\prime} \geq \tau}
                          \!\!\!\! d\tau^{\prime}
           \delta\left(\tau^{\prime}-\vert \Delta x^{\,\prime}\vert/\vert v_x\vert\right) \nonumber \\
&&  \ \ \ \ \ \ \ \times  q_{\Delta x}(\Delta x^{\,\prime})  .
\label{rwPhi1Da}
\end{eqnarray}
This form of $\Phi_{\Delta x}^{(mag)}$ is adequate for magnetized plasmas,
where particles cannot travel along straight lines in the
direction of their velocity.
The factor $1/2$ appears since we consider the symmetric 
two-sided distribution
$q_{\Delta x}(\Delta x^{\prime)}$, with positive and negative arguments:
$\Phi_{\Delta x}(\Delta  x,\tau;v_x)$ should express
the probability to make a jump
larger than $\vert \Delta x^\prime\vert$ in the direction of $\Delta x^\prime$, which 
is half of the probability to make a jump larger than 
$\vert\Delta x^\prime\vert$ to either side for a two-sided distribution of 
increments that is assumed to be symmetric.

Second, the jump might be in the direction of the instantaneous
velocity, i.e.\ along $v_x/\vert v_x\vert$. In this case, we 
consider a one-sided pdf of increments 
$q_{\vert \Delta x\vert}(\vert \Delta x^{\,\prime}\vert)$
only for the length $\vert \Delta x^{\,\prime}\vert$ of the jump, 
and we define $\Phi_{\Delta x}^{(free)}$ as
\begin{eqnarray}
\Phi_{\Delta x}^{(free)}(\Delta  x,\tau;v_x) &=& 
\delta(\Delta x - v_x\tau)
                                  \nonumber\\
&\times&\!\!\!\! 
\int\limits_{\vert \Delta x^{\prime}\vert \geq 
             \vert \Delta x\vert}
                                    \!\!\!\!\!\!\!\!
                     d \vert \Delta  x^{\,\prime}\vert 
                         \nonumber \\
&\times&       \int\limits_{\tau^{\prime} \geq \tau}
                          \!\!\!\! d\tau^{\prime}
           \delta\left(\tau^{\prime}-\vert\Delta x^{\,\prime}\vert/\vert v_x\vert \right)    \nonumber \\
&&\ \ \ \ \ \ \ \times   
         \,q_{\vert \Delta x\vert }(\vert \Delta x^{\,\prime}\vert)   .
\label{rwPhi1Db}
\end{eqnarray}
In this case, everywhere $\Delta x = \vert\Delta x\vert \cdot v_x/\vert v_x\vert$ is understood,
i.e.\ also in the expressions for $Q(x,p_x,t)$ and $P(x,p_x,t)$, Eqs.\ 
(\ref{rwQ1D}) and (\ref{rwP1D}), respectively.
This form of $\Phi_{\Delta x}^{(free)}$ is adequate for particles in free 
space or in an unmagnetized plasma, where the particles travel along straight lines in the direction 
of their velocity. This form is also useful in the case where the velocities do not represent
the thermal velocities, but e.g.\ the set of possible drift velocities
in a magnetized plasma.
No factor $1/2$ appears in Eq.\ (\ref{rwPhi1Db}), since the one-sided
distribution of increments 
$q_{\vert \Delta x\vert }(\vert \Delta x^{\,\prime}\vert)$ is considered.

\subsection{The final equations}

To calculate the $t^\prime$-integral in the expression for $Q$, 
Eq.\ (\ref{rwQ1D}),
we note that the delta function implies that 
$t^\prime=t-\vert x-x^\prime\vert/\vert v\vert$, as long as 
$t-\vert x-x^\prime\vert/\vert v\vert$ is in the $t^\prime$-integration range,
which is obviously the case if $\vert x-x^\prime\vert/\vert v\vert\leq t$.
If $x-x^\prime> 0$, then this condition reduces to 
$x^\prime\geq x-\vert v\vert t$, and if $x-x^\prime< 0$ then it must hold that 
$x^\prime\leq x+\vert v\vert t$, which
are just the conditions imposed by the integration limits of the 
$x^\prime$-integral. In the case where the upper integration 
limit is $L$, it again 
holds that  
$x^\prime\leq L\leq x+\vert v\vert t$,
and the like if the lower integration limit is $-L$.
It thus follows that the equation for $Q$ writes as 
\begin{eqnarray}
Q(x,p_x,t) &=& \int d p_x^\prime    
    \int\limits_{\max[x-\vert v_x\vert t,\,-L]}^{\min[x+\vert v_x\vert t,\,L]}
                         d x^\prime  \nonumber \\
&& \ \times\, Q(x^\prime, p_x^\prime,
                  t-\vert x- x^\prime\vert /\vert v_x(p_x)\vert) 
                              \nonumber \\
&& \ \times\, q_{\Delta x}(x- x^\prime)
               \,  q_{\Delta p_x}(p_x-p_x^\prime)
                        \nonumber \\
&+&\delta(t)\, P(x,p_x,t=0) + S(x,p_x,t)   .
\label{rwQ1Db}
\end{eqnarray}

Also in any of the two cases of Eqs.\ (\ref{rwPhi1Da}) and (\ref{rwPhi1Db})
for $\Phi_{\Delta x}^{(.)}(\Delta  x,\tau;v_x)$, the temporal integral is 
trivial, as long 
as $\vert\Delta x^{\,\prime}\vert/\vert v_x\vert$ is in the 
integration range of the $\tau^{\prime}$ integral, i.e.\ 
if $\vert\Delta x^{\,\prime}\vert/\vert v_x\vert\geq \tau$.
This is always the case, since the lower limit of the 
$x^{\,\prime}$-integration implies that
$\vert\Delta x^{\,\prime}\vert/\vert v_x\vert \geq
\vert\Delta x\vert/\vert v_x\vert$
and the delta function in front of the 
integrals ensures the relation $\vert\Delta x\vert/\vert v_x\vert
=\tau$, so that the inequality
$\vert\Delta x^{\,\prime}\vert/\vert v_x\vert
\geq \tau$ follows.
Eq.\ (\ref{rwPhi1Da}) thus turns to
\begin{eqnarray}
\Phi_{\Delta x}^{(mag)}(\Delta  x,\tau;v_x) &=& 
\frac{1}{2}
\delta(\vert \Delta x\vert - \vert v_x\vert\tau)
                                  \nonumber\\
&\times&\!\!\!\! 
\int\limits_{\vert \Delta x^{\prime}\vert \geq 
             \vert \Delta x\vert}
                                    \!\!\!\!\!\!\!\!
                     d \Delta  x^{\,\prime} \,
                     q_{\Delta x}(\Delta x^{\,\prime})    ,
\label{rwPhi1Dc}
\end{eqnarray}
and Eq.\ (\ref{rwPhi1Db}) becomes
\begin{eqnarray}
\Phi_{\Delta x}^{(free)}(\Delta  x,\tau;v_x) &=& 
\delta(\Delta x - v_x\tau)
                                  \nonumber\\
&\times&\!\!\!\! 
\int\limits_{\vert \Delta x^{\prime}\vert \geq 
             \vert \Delta x\vert}
                                    \!\!\!\!\!\!\!\!
                     d \vert \Delta  x^{\,\prime}\vert 
%                         \nonumber \\
%&&\ \ \ \ \ \ \ \times   
         \,q_{\vert \Delta x\vert }(\vert \Delta x^{\,\prime}\vert)  .
\label{rw1d7}
\end{eqnarray}

In the applications presented below, we will use the variant 
$\Phi^{(mag)}_{\Delta x}$ for the case of a magnetized plasma, we thus 
insert $\Phi^{(mag)}_{\Delta x}$ from Eq.\ (\ref{rwPhi1Dc}) into 
the expression for $P$ [Eq.\ (\ref{rwP1D})], which 
yields
\begin{eqnarray}
P(x,p_x,t) &=& 
    \int\!\! d p_x^{\prime}   
     \int\limits_{\max[x-\vert v_x\vert t,\,-L]}^{\min[x+\vert v_x\vert t,\,L]}
  \!\!\!\!\!\!\!\! d x^{\prime}
      \int\limits_0^t\!\! dt^\prime 
                                                   \nonumber \\
  &&\ \ \ \ \ \times Q(x^{\prime},
                     p_x^\prime,t^\prime) 
                                                        \nonumber\\
  &&\ \ \ \ \ \times 
\delta\left(\vert x-  x^\prime\vert - \vert v_x\vert (t-t^\prime)\right)
                                  \nonumber\\
&\times&\!\!\!\! \frac{1}{2}
\int\limits_{\vert  x^{\prime\prime}\vert \geq 
             \vert x -  x^\prime\vert}
                                    \!\!\!\!\!\!\!\!
                     d   x^{\,\prime\prime} \,
                     q_{\Delta x}( x^{\,\prime\prime})
                                                        \nonumber\\
  &&\ \ \ \ \ \times    q_{\Delta p_x}(p_x- p_x^{\prime}) .
\label{rwP1Db}
\end{eqnarray}
On doing the $t^\prime$-integration, the delta function would 
impose that $t^\prime=t-\vert x-x^\prime\vert/\vert v\vert$, if this 
is contained in the $t^\prime$-integration range. Obviously,
$t-\vert x-x^\prime\vert/\vert v\vert\leq t$, since a positive term is 
subtracted from $t$. On the other hand, 
$t-\vert x-x^\prime\vert/\vert v\vert\geq 0$ if  
$\vert x-x^\prime\vert/\vert v\vert\leq t$, which is though exactly the 
condition imposed by the $x^\prime$-integration limits.
The $t^\prime$ integration is thus trivial and can be done e.g.\ 
by substituting $\bar t:=\vert v_x\vert t^\prime$, which yields
\begin{eqnarray}
P(x,p_x,t) &=& 
    \int\!\! d  p_x^{\prime}   
     \int\limits_{\max[x-\vert v_x\vert t,\,-L]}^{\min[x+\vert v_x\vert t,\,L]}
  \!\!\!\!\!\!\!\! d x^{\prime} \,
         \frac{1}{\vert v_x\vert  }
                                                   \nonumber \\
  &&\ \ \ \ \ \times Q( x^{\,\prime},
                     p_x^\prime,t-\vert x-  x^\prime\vert/\vert v_x\vert) 
                                                        \nonumber\\
&\times&\!\!\!\! \frac{1}{2}
\int\limits_{\vert  x^{\prime\prime}\vert \geq 
             \vert x -  x^\prime\vert}
                                    \!\!\!\!\!\!\!\!
                     d   x^{\,\prime\prime} \,
                     q_{\Delta x}( x^{\,\prime\prime})
                                                        \nonumber\\
  &&\ \ \ \ \ \times    q_{\Delta p_x}(p_x-p_x^{\prime}) .
\label{rwP1Dc}
\end{eqnarray}
Last, we introduce the abbreviation 
\beq
\Psi_{\Delta x}(\Delta x):=\frac{1}{2}
\int\limits_{\vert  x^{\prime\prime}\vert \geq 
             \vert \Delta x \vert}
                                    \!\!\!\!\!\!\!\!
                     d   x^{\,\prime\prime} \,
                     q_{\Delta x}( x^{\,\prime\prime})  ,
\eeq
so that the propagator equation takes the final form
\begin{eqnarray}
P(x,p_x,t) &=& 
    \int\!\! d  p_x^{\prime}   
     \int\limits_{\max[x-\vert v_x\vert t,\,-L]}^{\min[x+\vert v_x\vert t,\,L]}
  \!\!\!\!\!\!\!\! d x^{\prime} \,
         \frac{1}{\vert v_x\vert  }
                                                   \nonumber \\
  &&\ \ \ \ \ \times Q( x^{\,\prime},
                     p_x^\prime,t-\vert x-  x^\prime\vert/\vert v_x\vert) 
                                                        \nonumber\\
&\times&\!\!\!\!  
             \Psi_{\Delta x}(\vert x -  x^\prime\vert)
            \, q_{\Delta p_x}(p_x-p_x^{\prime}) .
\label{rwP1Dd}
\end{eqnarray}

The marginal distributions for the density and the momentum, as well
as the kinetic energy distribution and the temperature profile 
are determined completely analogous to the way described in Sec.\ \ref{margi},
for the case of here only one spatial and one momentum coordinate.
The only modification we make is that the one-dimensional version
of the relation of the temperature to the mean kinetic energy
is used, $\frac{1}{2}k_BT= \langle E_{kin}\rangle$.

\subsection{Connection to electric fields}

In a plasma, acceleration events are usually connected to the appearance
of electric fields, and we consider the increments in momentum $\Delta p_x$ 
to be caused by 
electric fields, assuming that  
\beq
\Delta p_x=eE_x\langle t_{acc}\rangle  ,
\label{rw1d12}
\eeq
with $e$ the particle charge, $E_x$ the electric field, and $\langle t_{acc}\rangle$
an acceleration time assumed to be constant (in this way, 
the statistics of the 
duration of the acceleration process is 
absorbed in the statistics of the effective electric field).
Since $q_{\Delta p_x}(\Delta p_x)d\Delta p_x=q_{E_x}(E_x)dE$ must hold, the connection 
of the pdf of $\Delta p_x$ to the pdf of the electric fields $q_{E_x}(E_x)dE$ 
is given by 
\beq
q_{\Delta p_x}(\Delta p_x)
=q_{E_x}(E_x)\frac{dE_x}{d\Delta p_x}
= q_{E_x}\left(\frac{\Delta p_x}{e\langle t_{acc}\rangle}\right)
\frac{1}{e\langle t_{acc}\rangle}  .
\label{rw1d13}
\eeq

\subsection{Numerical solution of the CTRW equations}

We solve the system of Eqs.\ (\ref{rwQ1Db}) and (\ref{rwP1Dd}) numerically
with a pseudospectral method based on Chebyshev polynomials. In this 
method, the unknown functions $Q(x,p_x,t)$ and $P(x,p_x,t)$ are expanded in terms of 
Chebyshev polynomials
in the $x$-, $p_x$-, and $t$-direction, and in case where the integral equations
are linear (which basically implies that the distributions of increments $q_{\Delta x}$
and $q_{\Delta p_x}$ do not contain information on $P(x,p_x,t)$), 
the integral equations turn into a system of linear algebraic equations, 
%for the expansion coefficients,
which can be solved with a standard linear system 
solver. In App.\ \ref{AppA}, a more detailed description of the numerical
method is given.

The critical gradient model, which will be 
introduced below 
(Sect.\ \ref{cragra}), is non-linear, the spatial jump increments contain 
information on the density 
gradient, so that in principle the pseudospectral method is applicable
also to this case, it yields though a system of non-linear algebraic equations,
for which we have not yet implemented a numerical method to solve it
(see the more technical explanations in App.\ \ref{AppA}),
so that the results presented here for the critical gradient model are all 
derived
with Monte Carlo simulations. 

\subsection{Monte Carlo simulations}

In parallel to the numerical solution of the CTRW equations, 
Monte Carlo simulations are performed, in
two different versions. 
In the first version, an individual particle has no information
during its evolution on the rest of the particles. This version is
very fast in what computing time is concerned, and it is
used to verify results of the mixed model (introduced in Sect.\ \ref{mimo}) 
and to determine the
mean velocity profile in Sect.\ \ref{pahef}.
In the second version, an individual particle knows where
all the other particles are during its evolution, and it is 
influenced by local statistical properties, such as the density
gradient. This version has to be used
for the critical gradient model (introduced in Sect.\ \ref{cragra}), 
and it is very demanding in computing
time, so that statistically sensitive quantities such as the
mean velocity profile, and therewith the particle and heat flux (see Sect.\ \ref{pahef}),
could not be determined reliably enough and are not presented in this 
article.

\section{Application \label{SecIII}}

\subsection{overview}

We consider a plasma in a toroidal confinement device such as a tokamak, and 
our purpose is to study the particle and heat transport phenomena 
in the direction of the minor radius of the torus, perpendicular to the 
confining magnetic field. This
direction is of major interest in tokamaks, since heat or particle losses
in this direction determine the quality of the achieved confinement.

\subsection{Parameters and distribution of increments}
\label{params}

We assume a finite spatial range $[-L,L]$, with $L=200\,$cm.
For numerical reasons, we have to truncate also momentum space, 
we usually assume the largest positive momentum to be $p_2=10\,p_{th}$,
except in the case of pure power-law increments in $p_x$, where we use 
$p_2=50\,p_{th}$, with 
$p_{th} := \sqrt{m k_B T}$ the 
thermal momentum.
%smallest positive momentum $p_s=10^{-5}\,p_{th}$.
In all applications below, the system initially is empty, $P(x,p_x,t=0)=0$,
there is only a source term active that is constant in time.
The momentum distribution of the particles at injection into the system 
is throughout a Gaussian distribution, with plasma temperature 
$k_BT = 8\,$keV 
for the uniform
background source, so that 
$p_{th}=3.4\times 10^{-18}\,$g cm/s ($v_{th}=3.8\times 10^{9}\,$cm/s) for electrons.

The distribution of increments we use below are a combination of two
basic distributions,
one which allows only small increments 
compared to the system size, and one which allows also large increments, of the
order of system size or even larger. For the former, a natural choice are Gaussian 
distributions,
\beq
q_{\Delta x}^{(Gauss)}(\Delta x) = \frac{1}{\sqrt{2\pi}\sigma_{\Delta x}}
              e^{-\frac{\Delta x^2}{2 \sigma_{\Delta x}^2}}
\eeq
for position increments, and
\beq
q_{\Delta p}^{(Gauss)}(\Delta p) = \frac{1}{\sqrt{2\pi}\sigma_{\Delta p}}
              e^{-\frac{\Delta p^2}{2 \sigma_{\Delta p}^2}}
\eeq
for momentum increments,
and we 
usually use $\sigma_{\Delta x}=L/40=5\,$cm, and 
$\sigma_{\Delta p} = p_{th}/40 = 8.5 \times10^{-20}\,$ g cm/s, respectively.

As distribution for large increments, 
we use Levy-like distributions with power-law tails, 
\beq
q_{\Delta x}^{(pl)}(\Delta x) = 
\left\{
\begin{array}{ll}
A \Delta \vert x\vert ^{-\alpha}, &  {\rm if }\ \left\vert \Delta x\right\vert \geq \Delta x_1,  \\
A \Delta \vert x_1\vert ^{-\alpha},     &  {\rm if }\ \left\vert \Delta x\right\vert < \Delta x_1
\end{array}
\right.
\eeq
in position space, with $x_1=L/200=1\,$cm and a standard value for the power-law index 
$\alpha = 1.2$, as it corresponds to a random 
walk through a fractal environment with fractal dimension $D_F=1.8$ (in 3-D space,
$\alpha=3-D_F$, see \cite{Isliker03}).
In momentum space, we analogously choose
\beq
q_{\Delta p}^{(pl)}(\Delta p) = 
\left\{
\begin{array}{ll}
B \vert \Delta p\vert ^{-\beta}, &  {\rm if }\ \left\vert \Delta p\right\vert \geq \Delta p_1,  \\
B \vert \Delta p_1\vert ^{-\beta},     &  {\rm if }\ \left\vert \Delta p\right\vert < \Delta p_1,
\end{array}
\right.
\eeq
with $p_1=0.1\,p_{th}$ and power-law index $\beta =2.5$, chosen such that there is not 
unrealistically strong heating of the plasma.

We just note that for the purpose of studying non-local effects on transport, 
one could, instead of power-law distributed increments, also apply
Gaussian distributed increments that are large, where large is understood in the sense of being 
comparable to the system-size.

The CTRW approach has the possibility to model a wide variety of anomalous 
transport phenomena, whose concrete form is determined by the 
the choice of the distributions of increments. In order the random walk model
to become relevant for toroidally confined plasmas, 
the increment 
distributions must incorporate some essential properties of the physical 
system. Here, we make two different choices for the distribution of position
increments.

\subsubsection{The mixed model\label{mimo}}

First, we implement the observation that
anomalous diffusion is more active to-wards the edges of the plasma, whereas the center 
exhibits
more normal diffusive behaviour, the plasma is better confined near the center 
(see the remark in the following Sec.\ \ref{cragra}).
We thus let in the mixed model the distribution of position increments
be spatially dependent on $x$ such that it is a weighted mixture of a power-law and a Gaussian 
distribution,
\bea
q_{\Delta x}^{(mixed)}(\Delta x,x) &=& 
f_1(x)\,q_{\Delta x}^{(Gauss)}(\Delta x) \nonumber \\
&&+ (1-f_1(x))\,q_{\Delta x}^{(pl)}(\Delta x) .
\eea
The function $f_1(x)$ equals one in the center and falls off linearly to-wards the edges,
\beq
f_1(x) = 
\left\{
\begin{array}{ll}
1, &  {\rm if }\ \left\vert x\right\vert < x_f  \\
1-\epsilon \frac{\vert x \vert - x_f}{L-x_f},     &  {\rm if }\ \left\vert x \right\vert \geq x_f 
\end{array}
\right.   ,
\eeq
where we use the basically arbitrary values $x_f=20\,$cm and $\epsilon=0.2$, which are chosen 
by trial and error in order to achieve a good confinement of the particles in the system.

\begin{figure*}
\resizebox{18truecm}{!}{\includegraphics{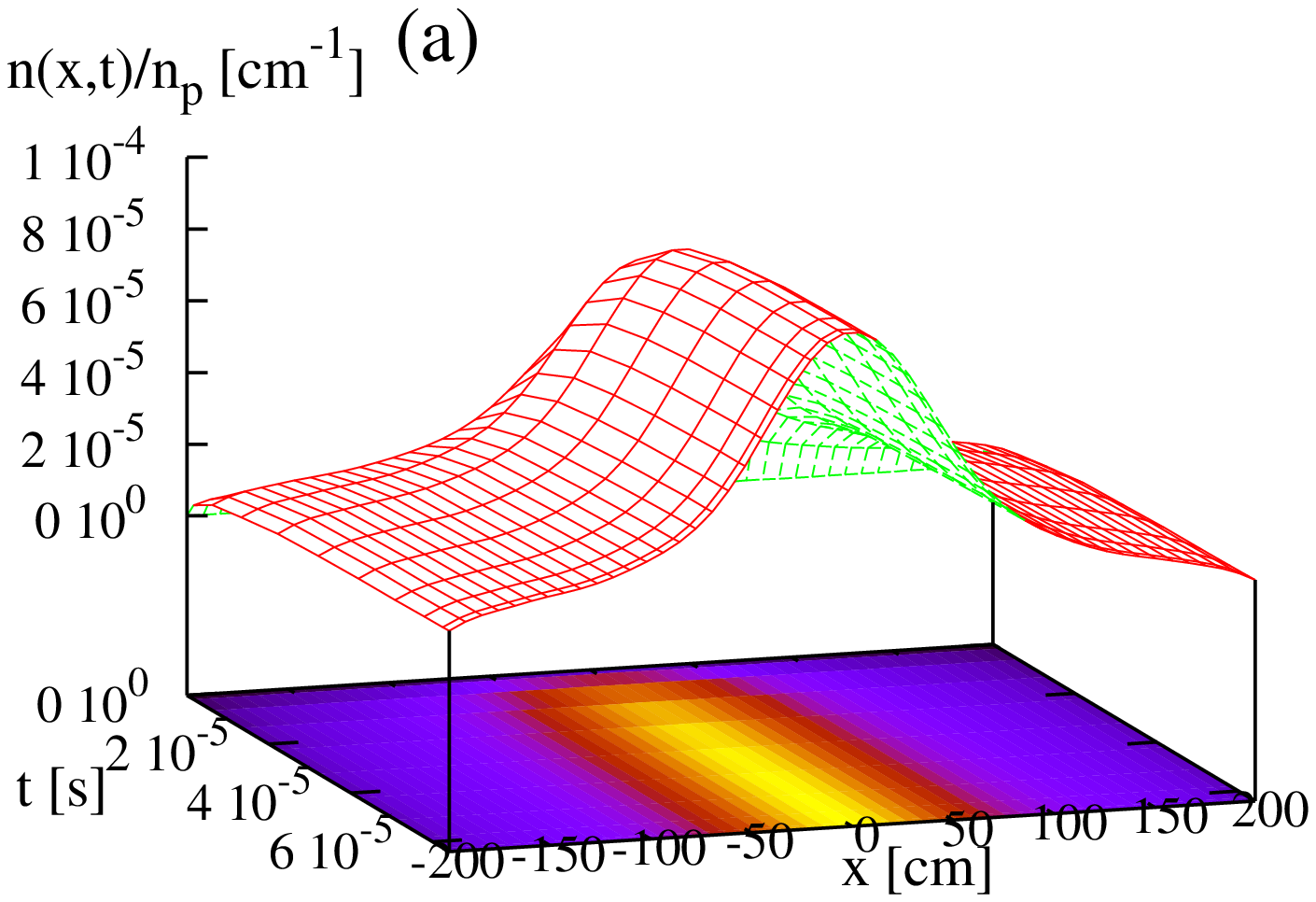},\includegraphics{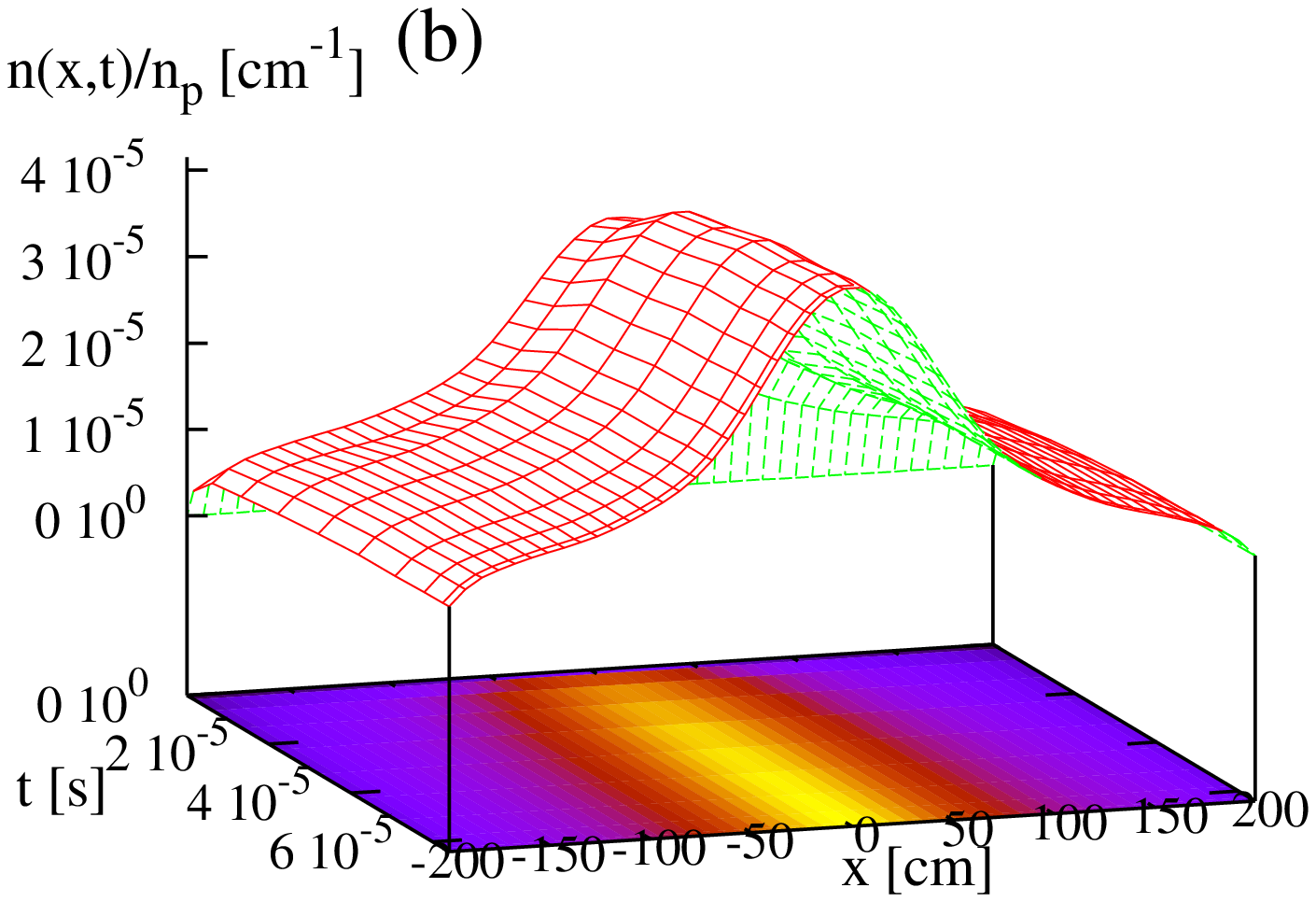}}
%\resizebox{18truecm}{!}{\includegraphics{fq11.ps},\includegraphics{fc125.ps}}
%\resizebox{9truecm}{!}{\includegraphics{fq14.ps}}
%\resizebox{18truecm}{!}{\includegraphics{fq13.ps},\includegraphics{fc126.ps}}
%%\resizebox{9truecm}{!}{\includegraphics{fq114.ps}}
%\resizebox{9truecm}{!}{\includegraphics{fq1134.ps}}
\caption{{\it Uniform injection in space and time:} Number density $n(x,t)/n_p$ as 
a function of space $x$ and time $t$, for the mixed model with 
Gaussian (a) and with power-law (b) distributed momentum increments.
\label{fq11}
}
\end{figure*}

\subsubsection{The critical gradient model\label{cragra}}

In the second choice for the distribution of position increments we follow 
the critical
gradient model, suggested originally by \cite{Imbeau01} (see also 
\cite{Garbet04b}) in the frame of classical diffusion 
(Fick's or Fourier's law), making the basic assumption that
the heat (particle) diffusivity depends critically on the density or 
temperature gradient, in the sense that transport becomes efficient only
if the respective gradient exceeds a threshold. A variant of the critical gradient model 
was implemented in the frame of CTRW by \cite{vanMil04}, \cite{vanMil04b}, with 
the distribution of position increments depending on the local density gradient. We
thus set
\beq
q_{\Delta x}^{(crit)}(\Delta x) = 
\left\{
\begin{array}{ll}
q_{\Delta x}^{(Gauss)}(\Delta x), &  {\rm if }\ \left\vert \frac{dn(x,t)}{dx}\right\vert < c_{cr},  \\
q_{\Delta x}^{(pl)}(\Delta x),     &  {\rm if }\ \left\vert \frac{dn(x,t)}{dx}\right\vert \geq c_{cr},
\end{array}
\right.
\eeq
with $n(x,t)$ the particle density and $c_{cr}$ the threshold,
so that the distribution of increments
is a power-law in regions where a large density gradient has developed, and it is Gaussian in 
regions where the gradients are small. 

Remarks: (i) The general CTRW formalism we introduced would also allow 
to use different criticality conditions, the distribution of position 
increments could e.g.\ 
be made critically dependent on the temperature gradient, or on a combination
of the density and the temperature gradient, and also in the distribution of 
momentum increments a critical dependence could be introduced. 
%As \cite{...} note, 
(ii) The critical gradient model in Refs.\ \cite{vanMil04}, \cite{vanMil04b} 
is different from the critical gradient model used here in that waiting times and not free
flight times are used in
Refs.\ \cite{vanMil04}, \cite{vanMil04b}, which moreover are assumed 
to always follow an exponential distribution, and of course momentum space
is not included. 
(iii) For symmetry reasons, one would expect that in the critical gradient 
model the density gradient is
 small at the center and becomes larger to-wards the edges, 
so that transport is more normal near the center and more anomalous 
to-wards the edges, which is just the scenario we explicitly
implement in the mixed model. 
(iv) The density gradient 
dependence of the position increments 
introduces a non-linearity into the integral 
equations of the CTRW, %for the critical gradient model, 
whereas the equations of the mixed model are linear in $P(x,p_x,t)$.

\subsubsection{The momentum increments}

As distribution of momentum increments we use either a pure Gaussian
distribution or a pure power-law distribution. These two cases 
correspond to low and high level activity (heating) in momentum space,
respectively,
and they allow us to explore the role of momentum space dynamics
for the system evolution
in the two extreme cases of interest, the almost local and the non-local
momentum transport, respectively.

\subsection{Uniform injection in space and time}

\begin{figure}
\resizebox{9truecm}{!}{\includegraphics{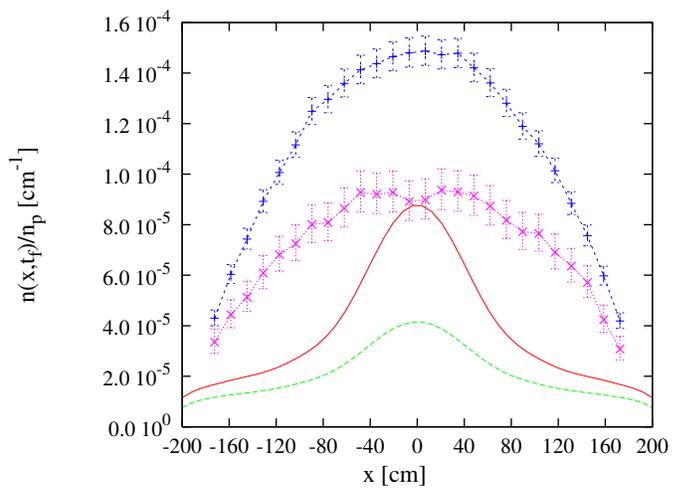}}
\caption{{\it Uniform injection in space and time:} Number density $n(x,t)/n_p$  as 
a function of space $x$ at final time $t=t_f$,  for the 
mixed model with Gaussian (solid - red) and with 
power-law (long dashes - green) distributed momentum increments, and for the 
critical gradient model 
with Gaussian (short dashes - blue, with error-bars, and scaled with a factor $1/3$ for 
better visualization) and  with 
power-law (dotted - violet, with error-bars) distributed momentum increments.
\label{fq11nn}
}
\end{figure}

In the first application, we consider the case where the plasma is uniformly 
and continuously injected into the system, 
with the initial position 
and injection time of the particles uniformly distributed in the 
spatial interval $[-L,L]$ and in the 
time interval 
$[0,t_f]$, respectively, with $t_f$ the final time up to which
the system is monitored. The initial momentum follows 
a thermal distribution with a fixed temperature of $8\,$keV. The source 
function thus
takes the form
\beq
S(x,p_x,t) = \frac{1}{\sqrt{2\pi}\sigma_p} 
\,e^{-\frac{p_x^2}{2\,\sigma_p^2}}
\,\frac{1}{2L} ,
\label{unifs}
\eeq
with $\sigma_p=p_{th}\equiv\sqrt{mk_BT}$, the thermal momentum. 

In Fig.\ \ref{fq11}, the density 
profiles as a function of time, normalized to the total number of injected
particles $n_p$, are shown for the mixed model
with Gaussian and power-law distributed 
increments in momentum,
respectively. The initial density is zero, according to the chosen set-up, 
and the 
system evolves to a stationary dynamic equilibrium state. 
Fig.\ \ref{fq11nn} shows the density profiles at the final time $t=t_f$, 
including now also the critical gradient model with 
again Gaussian and power-law distributed momentum increments, respectively.
In the cases of the mixed model and the critical gradient model with 
Gaussian momentum
increments, density profiles are formed that are peaked at the 
center --- within the statistical error for the critical gradient model ---, 
with the
characteristic difference that in the mixed model the density profile has 
flatter
wings to-wards the edges, the particles are more concentrated at the center. 
The critical gradient model with power-law distributed momentum increments
exhibits a broad plateau in the central region.
The highest
density, and thus best particle confinement, is achieved by the critical 
gradient
model with Gaussian distributed momentum increments (note that the density
profile in this case is scaled by a factor of $1/3$ in Fig.\ \ref{fq11nn}). 
In both the mixed and the critical gradient
model, respectively, the particle confinement 
deteriorates when power-law distributed momentum increments are used, the increased 
energy input in acceleration events leads to a faster loss of particles.

\begin{figure}
\resizebox{9truecm}{!}{\includegraphics{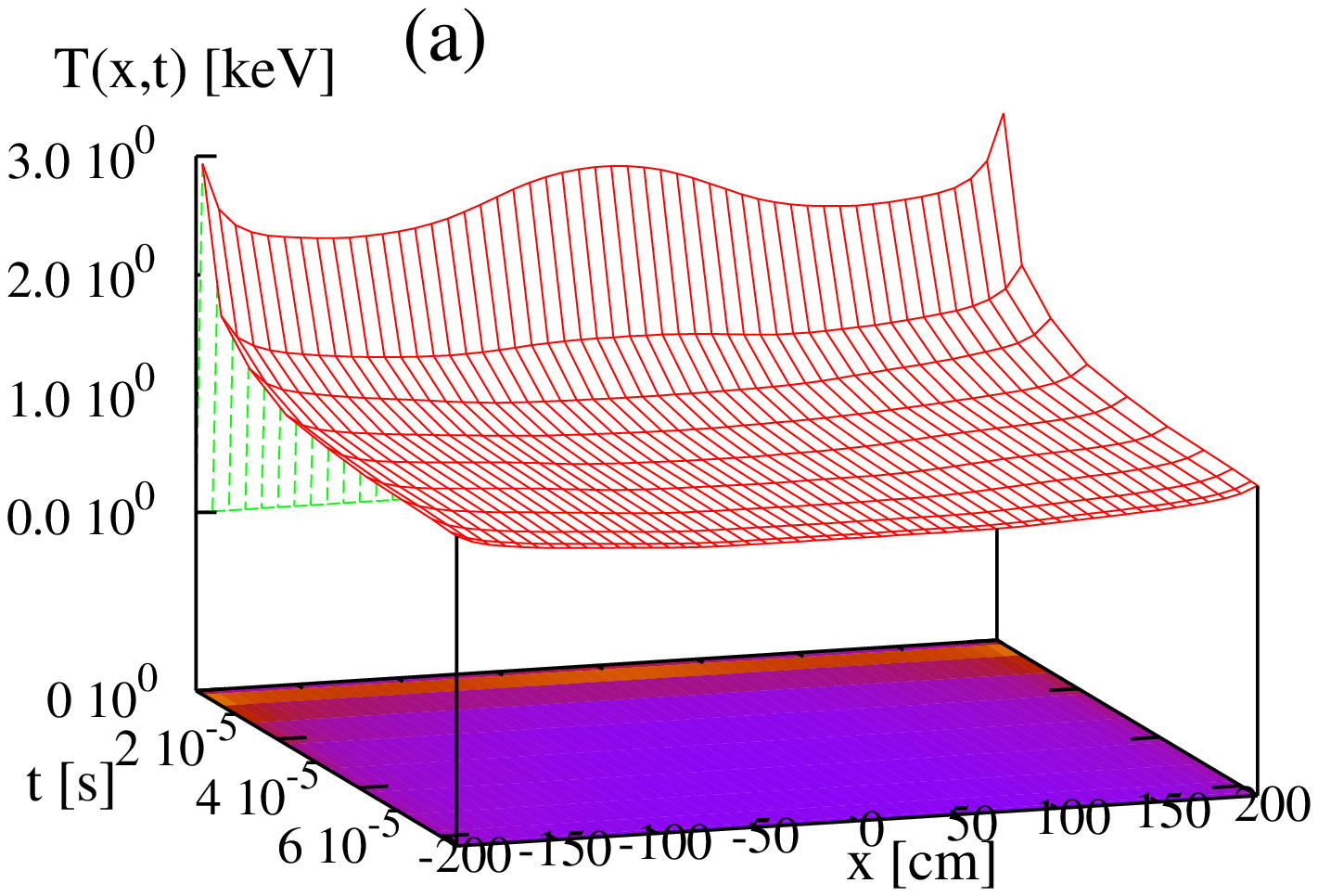}}
%\resizebox{18truecm}{!}{\includegraphics{fq11T.ps},\includegraphics{fc125T.ps}}
%%%\resizebox{9truecm}{!}{\includegraphics{fq14T.ps}}
\resizebox{9truecm}{!}{\includegraphics{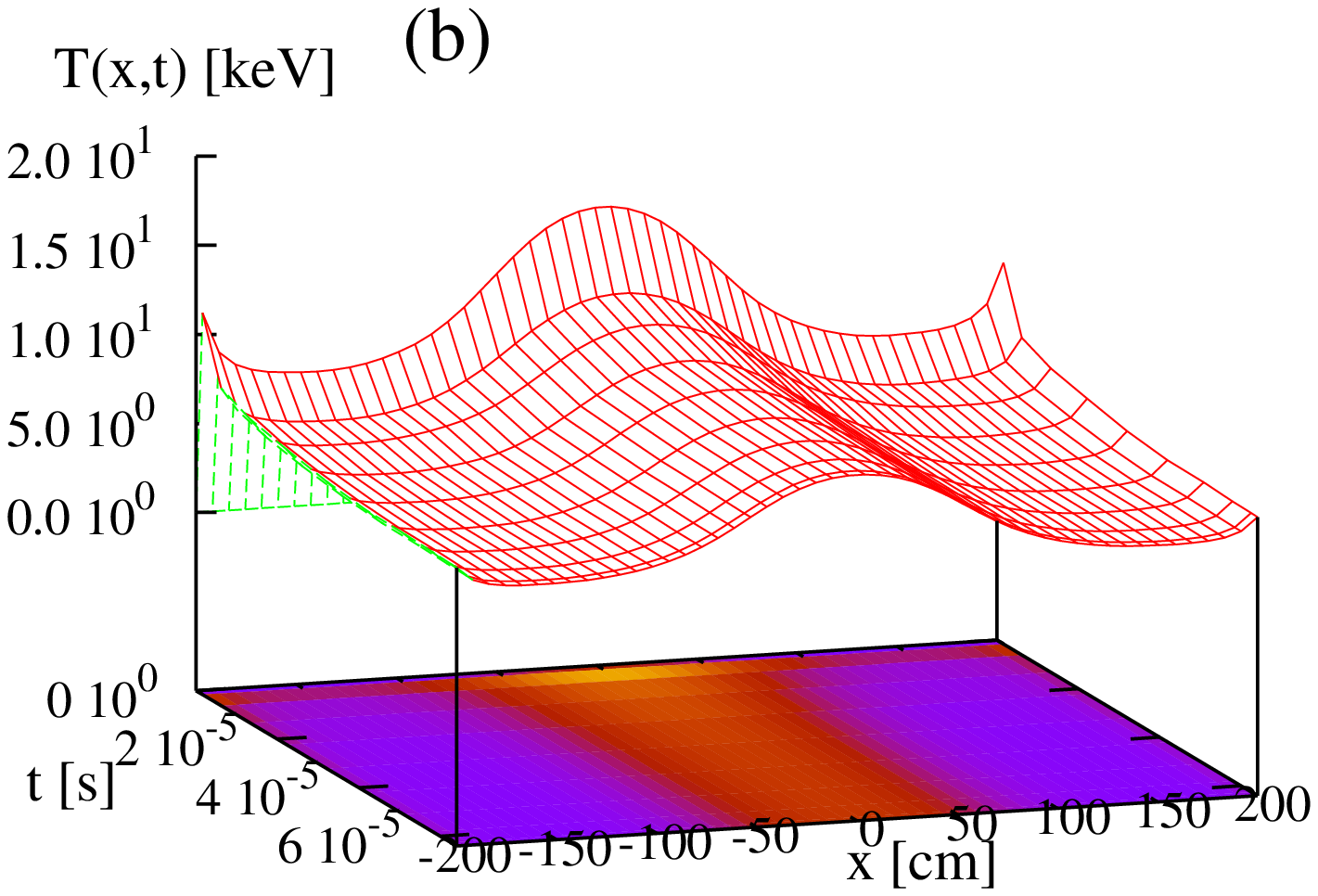}}
%\resizebox{18truecm}{!}{\includegraphics{fq13T.ps},\includegraphics{fc126T.ps}}
%%%\resizebox{9truecm}{!}{\includegraphics{fq14T.ps},\includegraphics{fq13T.ps}}
%%%\resizebox{9truecm}{!}{\includegraphics{fq114T.ps}}
%%%\resizebox{9truecm}{!}{\includegraphics{fq1134T.ps}}
\caption{{\it Uniform injection in space and time:}
Temperature distribution $T(x,t)$ 
in case of the mixed model for Gaussian (a) and power-law (b) distributed increments 
in momentum.
\label{fq1T}}
\end{figure}

Fig.\ \ref{fq1T} shows the spatial temperature profiles as a function
of time, for the mixed model only, and for Gaussian and power-law distributed 
momentum increments. (Temperature distributions from Monte Carlo simulations
have a much larger statistical error than the respective density distributions, so that
temperature profiles for the critical gradient
will only be shown below for the case of strong off-axis heating, where 
a substantially larger number of particles has been used.) 
For the mixed model then, the temperature profile is basically flat when
Gaussian momentum increments are used, with a small rise to-wards the edges.
With power-law momentum increments, the temperature profile is clearly 
peaked at the center, with again a small rise to-wards the edges. This small rise  
is not a numerical artifact but a property of the model, and it  
appears also in Monte-Carlo simulations.
It thus follows that low level energy input into the system leads to flat temperature
profiles, whereas intense heating naturally leads to temperature profiles that are 
peaked at the center, which is reminiscent of profile consistency. Due to the increased
energy injection in the case of power-law distributed momentum increments, the temperature
reached in this case is higher than the one in the Gaussian case, where the temperature
is even below the particles' injection temperature,
the high energy 
particles are very quickly lost from the system, and the bulk of particles that 
stays in the system is of low energy, on the average.

Common to the temperature profiles is that very soon after the
start of the simulation, the temperature assumes its highest values, 
and decays then to lower values until a stationary state is reached. 
This initial rise is not resolved in Fig.\ \ref{fq1T} and is seen just as 
an initial step from zero temperature (there is no initial population
of particles) to its peak 
value at a very early time. 
This behaviour is a consequence of our specific set-up 
of the random walk: When a particle is injected, it first performs a step
in momentum space, which on the average corresponds to heating. At very small
times, the particles did not have enough time yet to leave the system, except for those
very close to the edge, so that 
the heated population of particles is accumulated until time is large enough 
so that particles start to leave. This can also be seen from the relation between
the kinetic energy distribution $E_{kin}(x,t)$ and the temperature distribution,
$E_{kin}(x,t) = 1/2k_BT(x,t) n(x,t)$, so that 
$k_B T(x,t) =  2 E_{kin}(x,t) /n(x,t)$ holds. Both $E_{kin}(x,t)$ and $n(x,t)$ gradually 
increase initially from zero,
$E_{kin}(x,t)$ increases though faster than $n(x,t)$, which leads to a peak in temperature in 
the early phase where the number
of particles in the system is still very low (and the denominator $n(x,t)$ is small).

\begin{figure}
\resizebox{9truecm}{!}{\includegraphics{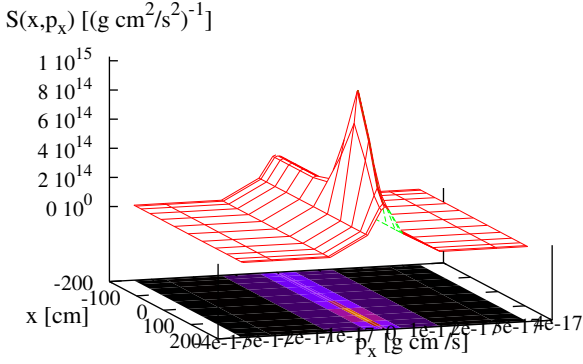}}
\caption{{\it Weak off-axis source:} 
Source function $S(x,p_x)$ for weak spatially 
localized off-axis loading with colder particles (at $x=100$) on top of uniform spatial loading.
\label{fq3bs}
}
\end{figure}

\subsection{Localized off-center loading}

In a second application, we consider the case of off-axis injection,
where two sources are active, a uniform one in the entire position space, 
and one spatially localized off-axis,
with a temperature lower by a factor of $1/10$ than the one of the 
uniform background source. The source function takes the form 
\bea
S(x,p_x,t) &=& 
\frac{1}{(1+\delta)\sqrt{2\pi}\sigma_p} 
\,e^{-\frac{p_x^2}{2\,\sigma_p^2}}
\,\frac{1}{2L} \nonumber \\
&+&
 \frac{\delta}{(1+\delta)\sqrt{2\pi}s_p} 
\,e^{-\frac{p_x^2}{2\,s_p^2}}\,
\frac{1}{\sqrt{2\pi}\sigma_r} 
\,e^{-\frac{(x-x_c)^2}{2\,\sigma_r^2}}  ,
\label{offax}
\eea
with $\sigma_p=\sqrt{mk_BT}$, $s_p=\sqrt{0.1\,mk_BT}$, $\sigma_r=0.2 L=40\,$cm, $x_c=100\,$cm,
and $k_BT=8\,$keV,
and where for both sources the injection is uniform in time. The relative strength
of the off-axis source is expressed by the factor $\delta$.
In the following, we consider two cases, the case of a weak and of 
a strong off-axis source, respectively.

\subsubsection{Weak off-axis fueling}

First, we consider a source that is relatively weak in comparison to the background source,
with $\delta=0.2$ in Eq.\ (\ref{offax}), so that the fraction 
of particles injected off-axis is $0.17$. 
The source function is shown in Fig.\ \ref{fq3bs}.

\begin{figure*}%[p]
\resizebox{18truecm}{!}{\includegraphics{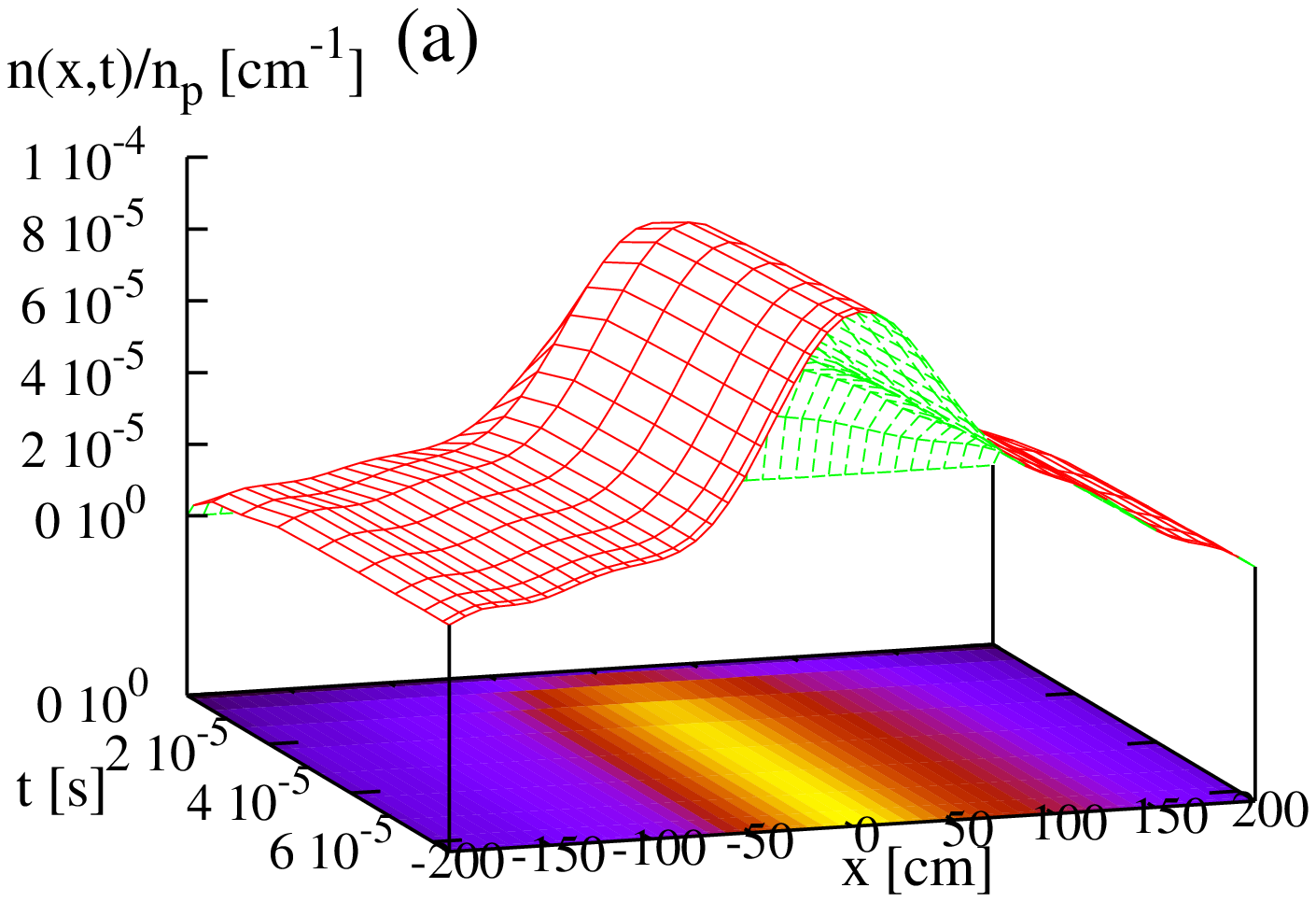},\includegraphics{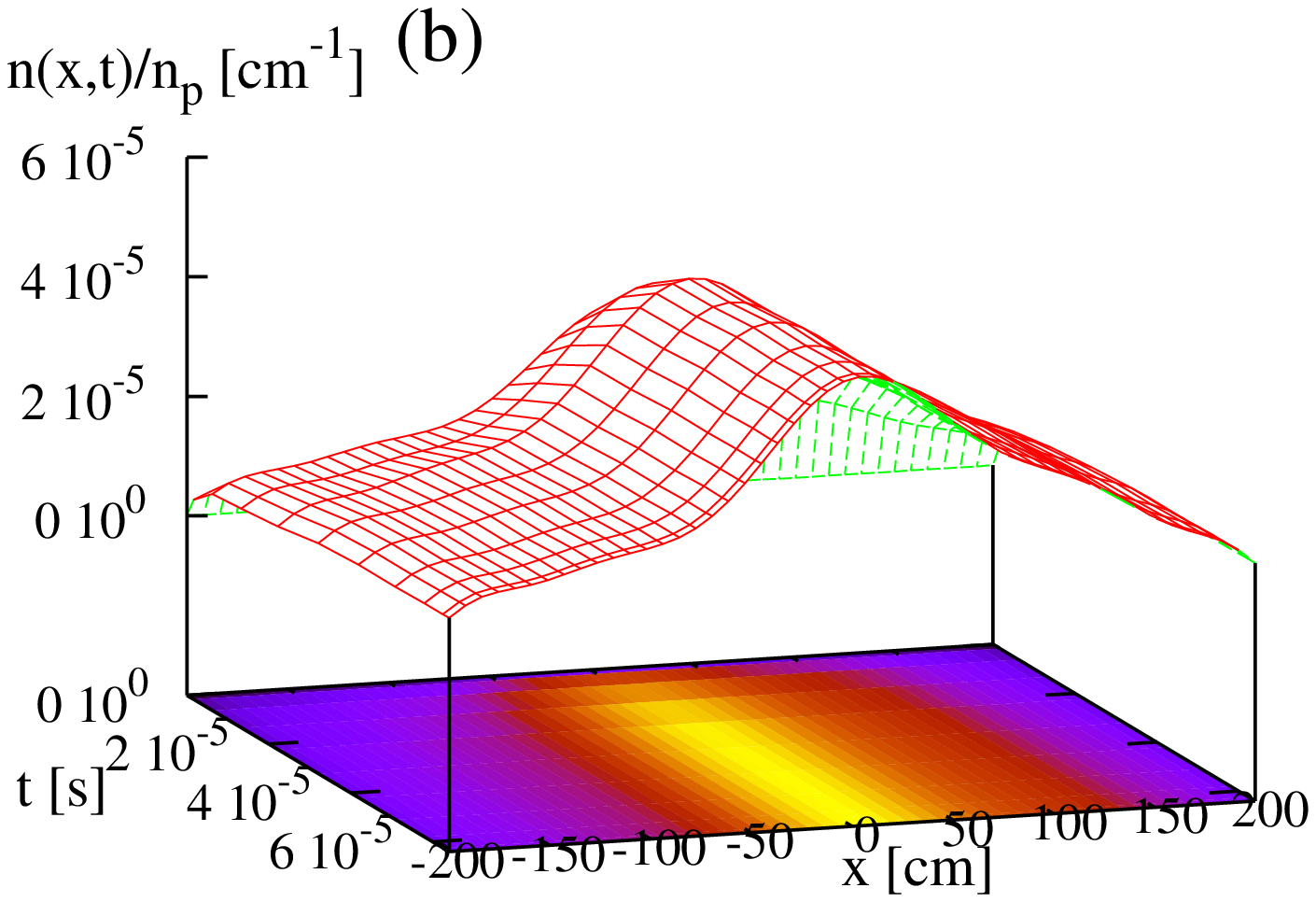}}
%\resizebox{18truecm}{!}{\includegraphics{fq3b1.ps},\includegraphics{fc132.ps}}
%\resizebox{9truecm}{!}{\includegraphics{fq3b4.ps}}
%\resizebox{18truecm}{!}{\includegraphics{fq3b3.ps},\includegraphics{fc133.ps}}
%%\resizebox{9truecm}{!}{\includegraphics{fq314.ps}}
%\resizebox{9truecm}{!}{\includegraphics{fq3b134.ps}}
\caption{{\it Weak off-axis source:}
Number density $n(x,t)/n_p$ as 
a function of space $x$ and time $t$, for the mixed model with 
Gaussian (a) and with power-law (b) distributed momentum increments.
\label{fq3b1}}
\end{figure*}

\begin{figure}%[p]
\resizebox{9truecm}{!}{\includegraphics{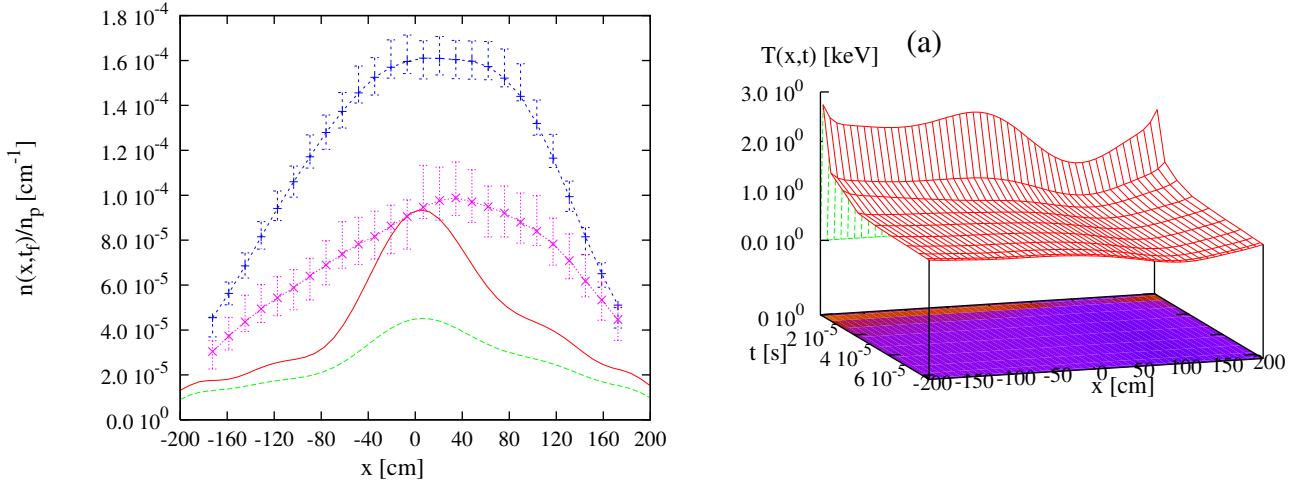}}
\caption{{\it Weak off-axis source:}
Number density $n(x,t)/n_p$  as 
a function of space $x$ at final time $t=t_f$,  for the 
mixed model with Gaussian (solid - red) and with 
power-law (long dashes - green) distributed momentum increments, and for the 
critical gradient model 
with Gaussian (short dashes - blue, with error-bars, and scaled with a factor $1/3$ for 
better visualization) and  with 
power-law (dotted - violet, with error-bars) distributed momentum increments.
\label{fq3b1nn}}
\end{figure}

In Fig.\ \ref{fq3b1}, the density profiles are shown as a function of time 
for the mixed model,
and Fig.\ \ref{fq3b1nn} presents the density profiles for the mixed
and the critical gradient model at final time. In case 
of the mixed model, the densities are still peaked at the center, with a slightly fatter
wing to-wards the side of the off-axis source, the picture remains similar to the one
of pure uniform injection, for both Gaussian and power-law distributed 
momentum increments.
The critical gradient model though shows now density peaks that are slightly 
off-center, more pronounced in case of power-law momentum increments, which
also has developed a peak now. 
Again, the critical gradient model with Gaussian momentum 
increments shows the highest density
and thus the best confinement, the mixed model, on the other hand, 
keeps the profiles more unaffected by the off-axis source and thus
shows a higher stiffness.

\begin{figure}
%\resizebox{18truecm}{!}{\includegraphics{fq3b1T.ps},\includegraphics{fc132T.ps}}
\resizebox{9truecm}{!}{\includegraphics{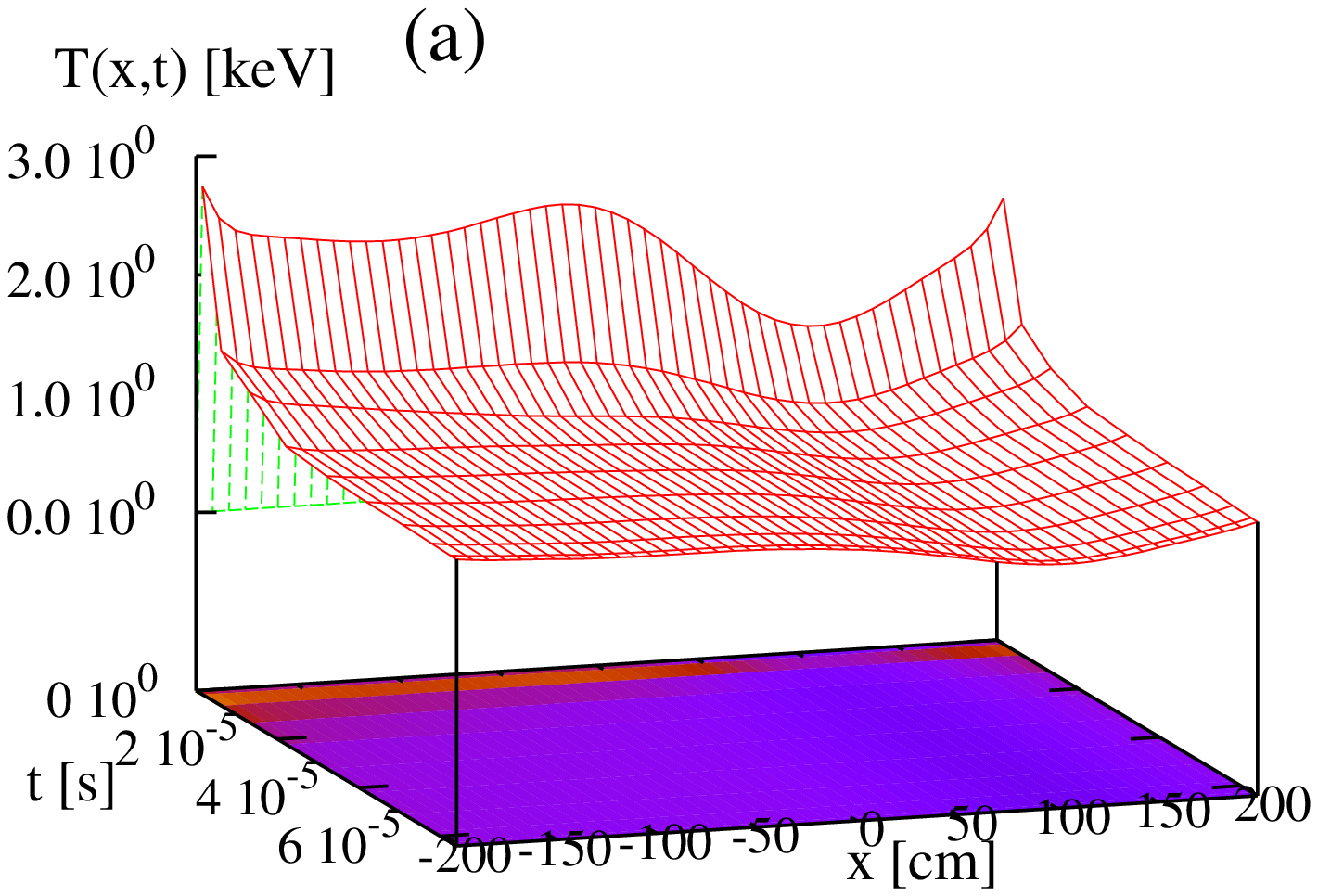}}
%%%\resizebox{9truecm}{!}{\includegraphics{fq3b4T.ps}}
%\resizebox{18truecm}{!}{\includegraphics{fq3b3T.ps},\includegraphics{fc133T.ps}}
\resizebox{9truecm}{!}{\includegraphics{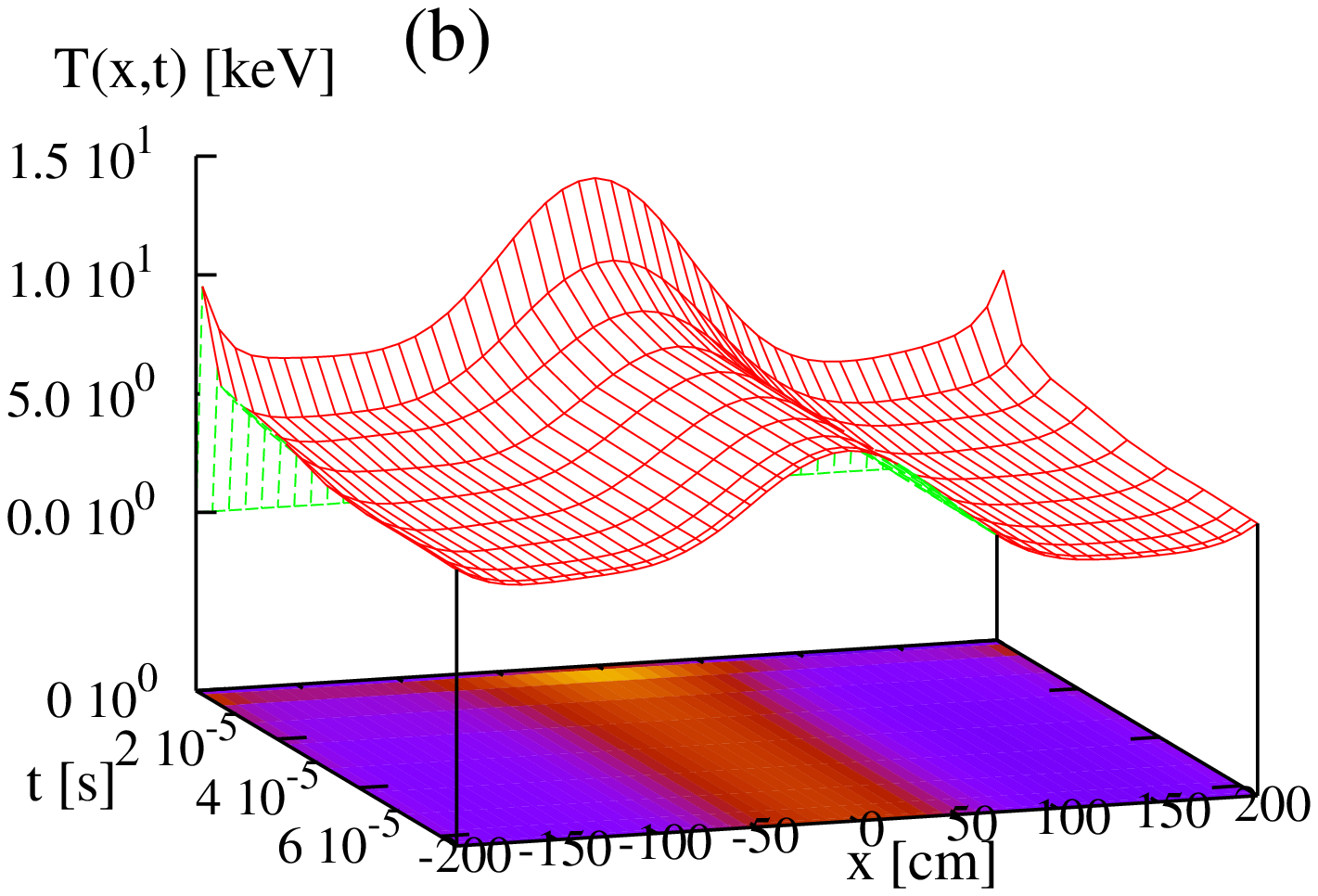}}
%%%\resizebox{9truecm}{!}{\includegraphics{fq314T.ps}}
%%%\resizebox{9truecm}{!}{\includegraphics{fq3b134T.ps}}
\caption{{\it Weak off-axis source:}
Temperature distribution $T(x,t)$ 
in case of the mixed model for Gaussian (a) and power-law (b) distributed increments 
in momentum.
\label{fq3bT}
}
\end{figure}

Fig.\ \ref{fq3bT} shows the evolution of the temperature profiles
as a function of time for the mixed model. The picture is similar 
to the case of pure uniform loading: With Gaussian momentum 
increments, the temperature profile is almost
flat, with a cooler region around the off-axis source, where cooler
material is injected. With power-law momentum increments, the temperature
remains peaked at the center, with a small asymmetry, the region around 
the off-axis injection is slightly cooler again. 
Again, the system behaviour is reminiscent of temperature profile
stiffness in case of power-law momentum increments.

\subsubsection{Strong off-axis fueling}

\begin{figure}
\resizebox{9truecm}{!}{\includegraphics{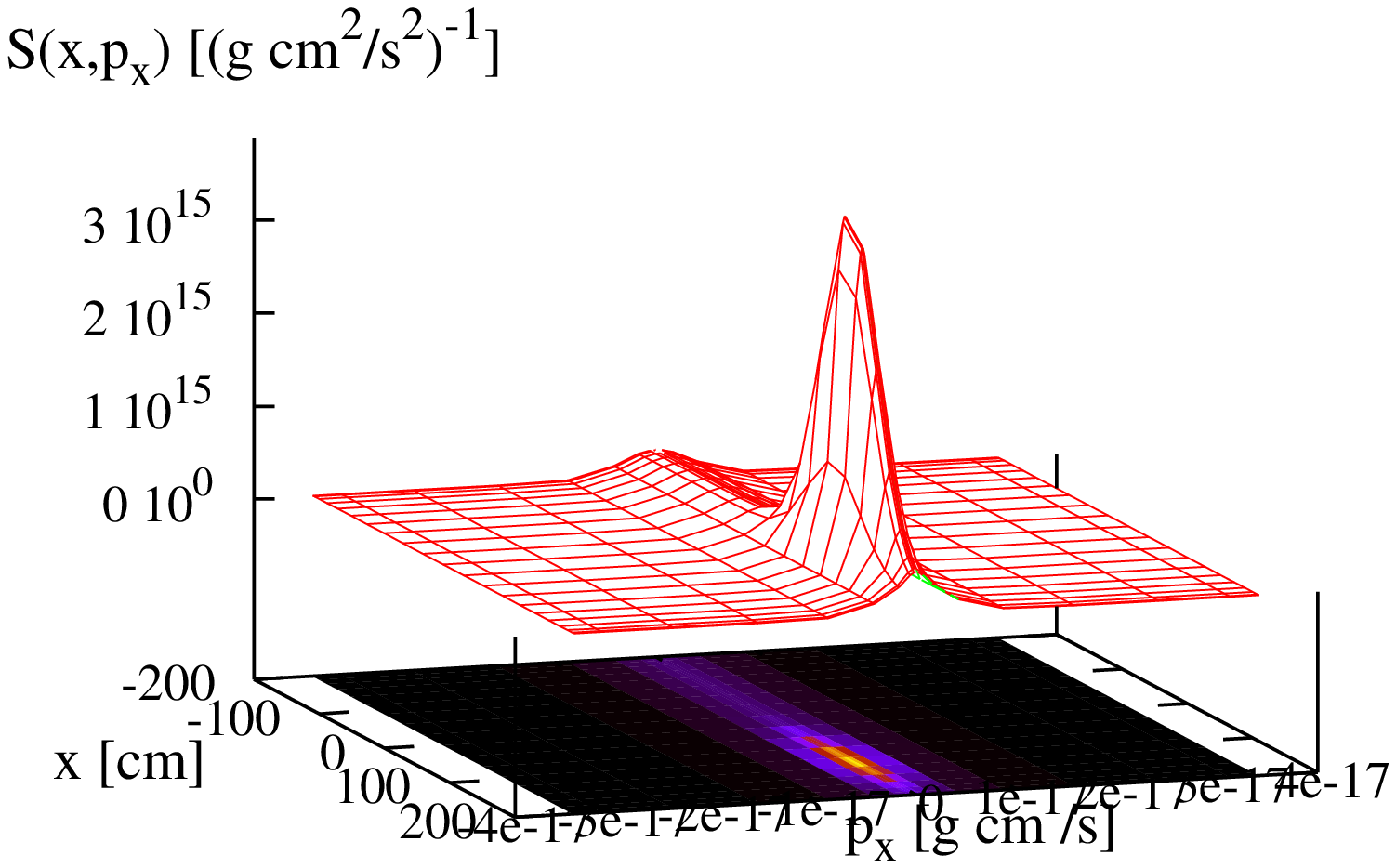}}
\caption{{\it Strong off-axis source:}
Source function $S(x,p_x)$ for strong spatially 
localized off-axis loading (at $x=100$) on top of uniform spatial loading.
\label{fq3s}}
\end{figure}

We now consider an off-axis source that is equally strong as the background 
source, with $\delta=1$ in Eq.\ (\ref{offax}). Fig.\ \ref{fq3s} shows the source term.

\begin{figure*}%[p]
\resizebox{18truecm}{!}{\includegraphics{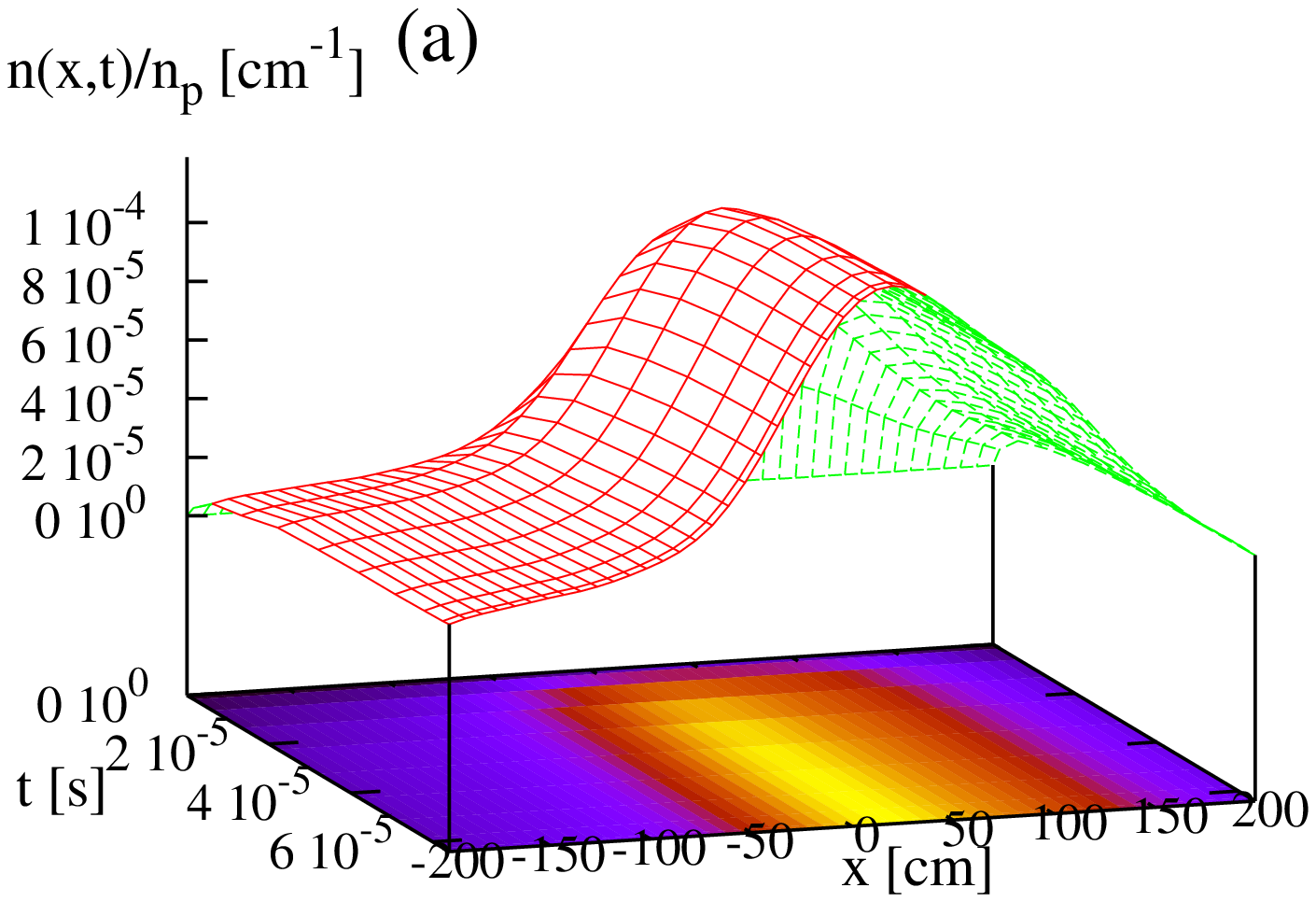},\includegraphics{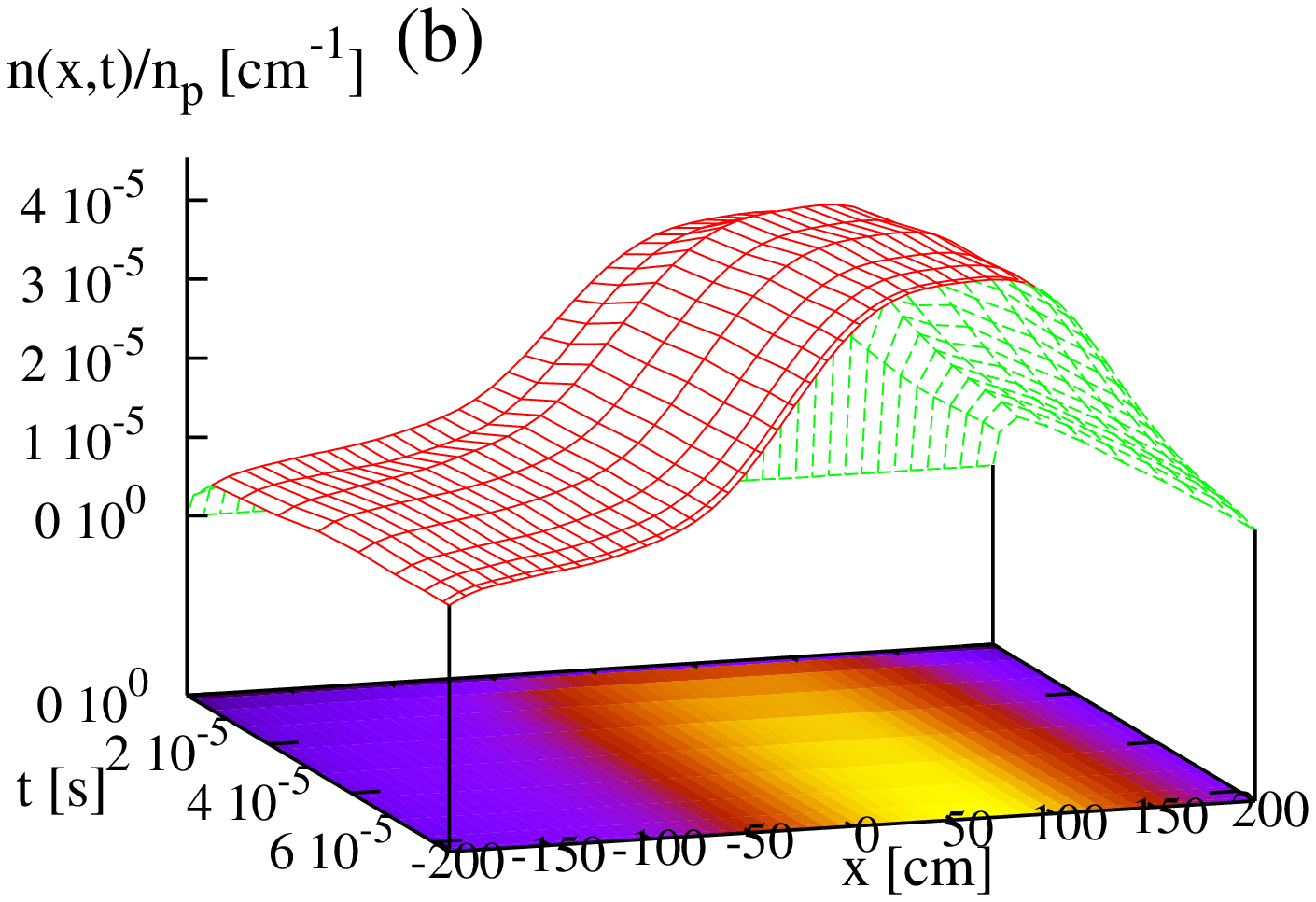}}
%\resizebox{18truecm}{!}{\includegraphics{fq31.ps},\includegraphics{fc17.ps}}
%\resizebox{9truecm}{!}{\includegraphics{fq34.ps}}
%\resizebox{18truecm}{!}{\includegraphics{fq33.ps},\includegraphics{fc18.ps}}
%%\resizebox{9truecm}{!}{\includegraphics{fq314.ps}}
%\resizebox{9truecm}{!}{\includegraphics{fq3134.ps}}
\caption{{\it Strong off-axis source:}
Number density $n(x,t)/n_p$ as 
a function of space $x$ and time $t$, for the mixed model with 
Gaussian (a) and with power-law (b) distributed momentum increments.
\label{fq31}
}
\end{figure*}

\begin{figure}%[p]
\resizebox{9truecm}{!}{\includegraphics{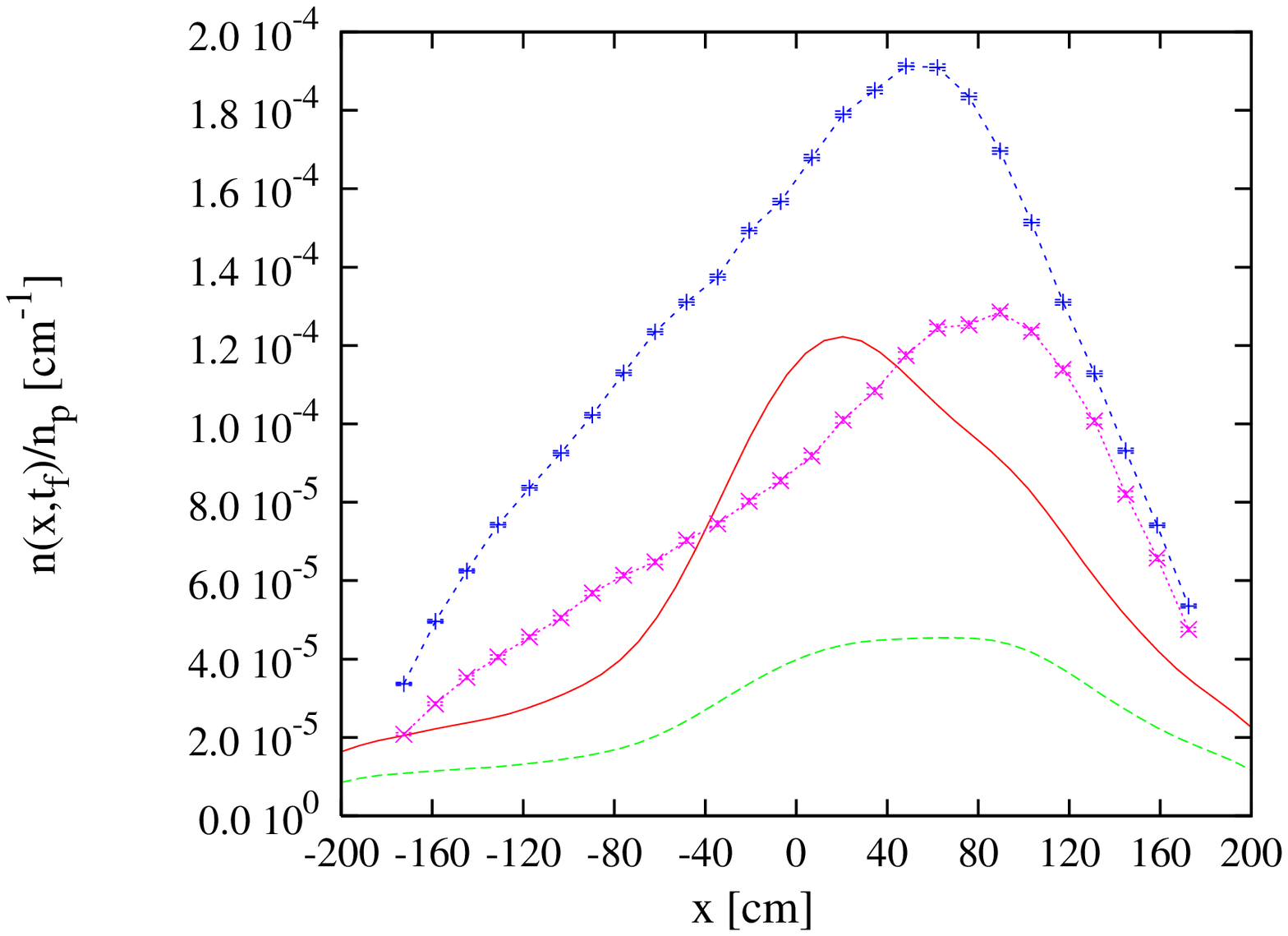}}
\caption{{\it Strong off-axis source:}
Number density $n(x,t)/n_p$  as 
a function of space $x$ at final time $t=t_f$,  for the 
mixed model with Gaussian (solid - red) and with 
power-law (long dashes - green) distributed momentum increments, and for the 
critical gradient model 
with Gaussian (short dashes - blue, with error-bars, and scaled with a factor $1/3$ for 
better visualization) and  with 
power-law (dotted - violet, with error-bars) distributed momentum increments.
\label{fq31nn}
}
\end{figure}

The temporal density evolution is presented in Fig.\ \ref{fq31} for the mixed 
model, and Fig.\ \ref{fq31nn} shows the density profiles at final time
for the mixed and the critical gradient model. 
All density profiles show now an asymmetry to-wards the off-axis source.
With Gaussian momentum increments, the mixed model shows a strong
(and the highest of all models) stiffness, the density peak is still
very close to the center, whereas for the critical gradient model the density
peak is located in between the center and the off-axis source, it is thus less stiff but 
exhibits a better confinement, i.e.\ a higher density.
With power-law momentum increments, the mixed model develops a plateau
region between the center and the off-axis source, and the critical
gradient model yields a density profile peaked at the location of off-axis source, it has
lost any stiffness.
The comparison to Fig.\ \ref{fq3b1nn} makes evident that the asymmetry of the profiles  
depends on strength of the off-axis source, the weaker the source, the less asymmetric 
the profiles obviously are.

\begin{figure}
\resizebox{9truecm}{!}{\includegraphics{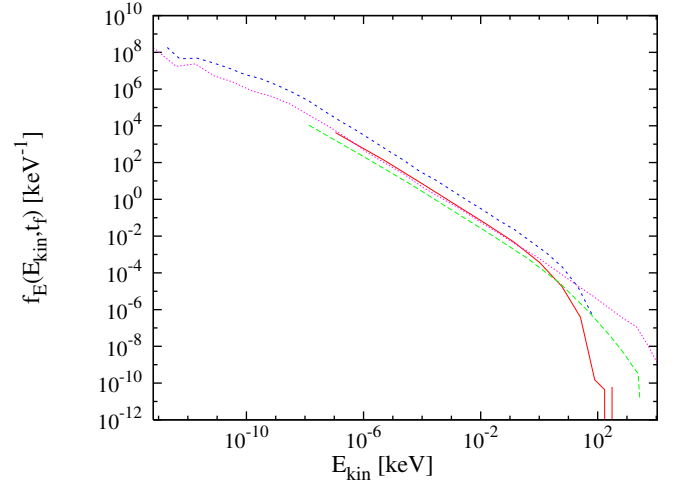}}
%\resizebox{9truecm}{!}{\includegraphics{fq31p.ps}}
\caption{{\it Strong off-axis source:}
Final kinetic energy distribution $f_E(E_{kin},t_f)$ for 
the mixed model with Gaussian (solid - red) and with 
power-law (long dashes - green) distributed momentum increments, and for the critical gradient model with 
Gaussian (short dashes - blue) and with 
power-law (dotted - violet) distributed momentum increments.
\label{fq3p}
}
\end{figure}

Fig.\ \ref{fq3p} shows the kinetic energy distributions at the final time $t_f$,
i.e \ at stationary state, for the mixed and
the critical gradient model.
A common feature of the distributions is the appearance of a quite extended and 
clear power-law scaling, with power-law index $-1$. 
Also the not shown energy distributions in case of the weak off-axis source and 
of the pure uniform loading are very similar to those shown in Fig.\ \ref{fq3p}.
The power-law with index $-1$ seems thus
to be a universal property of the kind of random walk considered.
It is interesting to note that the power-law appears also for the 
cases where the momentum increments are Gaussian distributed,
seemingly contradicting the central limit theorem.
The power-law must consequently originate from the coupling with position space, where 
in all variants of the models power-law increments
are always present, and there possibly is also a dynamic selection effect present. 
The cases with power-law momentum
increments differ from the cases with Gaussian distribution just in that the particles 
reach higher energies, due to the larger steps the 
particles' momentum is allowed to take. The low energy cut-off in Fig.\ \ref{fq3p}
in the cases of the mixed model corresponds to the lowest 
absolute value of the momentum in the numerical grid used. 
We just note that since the kinetic energy distributions are in all cases clearly 
non-thermal, 
the only temperature concept that makes sense is that of temperature defined through 
the mean kinetic energy, which is the definition we have throughout used.

\begin{figure}%[p]
%\resizebox{18truecm}{!}{\includegraphics{fq31T.ps},\includegraphics{fc17T.ps}}
\resizebox{9truecm}{!}{\includegraphics{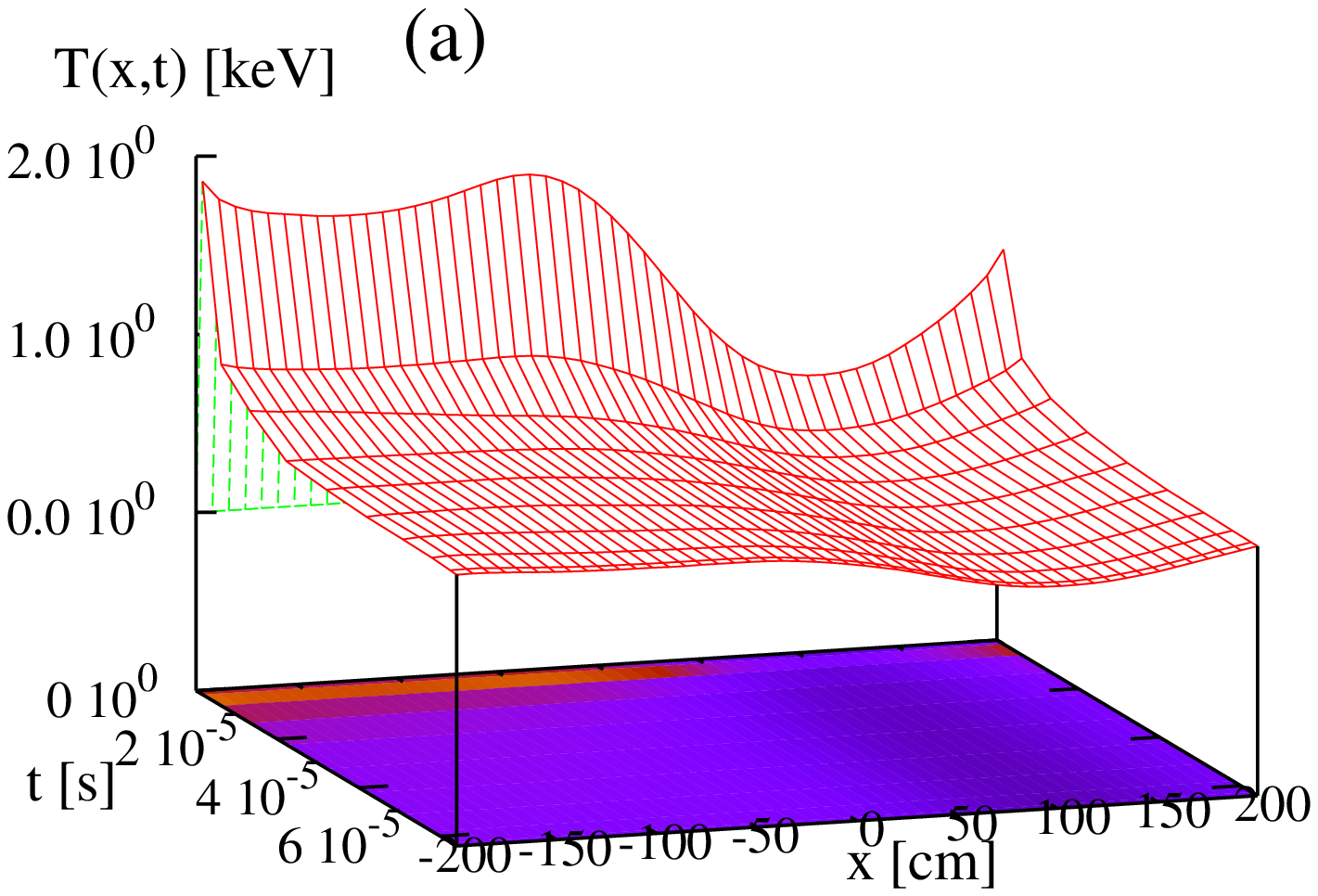}}
%%%\resizebox{9truecm}{!}{\includegraphics{fq34T.ps}}
%\resizebox{18truecm}{!}{\includegraphics{fq33T.ps},\includegraphics{fc18T.ps}}
\resizebox{9truecm}{!}{\includegraphics{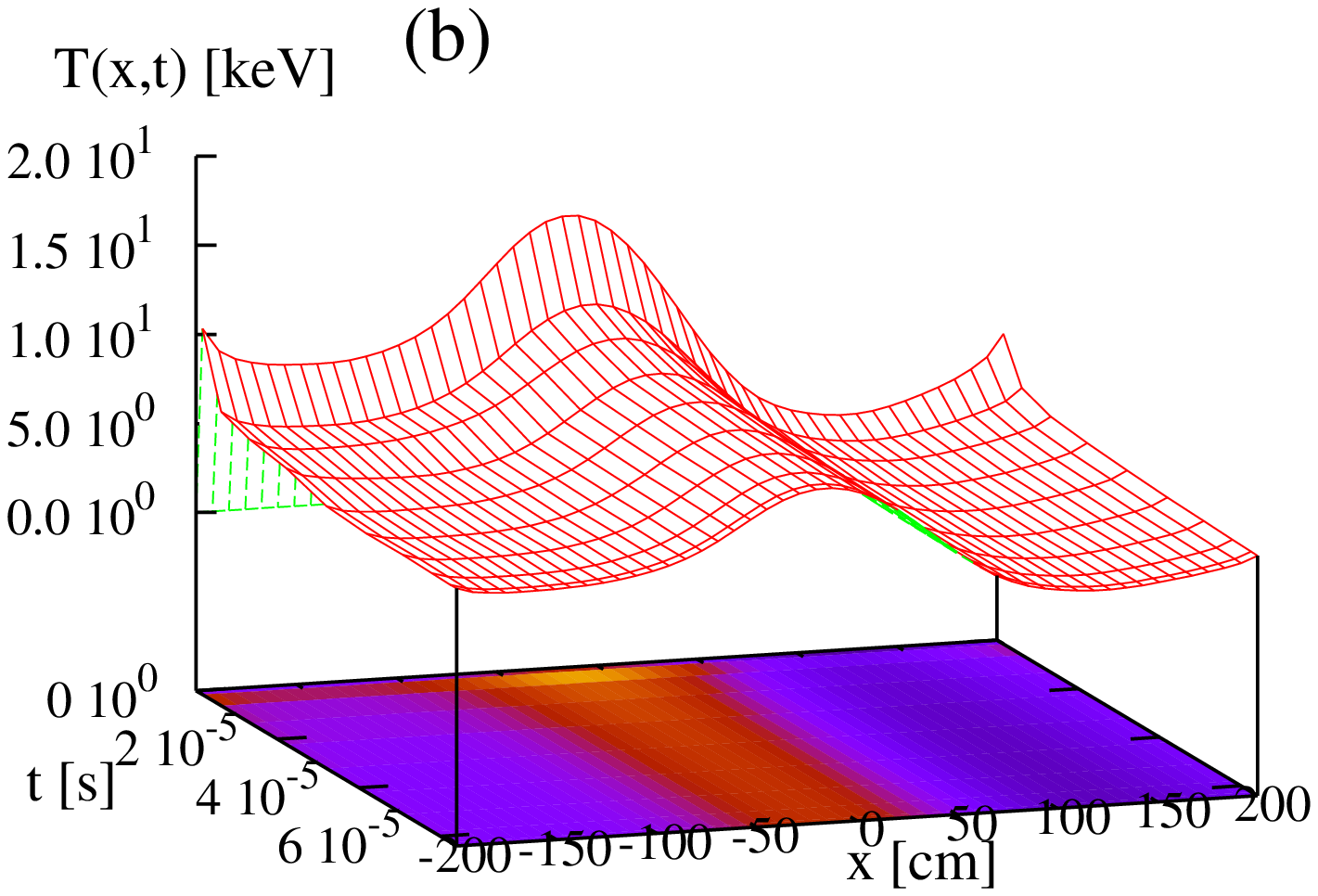}}
%%%\resizebox{9truecm}{!}{\includegraphics{fq314T.ps}}
%%%\resizebox{9truecm}{!}{\includegraphics{fq3134T.ps}}
\caption{{\it Strong off-axis source:}
Temperature distribution $T(x,t)$ 
in case of the mixed model for Gaussian (a) and power-law (b) distributed increments 
in momentum.
\label{fq3T}
}
\end{figure}

\begin{figure*}%[p]
\resizebox{18truecm}{!}{\includegraphics{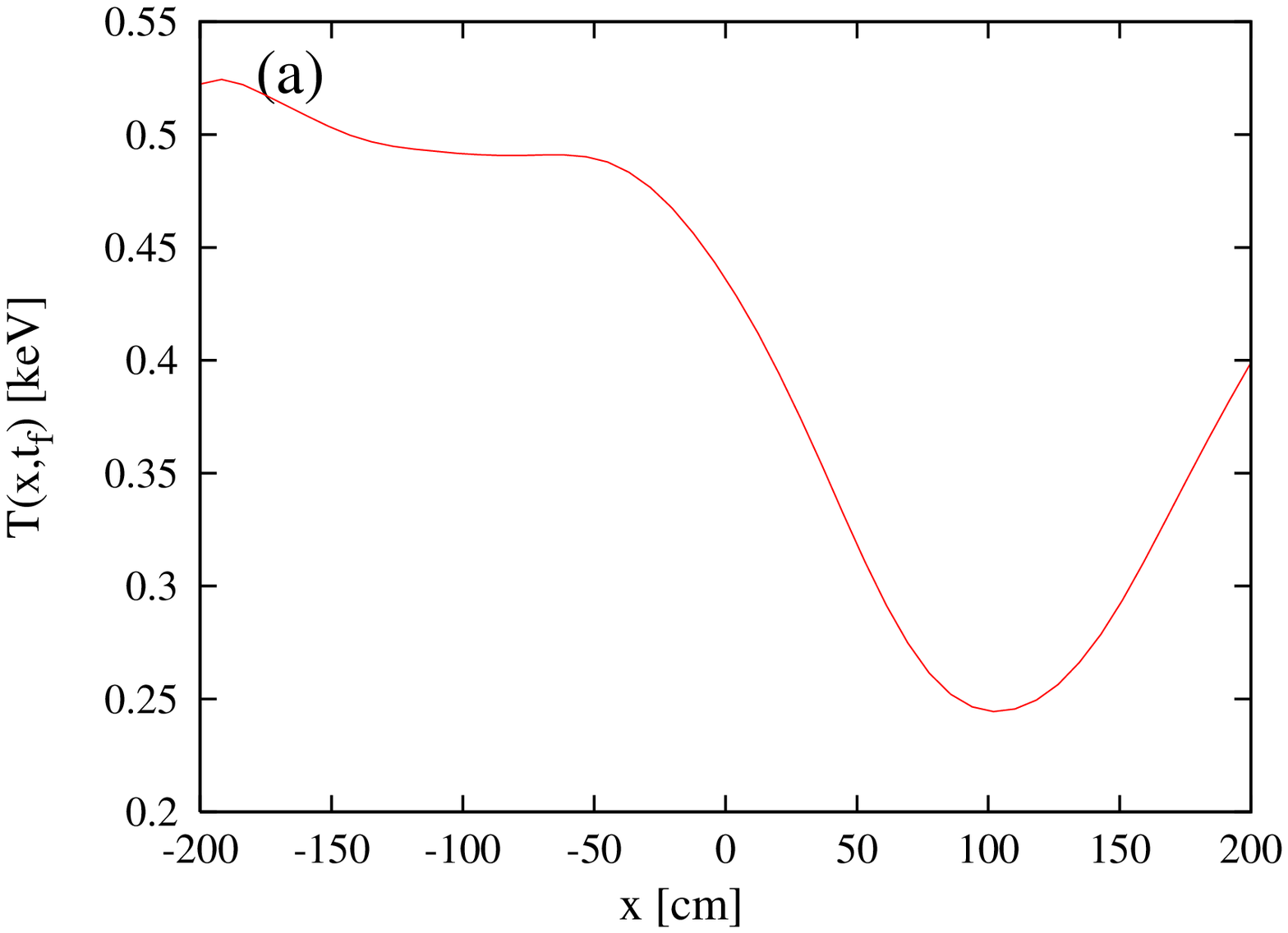},\includegraphics{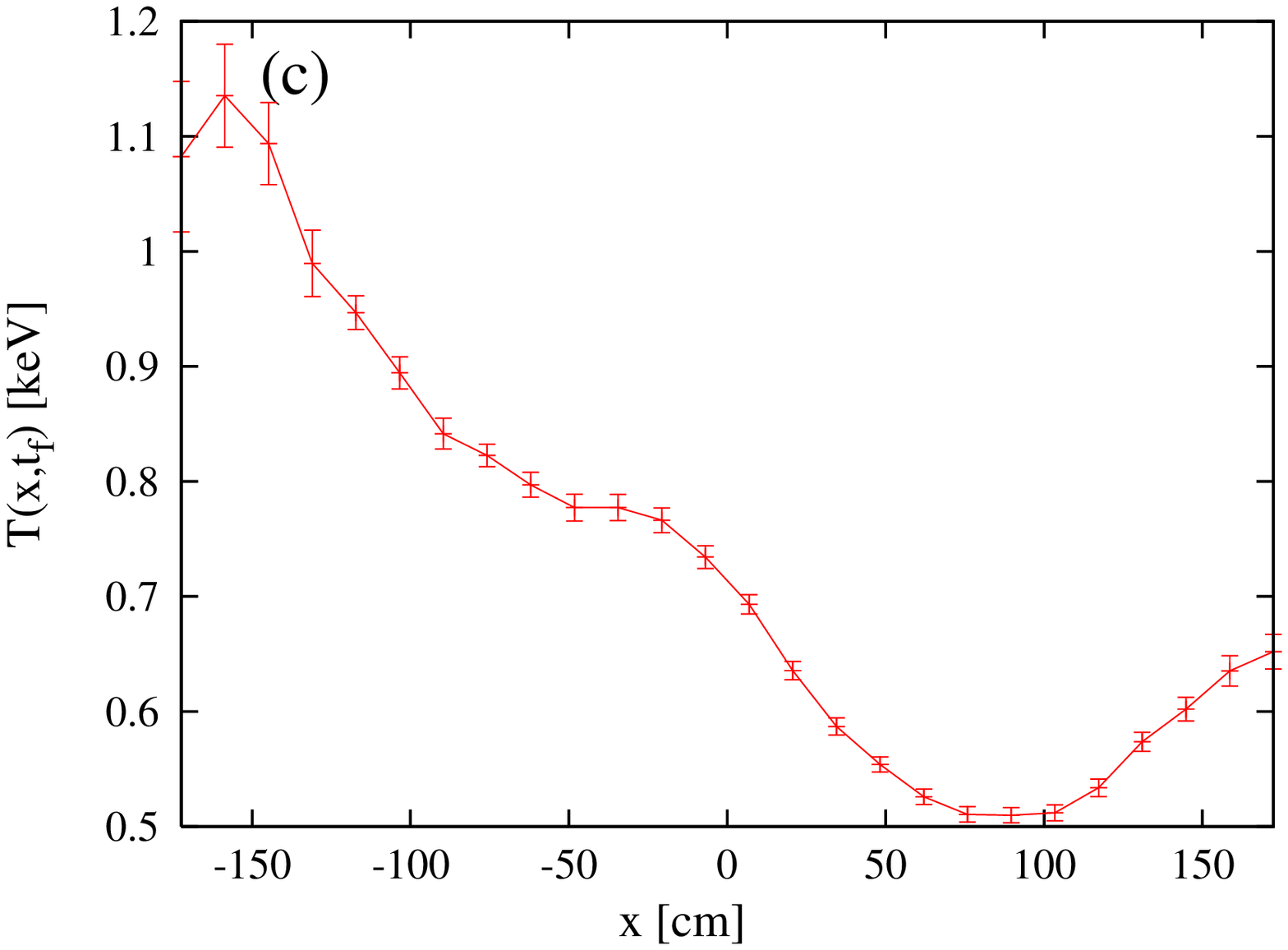}}
%\resizebox{9truecm}{!}{\includegraphics{fq34T.ps}}
\resizebox{18truecm}{!}{\includegraphics{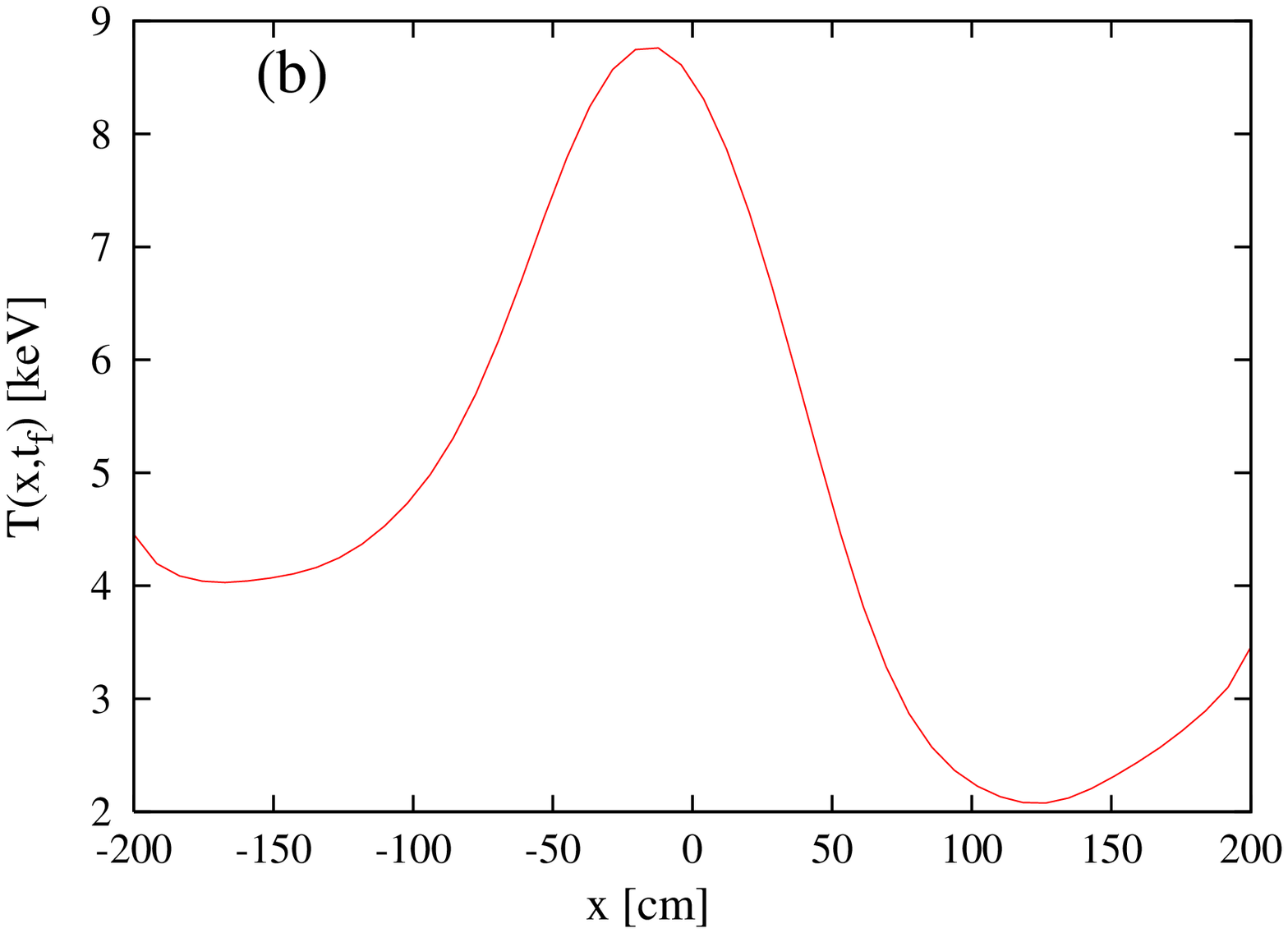},\includegraphics{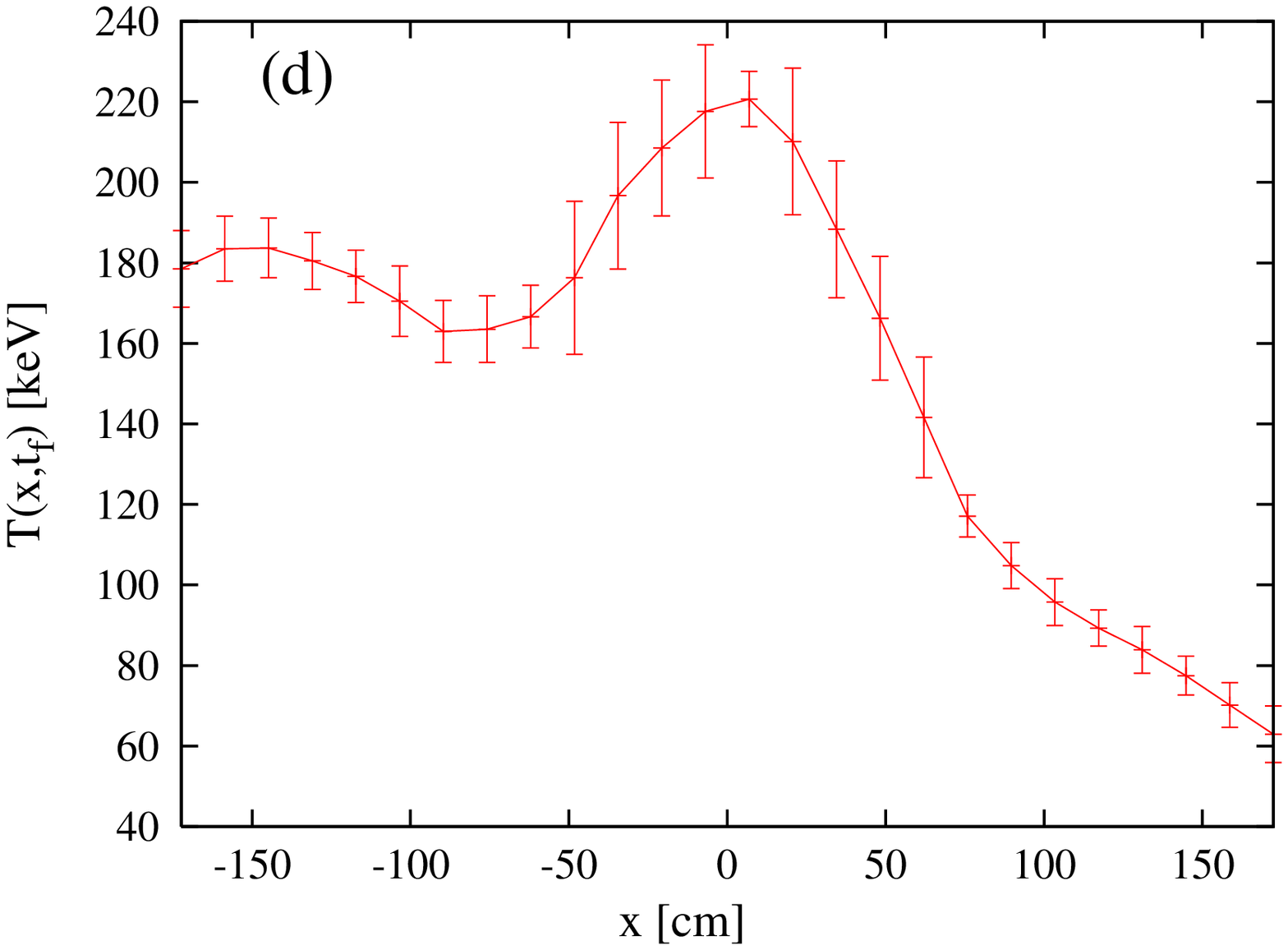}}
%%\resizebox{9truecm}{!}{\includegraphics{fq314T.ps}}
%\resizebox{9truecm}{!}{\includegraphics{fq3134T.ps}}
\caption{{\it Strong off-axis source:}
Temperature distribution $T(x,t_f)$ at final time $t_f$ for the 
mixed model with Gaussian (a) and with 
power-law (b) distributed momentum increments, and for 
the critical gradient model with Gaussian (c) and with power-law (d) distributed 
momentum increments.
\label{fq3Tline}
}
\end{figure*}

Fig.\ \ref{fq3T} shows the evolution of the temperature profile in time for the mixed 
model, and in Fig.\ \ref{fq3Tline} the temperature distributions at final time are presented 
for the mixed and the critical gradient model. 
In combination
with power-law momentum increments, the temperature profiles are clearly peaked at the center, for
both the mixed and the critical gradient model, respectively,
with an asymmetry that is more prominent in the critical gradient model, the injection region remains 
colder than the region on the left side of the central peak. 
The temperature values in the mixed model are close 
to the temperature
with which the particles are injected ($8\,$keV and $0.8\,$keV, respectively), whereas in the 
critical gradient model very
high temperatures are reached, high energy particles are more efficiently trapped in the system.
With Gaussian momentum increments,  
the temperature profiles are not peaked inside the system for both, the mixed and the 
critical gradient 
model, respectively,
the profiles basically reflect the temperature distribution of the two different injection sources.
Moreover, the temperature is in both models much lower than the injection temperature 
of the particles,
the energetic particles obviously leave the system efficiently.
We can thus conclude that power-law momentum increments give rise to high stiffness of the
temperature profile, whereas with Gaussian momentum increments there is basically no
stiffness of the temperature profile present. % of the 

\begin{figure}
\resizebox{9truecm}{!}{\includegraphics{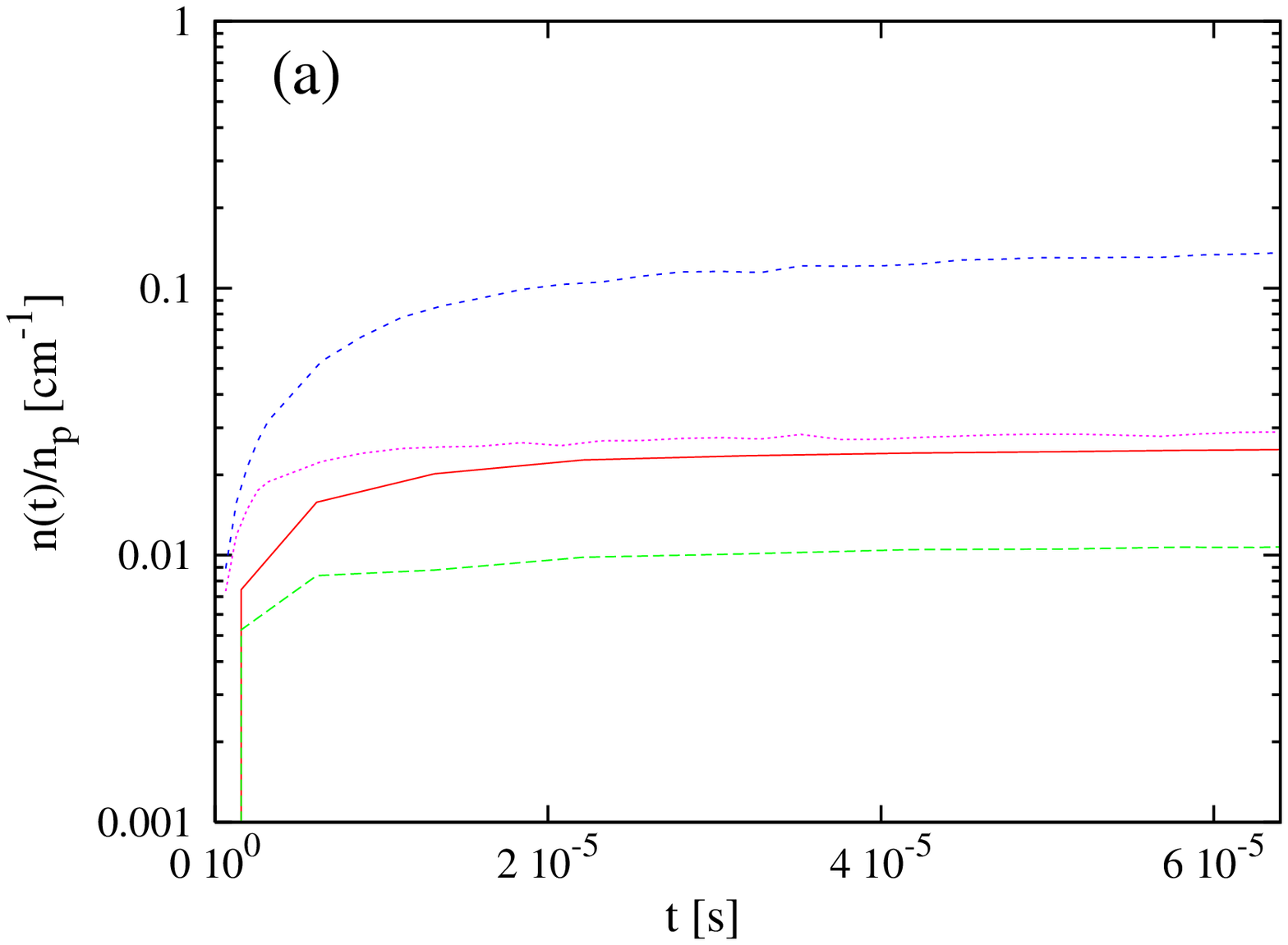}}
%\resizebox{9truecm}{!}{\includegraphics{fq31e.ps}}
\resizebox{9truecm}{!}{\includegraphics{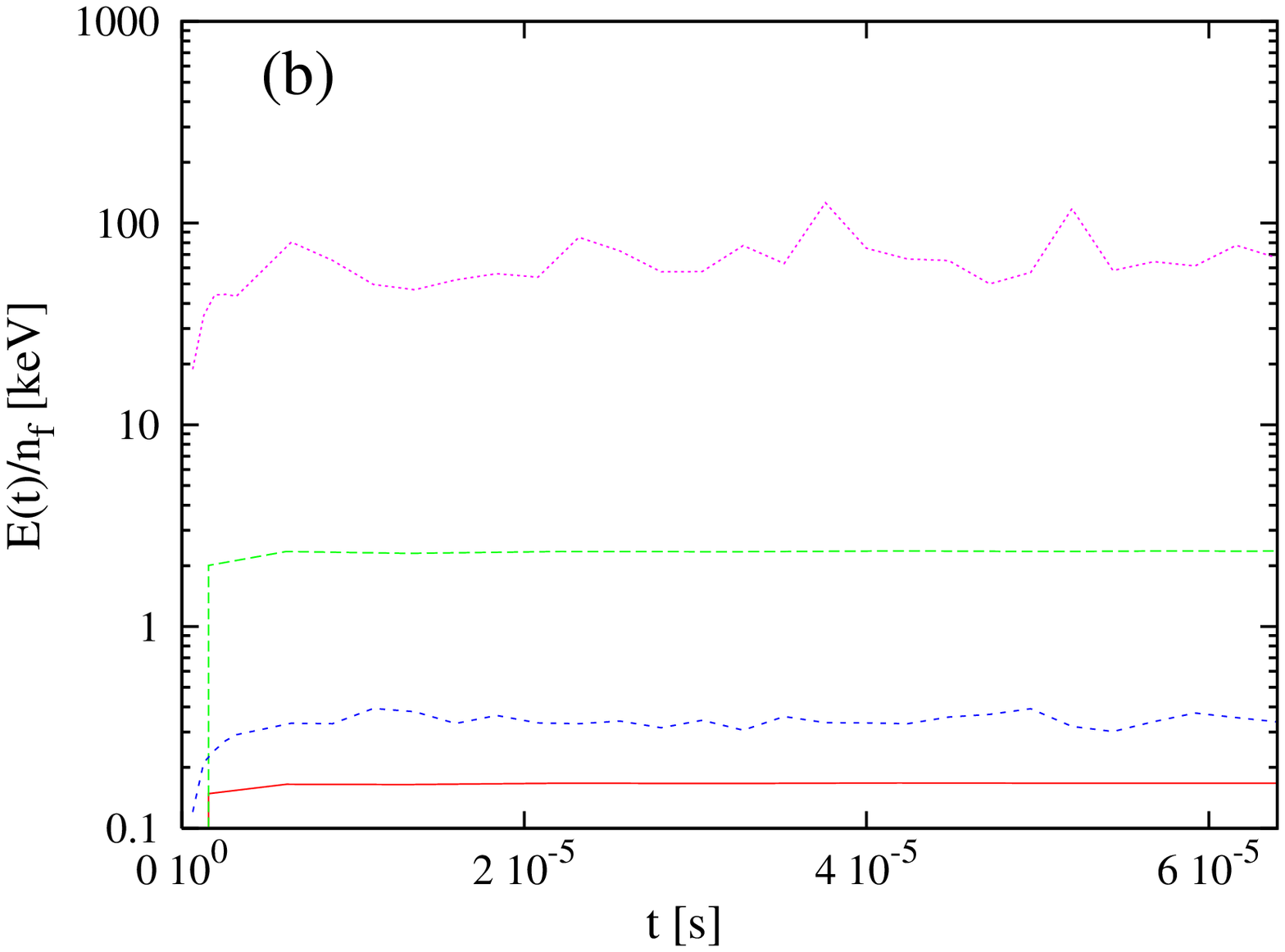}}
\caption{{\it Strong off-axis source:}
Total number density divided by the 
total number $n_p$ of injected particles, $n(t)/n_p$ (a), and total energy divided by the 
number $n_f$ of particles that are in the system at final time, $E(t)/n_f$ (b), 
both as a function of time,
for the mixed model with Gaussian (solid - red) and with 
power-law (long dashes - green) distributed momentum increments, and 
for the critical gradient model with Gaussian (short dashes - blue) and with
power-law (dotted - violet) distributed momentum increments.
\label{fq31ne}}
\end{figure}

In Fig.\ \ref{fq31ne}(a), the total density divided by the total number of injected 
particles $n_p$ is shown as a function of time. From the asymptotic
values it is seen that the particle confinement is in any case more effective for the critical 
gradient model than for the mixed model. 
In both the critical gradient and the mixed model, the particle confinement  
deteriorates when power-law distributed momentum increments are considered.
Last, Fig.\ \ref{fq31ne}(b) shows the mean kinetic energy per particle (total instantaneous kinetic
energy, divided by the 
number of particles that are in the system at final time $n_f$) as a function of time. 
Here now, power-law distributed momentum increments lead in both models to a higher
value of the energy per particle,  the possibly large momentum increments 
imply a stronger heating that is reflected in the mean energy per particle.
Particle and energy confinement times are discussed in Sect.\ \ref{parhea}.

%isis

\section{Discussion \label{SecIV}}

\subsection{Particle and energy confinement time}

\label{parhea}

For convenience, we have used the source term $S$ in the form 
normalized 
to one and independent of time, 
i.e.\ $\int dx\, dp\,S(x,p)=1$ [see Eqs.\ (\ref{unifs}) and (\ref{offax})], 
so that the injection rate is one particle
per unit time, and during a time interval of length $t_f$ the number of
particles injected is $\int dt\,dx\, dp\,S(x,p)=t_f$. To compare solutions 
of the CTRW equations 
with MC simulations, and to 
discuss confinement times, it is useful to 
explicitly allow a general injection rate $\nu_p$, and to replace the 
source function by $\nu_p S$. If an experiment lasts a time $t_f$ and a 
total number of particles $n_p$ is injected, then it holds that
$n_p = \nu_p \int dt\,dx\,dp\, S(x,p_x)=\nu_p t_f$,
so that the relation $\nu_p=n_p/t_f$ follows. In order then to directly 
compare solutions of the
CTRW equations with results from MC simulations, e.g.\ with respect to 
particle densities, we can for instance divide the densities yielded 
by the MC simulations by $n_p$.

\subsubsection{Particle confinement time}

To determine the particle confinement time during the stationary, dynamic 
equilibrium state, 
where the particle losses equal the injection of particles, we define 
first the 
total particle source rate 
$\Sigma_p:=\nu_p\int \! dx\,\int \!dp_x\,S(x,p_x)$, 
which, due to the chosen normalization, takes the form $\Sigma_p=\nu_p$. 
The total number of particles in the system at a given time is 
$n(t) = \int \! dx\, \int \! dp_x\, P_x(x,p_x,t)$, so that
we can define the 
particle confinement time as 
\beq
\tau_p := \frac{n(t)}{\Sigma_p}=\frac{n_f}{\nu_p}   ,
\label{parconf}
\eeq 
where $t$ is large enough so that a stationary dynamic equilibrium state is 
realized. For convenience, we focus on the final time of the experiment, 
$t=t_f$, and we denote by $n_f$ the number of particles at final time, 
$n_f:=n(t_f)$ ($t_f$ just marks an arbitrary instant during the 
stationary state, so that at any large enough time before $t_f$ there 
are $n_f$ particles in the system). 

Alternatively, we can determine the mean time the particles spend in the system until
they leave: 
%we first note that 
If
$n_p$ particles are injected in a total time $t_f$, uniformly distributed 
over time, i.e.\ with injection rate $\nu_p=n_p/t_f$, then the number of  
particles injected in a time interval 
$\Delta t$ is $n_p \Delta t/t_f\equiv \nu_p\Delta t$. Let now 
be $\bar t$ the mean time a particle stays
in the system. If all particles would stay the average time $\bar t$ in the 
system, then the $n_f$ particles that are in the 
system at final time $t_f$ would have 
been injected in the time interval $[t_f-\bar t,t_f]$, i.e.\ over a duration
$\Delta t=\bar t$, during which $n_p \bar t/t_f =n_f$
particles are injected, so that 
\beq
\bar t = t_f \frac{n_f}{n_p} .
\label{tbar}
\eeq
This relation allows to determine $\bar t$ easily in Monte Carlo 
experiments and from the solution of the CTRW equations. 
The mean time a particle stays in the system
$\bar t$ is actually identical to 
the particle confinement time, $\bar t=\tau_p$:
Inserting the definition of $\nu_p\equiv n_p/t_f$ into
Eq.\ (\ref{parconf}) yields $\tau_p=n_f/(n_p/t_f)$, and  
the further replacing of $n_f$ by $\bar t n_p/t_f$, according to 
Eq.\ (\ref{tbar}), leads to $\tau_p=\bar t$.

The values of $n_f/n_p$ and of $\tau_p$ are shown in Table \ref{table1}
in the case of strong off-axis loading, for the mixed and the critical
gradient model, with Gaussian and power-law distributed momentum increments,
respectively. The highest particle confinement time is achieved by the
critical gradient model with Gaussian momentum increments. It can also
be seen that in both, the mixed and the critical gradient model, power-law
momentum increments deteriorate the confinement times, i.e.\ strong heating
or the intense acceleration of particles deteriorates confinement, since
energetic particles leave more easily in both models. Due to
Eq.\ (\ref{tbar}), this behaviour 
is directly reflected in the values of $n(t)/n_p$ (shown in Fig.\ \ref{fq31ne}(a)), 
less particles are found in the system during stationary state
if the momentum increments are power-law distributed.

\subsubsection{Energy confinement time}

Every particle is initially 
injected with a certain energy, distributed according to a 
Maxwellian with a well defined temperature that corresponds to a mean 
kinetic energy $\langle E_{kin,0}\rangle$.  
The energy injection rate due to particle injection is thus given as 
$\Sigma_{E,inj}(t)
= \nu_p\int \! dx \int \! dp \,(\gamma(p)-1)\,mc^2 \,S(x,p,t))$,
and it obviously holds that $\Sigma_{inj}(t)= \nu_p \langle E_{kin,0}\rangle$.
A second source of energy is provided by the random walk in momentum space,
which can be considered either as heating or as acceleration, depending on the
distribution of momentum increments, and it gives rise  
to a heating power per unit time.
$\Sigma_{E,heat}$.
The total energy in the system at a given time is
given as $E(t)= \int \! dx \int \! dp \,(\gamma(p)-1)\,mc^2 \,P(x,p,t))$,
so that the energy confinement time can be defined as
\beq
\tau_E := \frac{E(t)}{\Sigma_{E,inj}+\Sigma_{E,heat}} 
= \frac{E(t)}{\nu_p \langle E_{kin,0}\rangle + \Sigma_{E,heat}}    .
\label{econf}
\eeq

$\Sigma_{E,heat}$ depends on the mean change 
in energy in a single acceleration event and on the number of acceleration
events per unit time, and we determine it from Monte-Carlo simulations, 
also in the cases of the mixed model, where we else solve the 
CTRW equations. Some care is needed in the analysis of the 
energy confinement time, since
mean values of different quantities have to be used, the mean values
are though not always very representative of the actual mean behaviour
when dealing with power-law distributed quantities.

In a first approach, we determine from Monte Carlo simulations the mean 
change in kinetic
energy $\langle \Delta e_{kin}\rangle$ that the particles undergo in one 
acceleration 
event (which would equal the analytically from the distribution of
increments calculated mean value, if the coupled random walk does not 
introduce
a selection effect, and in the cases where the increments are not 
power-law distributed), as well as the mean number of acceleration events 
$\langle n_{acc}\rangle$ 
a particle undergoes, which yields the 
number of acceleration
events per second as 
$\langle \nu_{acc}\rangle =\langle n_{acc}\rangle /\bar t$, 
and the energy increase per unit time 
$\langle\epsilon\rangle
=\langle\Delta e_{kin}\rangle \langle \nu_{acc}\rangle$. 
In this approach, the energy a particle attains on the average during its stay in the system 
is 
$\langle \epsilon\rangle \bar t$, which can directly be compared to 
the mean
energy $\langle e_{fin}\rangle$ of the particles at final time $t_f$,
as determined in Monte Carlo simulations, 
(and which should be 
representative for 
all intermediate time steps during stationary state), and it should hold 
that $\langle \epsilon\rangle \bar t = \langle e_{fin}\rangle$. We 
though find that 
$\langle \epsilon\rangle \bar t$ is too large by roughly a factor of 10, 
except for the 
mixed model with 
Gaussian momentum increments. This discrepancy must be attributed to the 
non-linearity
in the critical gradient model, and to the power-law distribution of momentum 
increments, in the respective cases where they are used, since 
in these cases mean values are not necessarily good representatives of the mean 
behaviour of the actual process. 
Thus, in order the estimates to be self-consistent,
we consider as mean change in kinetic energy
in one acceleration event 
$\langle \Delta e_{kin,eff}\rangle
=\langle e_{fin}\rangle/\langle n_{acc}\rangle$,
and as the energy injected per second and per particle 
$\langle \epsilon_{eff}\rangle =\langle e_{fin} \rangle/\bar t$.
Assuming then that at any time during stationary state there are 
$n_f$ particles
in the system, we determine the total heating rate as 
$\Sigma_{E,heat}= n_f\langle e_{fin} \rangle/\bar t$. Moreover, 
per definition it holds
that $E(t_f)=n_f\langle e_{fin}\rangle$, i.e.\  the final total energy equals
the number of particles at final time times the mean energy of the particles.
Inserting the latter two expressions for $\Sigma_{E,heat}$ and $E(t_f)$ into 
Eq.\ (\ref{econf}) yields
\beq
\tau_E 
= \frac{n_f\langle e_{fin}\rangle}{n_p E_{kin,0}/t_f+n_f\langle e_{fin} \rangle/\bar t} ,
\label{econf2}
\eeq
where we have also inserted the definition of $\nu_p$. 
According to Eq.\ (\ref{tbar}), we can
replace $n_f$ with $\bar t n_p/t_f$, so that Eq.\ (\ref{econf2}) turns into 
the simpler form
\beq
\tau_E 
= \frac{\bar t\langle e_{fin}\rangle}{E_{kin,0}+\langle e_{fin} \rangle}   ,
\label{econf3}
\eeq
where $\bar t\equiv \tau_p$, as shown above.

We again consider 
the example of the strong off-axis source, where one half of the particles is 
injected in the entire system with a kinetic energy of $4\,$keV, and the 
other half off-axis with a 
kinetic energy of $0.1\times 4\,$keV, so that the mean injection kinetic 
energy is $\langle E_{kin,0}\rangle = 2.2\,$keV. 
Table \ref{table1} shows the energy confinement times $\tau_E$ for the 
mixed and the
critical gradient model, for the two sub-cases of Gaussian and power-law
momentum increments. Here now, the critical gradient model 
shows a higher energy confinement time than the mixed model,
independent of the kind of momentum increments. Contrary to the particle
confinement time, the energy confinement time increases in both the mixed 
and the critical gradient model, respectively, if 
power-law momentum increments are applied. 
Remarkably, the number of acceleration events $\langle n_{acc}\rangle$ 
is an order of magnitude
larger in the critical gradient model than in the mixed model,
the particle dynamics inside the system is obviously more complex, the higher
collisionality is though directly reflected  only in the energy confinement
time, not in the particle confinement time.

\begin{table*}
\begin{tabular}{r|r|r|r|r|r|r|r|r}
model/momentum increments  & $n_f/n_p$ & $\tau_p$          & $\langle e_{fin}\rangle$       
             & $\langle n_{acc}\rangle$    
                  & $\langle \nu_{acc}\rangle$ 
                      & $\langle\epsilon_{eff}\rangle$   
                         & $\langle e_{fin}\rangle/\langle n_{acc}\rangle$ 
                             & $\tau_E$ \\
\hline
mixed/Gaussian &   0.028   & $2\times 10^{-6}$ &    $0.2$      & 75 
                                     & $4\times 10^7$   & $1\times 10^5$ 
                                       & 0.0027 &  $2\times 10^{-7}$ \\
mixed/power-law &   0.011   & $7\times 10^{-7}$ &    $3.3$      & 76  
                                     & $1\times 10^8$   & $5\times 10^6$ 
                                       & 0.0430 & $4\times 10^{-7}$ \\
critical gradient/Gaussian &   0.138   & $9\times 10^{-6}$ &    $0.4$      & 818 
                                     & $9\times 10^7$  & $4\times 10^4$ 
                                       & 0.0005 & $1\times 10^{-6}$\\
critical gradient/power-law &   0.030   & $2\times 10^{-6}$ &   $75.7$      & 1120 
                                     & $6\times 10^8$ & $4\times 10^7$ 
                                       & 0.0676 & $2\times 10^{-6}$
\end{tabular}
\caption{The table shows the fraction of particles $n_f/n_p$ that is in 
the system at final time $t_f$, the particle confinement time $\tau_p$, 
the mean kinetic energy $\langle e_{fin}\rangle$ of the particles at final
time, the number of acceleration events per second $\langle\nu_{acc}\rangle$ 
that a particle undergoes, the mean number of total acceleration events 
$\langle n_{acc}\rangle$ per particle, 
the mean gain of energy per unit time $\langle\epsilon_{eff}\rangle$, 
the mean energy gain in a single acceleration event 
$\langle e_{fin}\rangle/\langle n_{acc}\rangle$, 
and the energy confinement time $\tau_E$. 
The final time of the simulations is $t_f = 6.4\times 10^{-5}$.
\label{table1}}
\end{table*}

\subsection{Particle and heat fluxes and diffusivities\label{pahef}}

\subsubsection{Particle diffusion}

\begin{figure*}
%%\resizebox{18truecm}{!}{\includegraphics{diff33a.ps},
%%                         \includegraphics{diff141a.ps},\includegraphics{diff17a.ps}}
%%\resizebox{18truecm}{!}{\includegraphics{diff33b.ps},
%%                         \includegraphics{diff141b.ps},\includegraphics{diff17b.ps}}
%%\resizebox{18truecm}{!}{\includegraphics{diff33c.ps},
%%                         \includegraphics{diff141c.ps},\includegraphics{diff17c.ps}}
%%\resizebox{18truecm}{!}{\includegraphics{diff33d.ps},
%%                         \includegraphics{diff141d.ps},\includegraphics{diff17d.ps}}
%
\resizebox{18truecm}{!}{\includegraphics{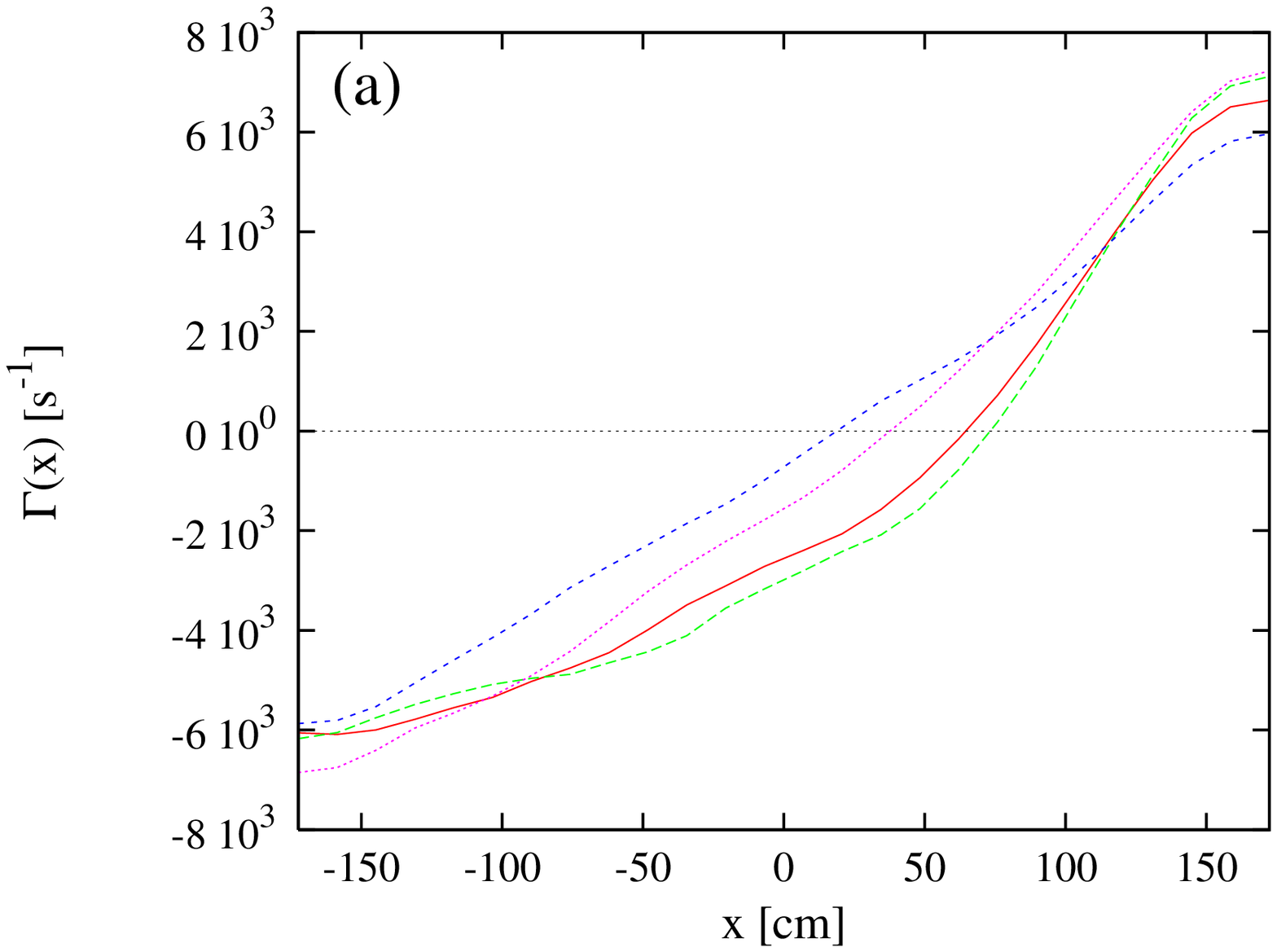},
                         \includegraphics{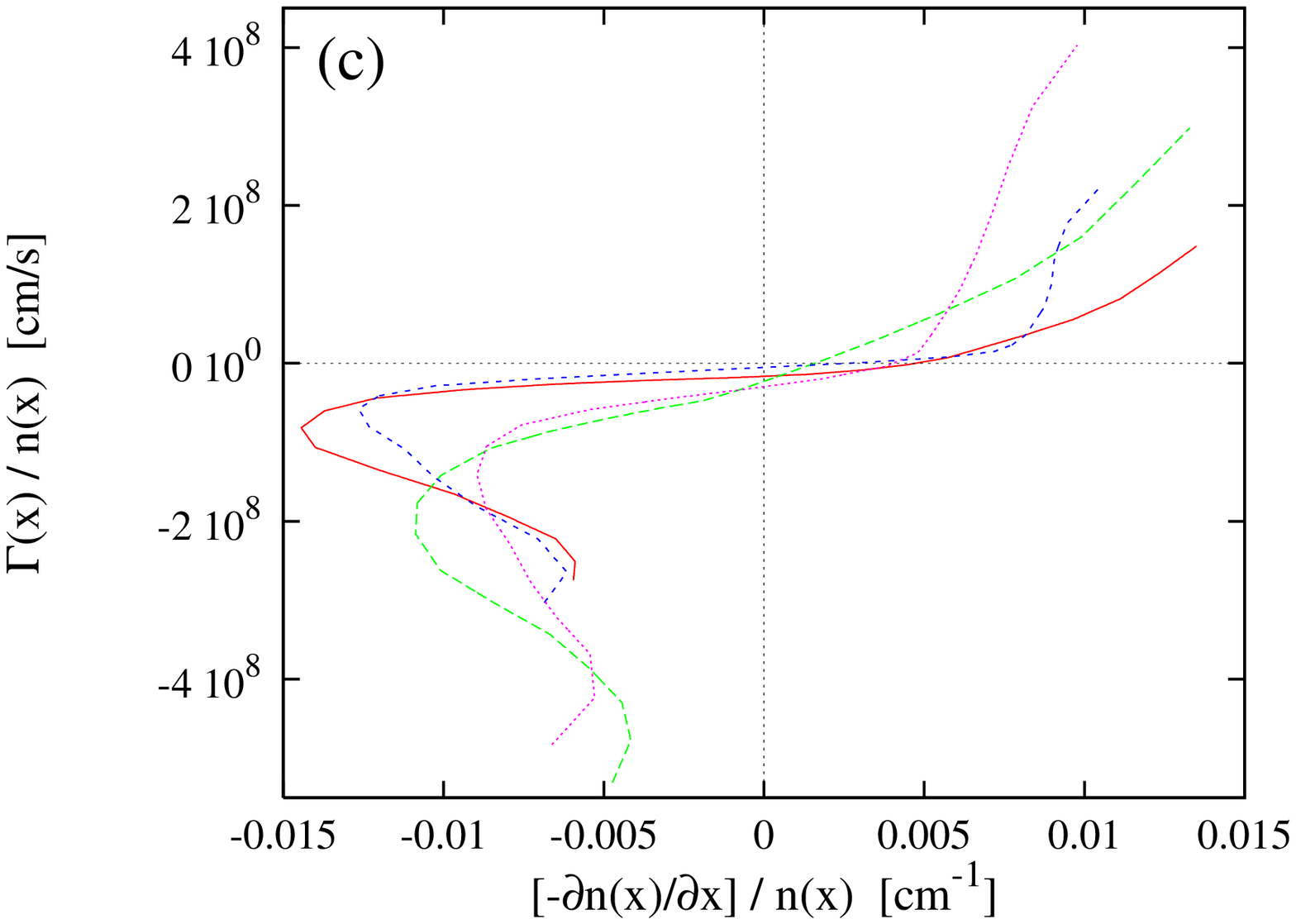}}
%\resizebox{18truecm}{!}{\includegraphics{diffnn4.ps},
%                         \includegraphics{diffnn10.ps}}
\resizebox{18truecm}{!}{\includegraphics{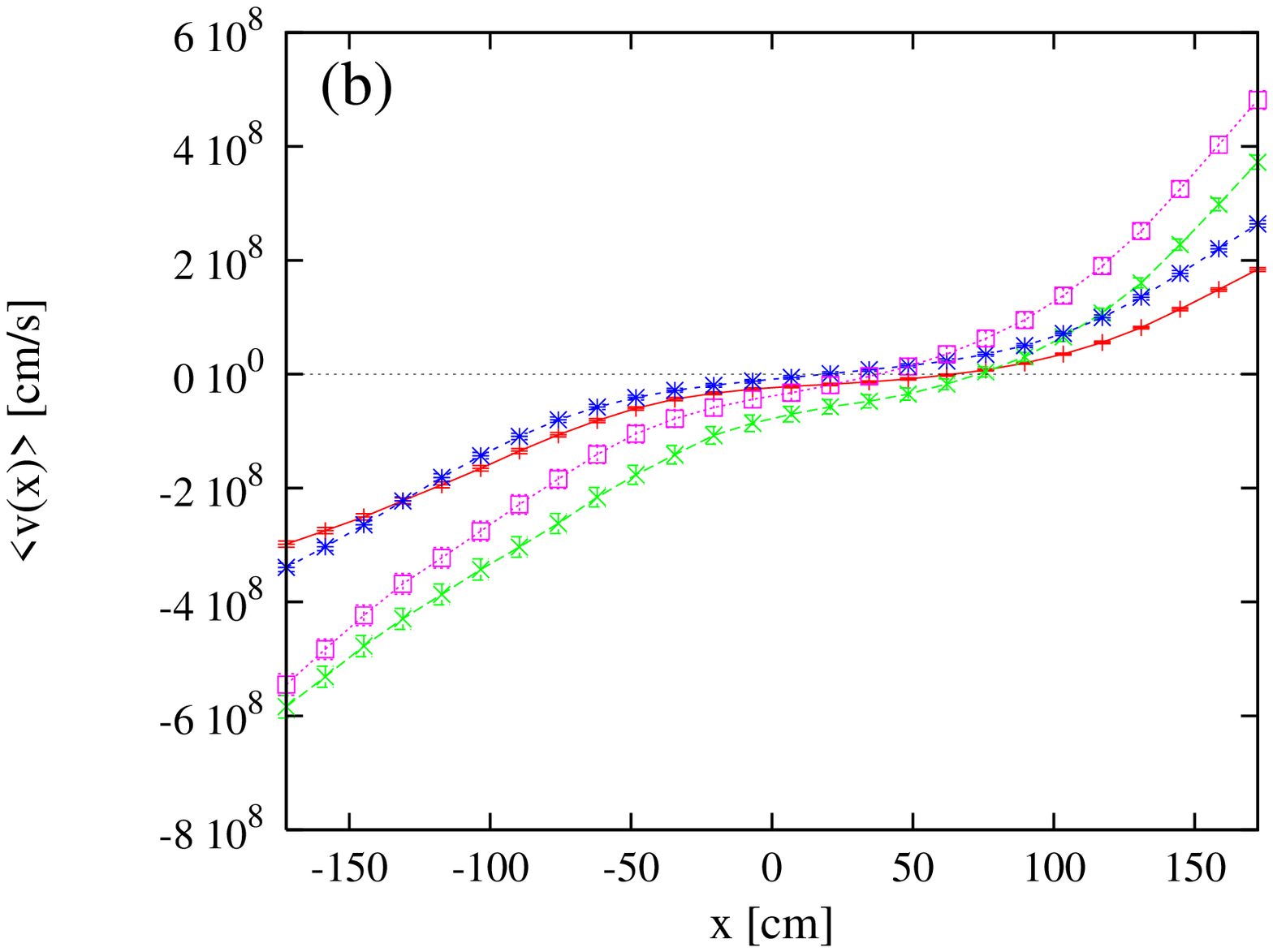},
                         \includegraphics{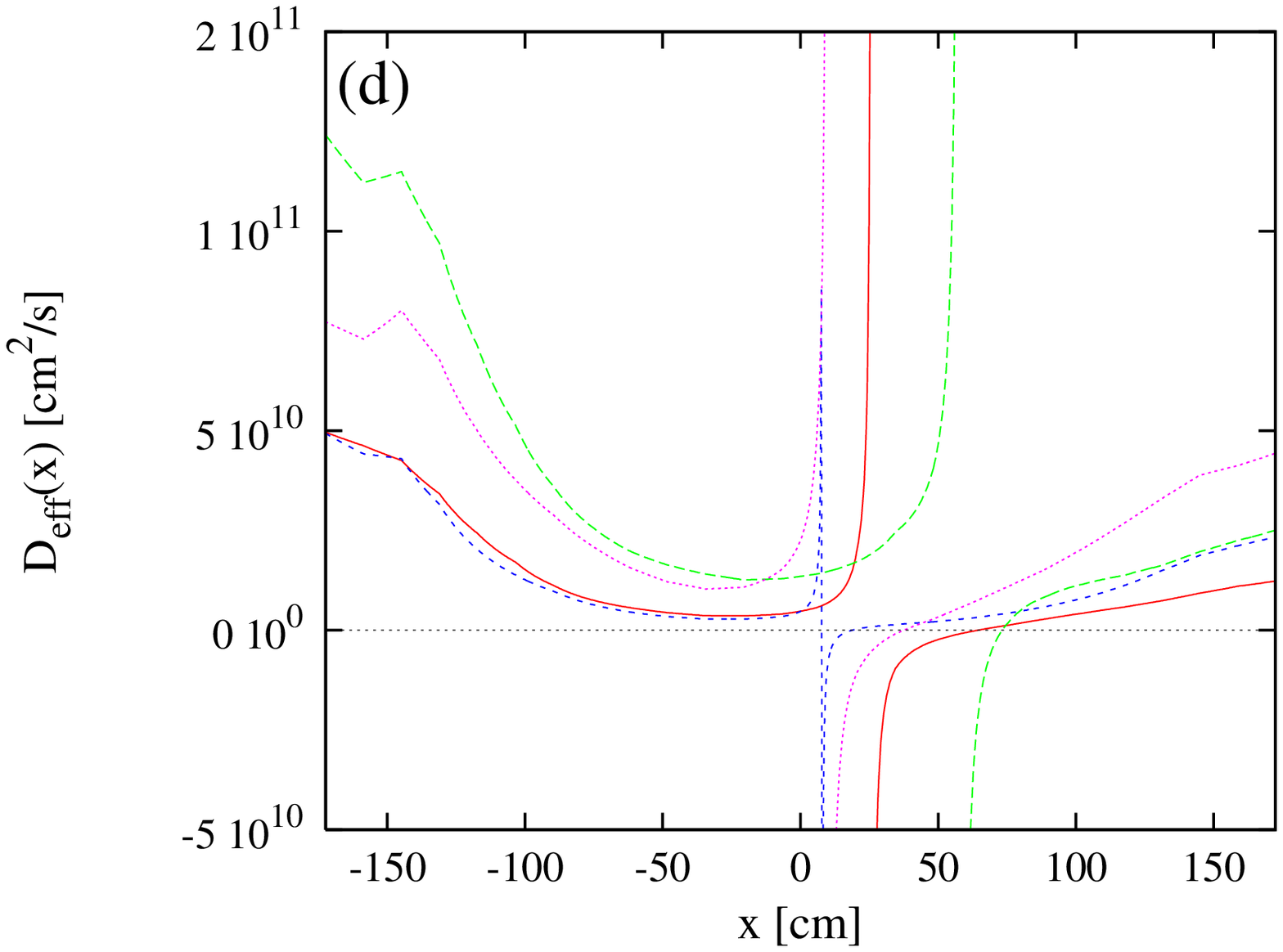}}
%\resizebox{18truecm}{!}{\includegraphics{diffnn8.ps},
%                         \includegraphics{diffnn11.ps}}
\caption{{\it Strong off-axis source:}
Shown are 
(a) $\Gamma(x)$ as a function of $x$;
(b) $\langle v(x)\rangle$ as a function of $x$ (with error-bars);
(c) $\Gamma(x)/n(x)$ as a function of $-\p_x n(x)/n(x)$; and
(d) $D_{eff}(x)$ as a function of $x$, 
%(e) $D(x)$ as a function of $x$; and
%(f) $V_{in}(x)$ as a function of $x$,
as yielded by the mixed model,
for the case of  
strong off axis loading
  with
    Gaussian  (solid - red) 
    and power-law (long dashes - green) distributed momentum increments, respectively,
and for the case of weak off axis loading
  again with
    Gaussian (short dashes - blue)
    and with power-law (dotted - violet) distributed momentum increments, respectively.
In all panels a horizontal reference line at the respective zero level is drawn
(small dashes - black), and in panel (c) a vertical reference line at $-\p_x n(x)/n(x)=0$ is
drawn (small dashes - black).
\label{diffg}
}
\end{figure*}

From a dynamical point of view, the particle flux is given as
\beq
\Gamma(x,t) = n(x,t) \, \langle v(x,t) \rangle ,
\eeq
where $\langle v(x,t) \rangle$ is the local average velocity
and $n(x,t)$ is as usual the particle density. In the following,
we focus on the final time, $t=t_f$, and we omit $t$ as
an argument. It turns out that the determination of 
$\langle v(x) \rangle$ is very sensitive to the occurrence
of large velocities, and due to the symmetry and finiteness of
the momentum space used in the solution of the CTRW equations,
the local mean found from the solution of the equations is
close to zero and cannot be considered meaningful. We thus
determine $\langle v(x) \rangle$ throughout with the 
use of Monte Carlo simulations, using typically $2\times 10^7$
particles to achieve a satisfying numerical precision. Such large 
numbers of particles are though too expensive in computing time
for the critical gradient model, so that we will present
only results for the mixed model.

Fig.\ \ref{diffg}(a) shows $\Gamma(x)$ as a function of $x$
in case of the
mixed model, for strong and weak off-axis loading, and
for Gaussian and power-law distributed momentum increments, 
respectively. 
The location where the flux changes direction (sign) is determined 
by the change of sign in $\langle v(x,t) \rangle$,
which is shown in Fig.\ \ref{diffg}(b). In all cases 
shown, there is basically a particle flux from a location
in between the density peak and the peak of the off axis source 
(at $x=100$) out-wards to the system boundaries, and both the outflow
velocity and the particle flux increase to-wards the boundaries. 
The particle flux is in all cases shown of the same order 
of magnitude.

In the classical approach to diffusion, particle transport is traditionally modeled
by the equation
\beq
\p_t n(x,t) = -\p_x \Gamma(x,t) + S_p(x,t) ,
\eeq
together with the assumption that the particle flux is of the form of 
Fick's law,
\beq
\Gamma^{(cl)}(x) = -D\, \p_x n(x) +V_{in} n(x) ,
\label{Gcl}
\eeq
with the diffusivity $D$, and where the pinch velocity $V_{in}$
is introduced to account for possible anomalous diffusion effects
(e.g.\ \cite{Lopez95}; the sign convention is such that if $V_{in}>0$ then the 
pinch is in the positive 
$x$-direction, i.e.\ motion to the right).
To see whether $\Gamma(x)$ is compatible with the functional form
of Eq.\ (\ref{Gcl}), we plot in Fig.\ \ref{diffg}(c) 
$\Gamma(x)/n(x)$ against $-\p_x n(x)/n(x)$. 
For negative values of $-\p_x n(x)/n(x)$, the particle flux 
has the same sign (direction) as $-\p_x n(x)/n(x)$, in accordance
with Eq.\ (\ref{Gcl}), there is though a clear non-uniqueness, where a 
density gradient can correspond to two, and, in a narrow range, even three different flux values,
which is in clear contradiction with Eq.\ (\ref{Gcl}). For positive
values of $-\p_x n(x)/n(x)$, the particle flux can be slightly negative,
opposite to the direction of $-\p_x n(x)$, and, depending
on the model, there is a sudden increase at large values of $-\p_x n(x)/n(x)$. 
Last, we note that the particle flux is close to 
but not zero when
the density gradient is zero, which is reminiscent of a small 'off-diagonal'
term, such as a pinch velocity.
We thus conclude that the diffusive behaviour is not 
compatible with the 
form of Eq.\ (\ref{Gcl}), even when assuming non-constant 
coefficients $D$ and $V_{in}$, or, in other words,
particle transport is not driven by density gradients in the 
models analyzed, and the characteristics of the particle fluxes
we find must be attributed to the
strong non-local nature of the diffusion process we study.
Similar results concerning non-uniqueness and the particular characteristic
shapes of the curves of $\Gamma(x)$ as a function of $-\p_x n(x)$
have also been reported by \cite{vanMil04b} for the critical 
gradient model in position space alone.

Despite the inadequateness of Eq.\ (\ref{Gcl}),
we can define
on the base of Eq.\ (\ref{Gcl}) an effective diffusivity,
\beq
D_{eff} = -\frac{\Gamma}{\p_x n(x)} ,
\label{Deff}
\eeq
whereby we neglect a possible pinch velocity.
Fig.\ \ref{diffg}(d) shows the spatial dependence of $D_{eff}$. In all cases,
the diffusivity increases to-wards the edges and it is lowest in the 
central region, mostly between the center of the box
and roughly the peak of the off-axis source at $x=100$. The singularities
appear since the particle fluxes do not vanish there where the density
gradient is zero, there is on off-set between the zeros of the fluxes
and the zeros of the density gradient. After all, the singularities 
are the clearest
feature in $D_{eff}$ that give a hint that the classical description might not be 
valid,
together with the slightly negative values of $D_{eff}$ in the
interval $[0,50]$, which indicate that the flux is not in the direction
opposite to the density gradient. Thus, the determination of
$D_{eff}$ yields a seemingly reasonable picture for the diffusion
process, with just minor irregularities, 
it is though
not suited to describe the diffusive processes we study,
since Eq.\ (\ref{Gcl}) is not an adequate description,
as revealed by Fig.\ \ref{diffg}(c).

Alternatively, the diffusivity can be determined through the 
more general definition that is based on the scaling of the mean square 
displacement $\langle \Delta r^2\rangle$ a particle undergoes in time $t$, 
as shown in Eq.\ (\ref{msd}). 
If we assume that a particle on the average travels
from the center of the system to the edge, i.e. a distance $L$, in a time
that equals the particle confinement time $\tau_p$, then on setting 
$\langle \Delta r^2\rangle = (L)^2$ and defining the diffusivity
as  $D_{MSD}=\langle \Delta r^2\rangle/\tau_p$ yields values that are roughly
one order of magnitude smaller
than $D_{eff}$ in Fig.\ \ref{diffg}(d). This approach is though
again problematic, since the definition of $D_{MSD}$ presupposes
that diffusion is of classical nature, such that 
$\langle \Delta r^2\rangle = D_{MSD}t$, which would though first have to be 
confirmed against the more general behaviour of 
$\langle \Delta r^2\rangle \propto t^\gamma$, with $\gamma$ characterizing
the basic nature of the transport process.

\begin{figure*}
%%\resizebox{18truecm}{!}{\includegraphics{qdiff20a.ps},
%%                         \includegraphics{qdiff141a.ps},\includegraphics{qdiff17a.ps}}
%%\resizebox{18truecm}{!}{\includegraphics{qdiff20b.ps},
%%                         \includegraphics{qdiff141b.ps},\includegraphics{qdiff17b.ps}}
%%\resizebox{18truecm}{!}{\includegraphics{qdiff20c.ps},
%%                         \includegraphics{qdiff141c.ps},\includegraphics{qdiff17c.ps}}
%%\resizebox{18truecm}{!}{\includegraphics{qdiff20d.ps},
%%                         \includegraphics{qdiff141d.ps},\includegraphics{qdiff17d.ps}}
%%\caption{Mixed power-law/Gaussian increments $\Delta x$, strong spatially 
\resizebox{18truecm}{!}{\includegraphics{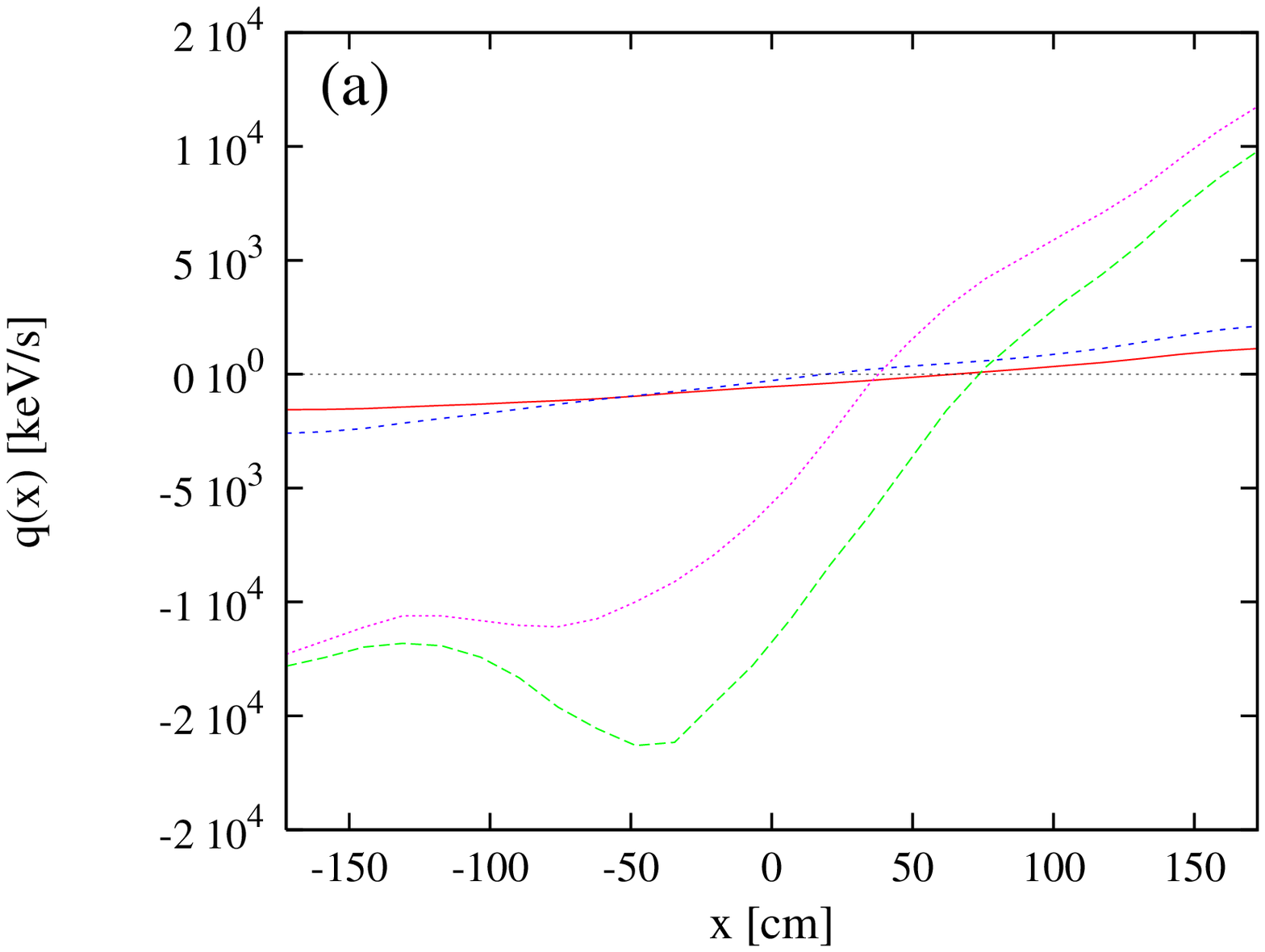},
                         \includegraphics{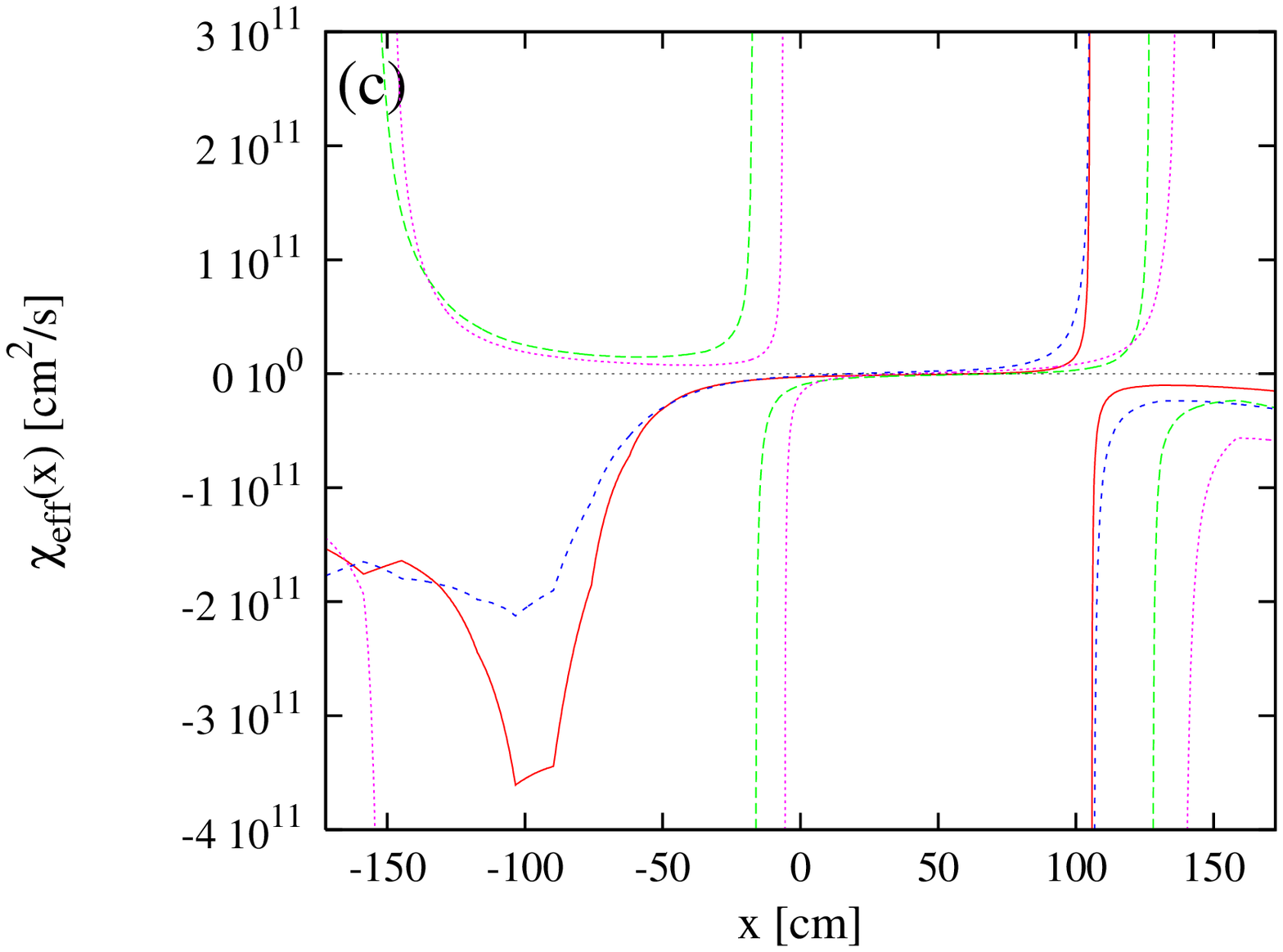}}
%%\resizebox{18truecm}{!}{\includegraphics{diffqq5.ps},
%%                         \includegraphics{diffqq9.ps}}
%\resizebox{18truecm}{!}{\includegraphics{diffqq5.ps},
%                         \includegraphics{diffqq6.ps}}
\resizebox{9truecm}{!}{\includegraphics{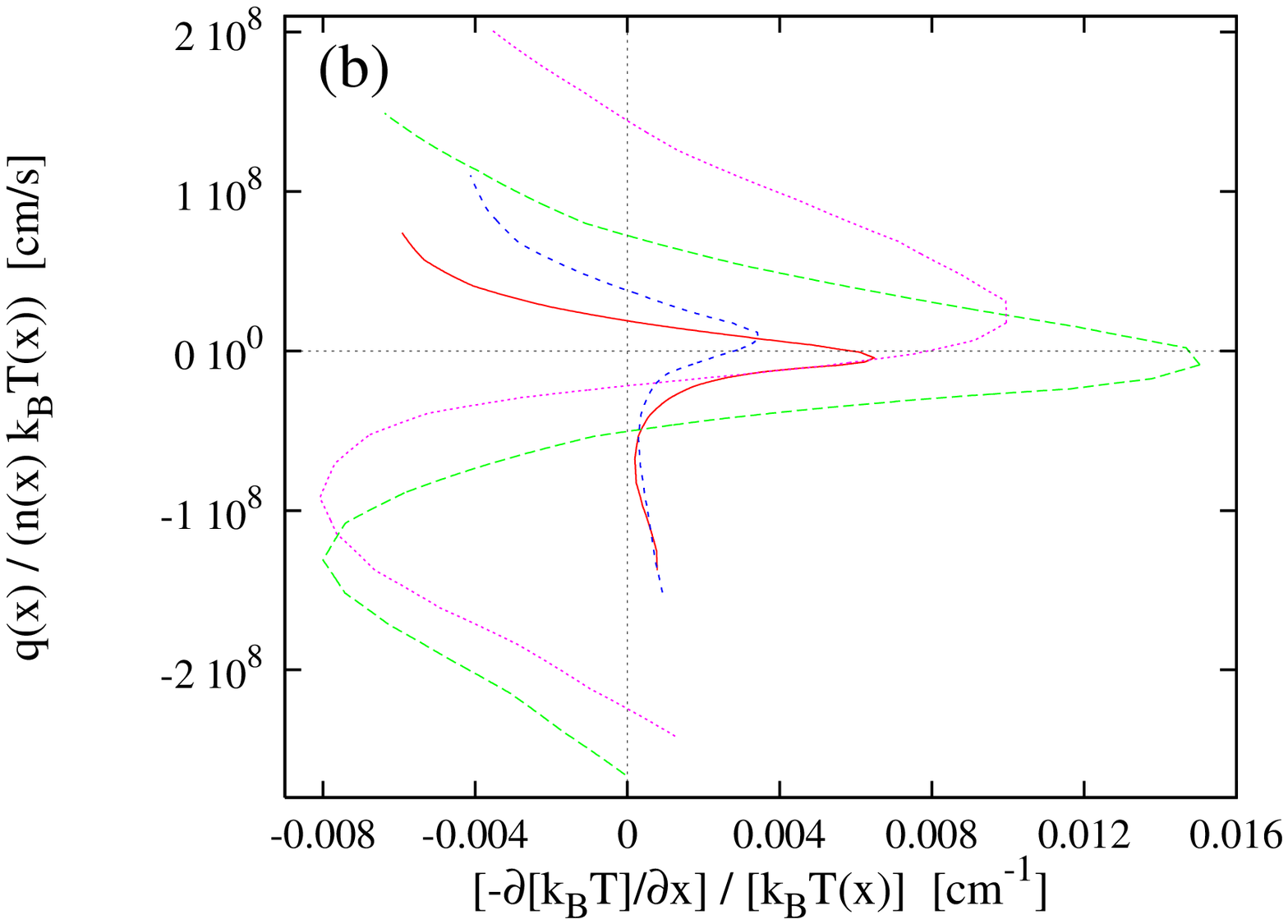}}
%%\resizebox{18truecm}{!}{\includegraphics{diffqq6.ps},
%%                         \includegraphics{diffqq10.ps}}
\caption{{\it Strong off-axis source:}
Shown are 
(a) $q(x)$ as a function of $x$;
(b) $q(x)/[n(x)\,k_B T(x)]$ as a function of $-\p_x [k_BT(x)]/[k_BT(x)]$; and
(c) $\chi_{eff}(x)$ as a function of $x$,
%(e) $\chi(x)$ as a function of $x$; and
%(f) $V_{T}(x)$ as a function of $x$,
as yielded by the mixed model,
for the case of  
strong off axis loading
  with
    Gaussian (solid - red) 
    and power-law (long dashes - green) distributed momentum increments, respectively,
and for the case of weak off axis loading
  again with
    Gaussian (short dashes - blue)
    and with power-law (dotted - violet) distributed momentum increments, respectively.
In all panels a horizontal reference line at the respective zero level is drawn
(small dashes - black), and in panel (b) a vertical reference line at $-\p_x k_BT(x)/T(x)=0$ is
drawn (small dashes - black).
\label{diffq}
}
\end{figure*}

\subsubsection{Heat diffusion}

The dynamic energy flux is determined as  
\beq
q(x) = \frac{1}{2} k_B T(x)\,n(x) \, \langle v(x) \rangle
\equiv \frac{1}{2} k_B T(x)\,\Gamma(x) ,
\eeq
where we again have assumed $t=t_f$ and suppress $t$ as an argument.
Fig.\ \ref{diffq}(a) shows $q(x)$ as a function of $x$ for 
weak
and strong off-axis fueling, as yielded by the mixed model for the cases of Gaussian and 
power-law distributed momentum increments, respectively.
The basic shape of the heat fluxes is similar 
to the ones of the particle fluxes, and the change in direction is again
at the locations where $\langle v(x)\rangle$ changes sign. Notably though,
the two cases of power-law distributed momentum increments lead to 
an almost one order of magnitude higher heat flux %to-wards the edges 
than the two cases of Gaussian distributed momentum increments, which must 
be attributed to the increased energy injection in the acceleration events.

In the classical context, heat diffusion is modeled
by 
\beq
\frac{1}{2}n(x,t) \,\p_t k_B T(x,t) = -\p_x q(x,t) + S_E(x,t) ,
\eeq
in combination with Fourier's law for the heat flux, 
\beq
q^{(cl)}(x) = -\chi \,n(x) \,\p_x [k_B T(x)] + V_H \,n(x)\, k_B T(x)  ,
\label{qcl}
\eeq
with $\chi$ the heat diffusivity and $V_H$ a thermal pinch velocity
as an additional 'off-diagonal' term to account for anomalous transport
effects (often though, $V_H$ is neglected; e.g.\ \cite{Lopez95}). 
As in the case of the particle flux, it is of interest to see 
whether the heat fluxes we find are compatible with the form of 
Eq.\ (\ref{qcl}).
In Fig.\ \ref{diffq}(b), we plot $q(x)/[n(x)T(x)]$ against 
$-\p_x [k_B T(x)]/[k_BT(x)]$. 
Obviously, there are again strong features of non-uniqueness: In case of 
power-law
distributed momentum increments, the non-uniqueness is clearly present
for negative and positive values of $-\p_x [k_B T(x)]/[k_BT(x)]$.
In case of Gaussian distributed momentum increments, the non-uniqueness
is restricted to positive values of $-\p_x [k_B T(x)]/[k_BT(x)]$, i.e.\ negative
gradients, which spatially occur in the region 
from the left edge of the simulation box to the location of the
peak of the source at $x=100$ (see Figs.\ \ref{fq3bT}(a), 
\ref{fq3T}(a), and \ref{fq3Tline}(a)), and at negative values of $-\p_x [k_B T(x)]/[k_BT(x)]$, 
the heat flux is 'uphill', i.e.\ opposite to the direction of
the negative temperature gradient. 
In none of the four cases the heat flux is zero when the
temperature gradient is zero, which in the frame of Eq.\ (\ref{qcl})
would be interpreted as the presence of an 'off-diagonal term', e.g.\ 
in the form of a pinch velocity.
Noteworthy, the non-uniqueness
in the cases of Gaussian distributed momentum increments
appears despite the fact that the random 
walk in momentum space is of local nature, and it must be
caused by the coupling of momentum space dynamics to the non-local
transport in position space.
The non-uniqueness, as in case of the particle fluxes,
is a result of the non-locality of the process, 
and it basically makes the transport incompatible with the 
structure of Eq.\ (\ref{qcl}).

Despite the incompatibility of Eq.\ (\ref{qcl}) with the transport
process we study,
we define, as in the case of particle diffusion, an effective heat diffusivity 
\beq
\chi_{eff} = - \frac{q(x)}{n(x)\p_x [k_B T(x)]} ,
\label{cheff}
\eeq
and where we again neglect a possible pinch term.
$\chi_{eff}$ is shown in Fig.\ \ref{diffq}(c). 
There occur again singularities, 
as in case of the particle diffusivity,
since the heat flux is not zero there where the temperature 
gradient is zero. The cases with power-law momentum increments
are qualitatively different from the cases with Gaussian momentum
increments, basically because the temperature profiles are qualitatively
different (see Fig.\ \ref{fq3Tline}). In the cases with power-law
momentum increments, the temperature is peaked near the 
center, and the temperature gradient has three zeros. The effective diffusivity
gives in these cases the following picture: diffusion is reduced in between the 
temperature peak and the peak of the off-axis injection source, and it increases outside this 
region to-wards the edges. In the cases of Gaussian momentum increments,
heat diffusion is again strongly reduced between the center 
and the off-axis injection site, and to-wards the edges
it is seemingly uphill, in the direction opposite to the
negative temperature gradient.
Again though it is clear that the picture given by 
the behaviour of $\chi_{eff}$ is not representative of the actual
diffusion process that is going on, since Eq.\ (\ref{qcl}) 
is itself not a valid approach.

\section{Summary and conclusion \label{SecV}}

We introduced a set of two coupled equations that describe 
the combined CTRW that includes, besides 
position space, also momentum space. The equations are of integral
form, and they are close to but formally not of the convolution type,
so that ways to treat the equations in Fourier Laplace space,
e.g.\ to solve them analytically, to cast them into a different form, 
%of a single integral equation or an integro-differential equation, 
or to construct a corresponding
fractional diffusion equation, are not obvious.

A way to solve the equations numerically was presented,
which is a variant of the pseudospectral method in three dimensional,
position-momentum-time space, and it is based on the expansion in terms of Chebyshev 
polynomials. The method works reliably for the case of linear equations, 
a method to solve non-linear equations has not yet been developed, so-far.
The method could thus be applied only to the mixed model, 
and not to the critical gradient model, which contains
a non-linearity, and for which the presented results were all 
derived with Monte-Carlo simulations. Monte Carlo simulations were 
also performed to verify the numerical solution of the combined CTRW equations
in case of the mixed model.

In the application to toroidally confined plasmas, 
we considered an off-axis fueling experiment, with a weak 
and a strong cold off-axis source, which is known experimentally
to give rise to anomalous transport phenomena.
Two variants of the combined CTRW have been considered, constructed 
with the purpose for them to adequately model
the specific characteristics 
of toroidally confined plasmas:
(i) the mixed model, where the spatial
increments depend on the position of the particles, 
and (ii) the critical gradient model, 
in which the position increments depend on the local density gradient.

The main scope in the applications was to reveal what new aspects
can be modeled and explained if momentum space is included
in a model of non-local, anomalous transport.
To explore the role of momentum space dynamics, all applications 
were done in two versions, one with small, Gaussian distributed momentum increments,
which can be interpreted as low-level heating, and one with
occasionally large, power-law distributed momentum increments, 
which corresponds
to the case of intense heating.
The results can be summarized as follows:

(i) All variants of the mixed and the critical gradient model basically
yield peaked density profiles, with a few exceptions of plateau shaped
densities. The density profiles exhibit different degrees of 
stiffness, in the sense of to what degree a model is able 
to maintain the density peak close to the center in the presence
of off-axis
sources. The highest stiffness is realized in the mixed model.
Generally, intense heating reduces the density profile stiffness.

(ii) The temperature profiles show no stiffness at all
with low-level heating,
the profiles become though very stiff with intense heating.
Intense heating has thus on
the temperature profile stiffness
the opposite effect it has on the density profile stiffness.

(iii) We find for both models in all variants and independent 
of the source term an universal distribution of the kinetic energy, 
namely a power-law
with index $-1$, also in the cases where the momentum
increments follow a Gaussian distribution. 

(iv) The particle confinement time is directly proportional 
to the density, and it is largest for the critical gradient model.
Intense heating reduces in all cases the particle
confinement time, and therewith the particle density in the system. 

(v) The energy confinement time is larger in the critical gradient 
model than in the mixed model, and, opposite to its effect on the particle
confinement, 
intense heating leads to
an increase of the energy confinement time.

(vi) The dynamic particle flux is of simple shape, plotting it though
against the density gradient reveals clear features of non-uniqueness, 
which are a consequence of the non-locality of the transport process, and 
which make the process incompatible with a classical approach and Fick's
law [Eq.\ (\ref{Gcl})], also when extended with a pinch velocity.
The effective particle diffusivity, defined through Fick's law, 
gives nonetheless 
a seemingly reasonable picture.

(vii) The dynamic heat flux is of similar shape as the particle flux,
it is though increased by roughly an order of magnitude 
in the case of intense heating.
The heat flux as a function
of the temperature gradient also exhibits clear features of non-uniqueness,
so that also in this case the classical Fourier's law [Eq.\ (\ref{qcl})] 
is incompatible with the actual heat transport process. This holds true
also in the cases where the momentum increments follow a Gaussian 
distribution and the random walk in momentum space is of classical, quasi-local nature.
The effective heat diffusivity determined according to 
Fourier's law gives a less reasonable picture than the particle
diffusivity, with more 
singularities and negative values (seeming uphill transport).

We thus conclude that particle and heat diffusivities or pinch velocities determined 
through a generalized Fick's 
and Fourier's 
law, respectively, 
are not adequate tools to characterize a transport process that is 
of non-local nature in at least the position space, or  
in position and  momentum space.

After all, the inclusion of momentum
space truly extends the CTRW formalism, it allows to model 
new features that belong to momentum space itself, and it substantially modifies 
and brings forth new 
aspects of position space dynamics.

\begin{acknowledgments}
The author is grateful to K.\ Arzner and L. Vlahos for helpful discussions. 
This work was supported under the Contract of Association ERB 5005 CT 99 0100 between 
the European Atomic Energy Community and the Hellenic Republic. The sponsors do not 
bear any responsibility for the contents of this work.
\end{acknowledgments}

\begin{widetext}

\appendix

\section{The pseudospectral method for the solution of integral equations\label{AppA}}

The pseudospectral method we apply for the solution of the 
integral equations follows the method in Ref.\ \cite{Mihaila02} for
one-dimensional integral equations, which is extended here to the three dimensional case.
Detailed properties of Chebyshev polynomials and a description of the 
pseudospectral method for differential equations can be found e.g.\ in Ref.\ \cite{Peyret02}.
The Chebyshev polynomials $T_j$, which we use as an expansion base, are defined as
\beq
T_j(z^T) = \cos(j \arccos(z^T)) ,
\eeq
for $j=0,1,...,N$, and where $z^T\in[-1,1]$.
To use variables $z$ in a general interval, $z\in [a,b]$, we  
need to establish a transformation between $z$ and $z^T$.
For the cases of position ($z=x\in [-L,L]$), and 
time ($z=t\in[0,t]$), respectively, we choose the simple form
\beq
z = z(z^T) = z^T B_z +A_z ,
\label{trans1}
\eeq
with $A_z=\frac{1}{2}(b+a)$ and 
$B_z=\frac{1}{2}(b-a)$. 
A different choice must be made for the momentum,
since the unbounded momentum space should be sufficiently covered   
with grid points from small to very large momentum values. 
Evaluating different functional forms, we found best coincidence with 
Monte-Carlo simulations when introducing a finite momentum interval,
$p\in[-p_2,p_2]$ and using the transformation 
\beq
p = p(p^T) = B_p\, \left[{\rm exp}\left(A_p \, p^T\right) - 1 \right]
\label{trans2}
\eeq
for $p^T\geq 0$, and $p = -B_p\, \left[ {\rm exp}\left(A_p \, \vert p^T\vert \right) -1\right]$
for $p^T < 0$, and where $p^T\in[-1,1]$.
The parameters $A_p$ and $B_p$ are adjusted by prescribing the 
largest ($p_2$) and the smallest ($\min \{\vert p_k\vert\}$)
momentum value, which in turn are chosen such that 
good coincidence with the results from Monte-Carlo simulations is achieved.

The grid points we use are the extrema of $T_N$ in the interval $[-1,1]$ 
(Gauss-Lobatto grid),
\beq
z^T_{k} = -\cos\left(\frac{k\pi}{N}\right),\ \ \ \ k=0,1,2,...,N, 
\eeq
from which the grid-points $z_k$ in the actual position, time, and momentum interval
are determined 
by means of the transformations in Eqs.\ (\ref{trans1}) and (\ref{trans2}), respectively,
\beq
z_k=z(z^T_k) .
\eeq

Before turning to the three dimensional case, we illustrate the Chebyshev expansion
in a one-dimensional example:
The expansion of a 
function $f(z)$ 
writes
\beq
f(z)=\sum\limits_{j=0}^N{}^{\prime\prime} b_j T_j(z^T(z)) ,
\label{fC1}
\eeq
where $z^T(z)$ is defined as the inverse of the relations in Eqs.\ (\ref{trans1}) (for 
$z=x$ or $z=t$) and (\ref{trans2}) (for 
$z=p$), respectively. The double prime on the sum means that the first and the last term of the 
sum are multiplied by a factor $1/2$. The expansion coefficients are given as 
\beq
b_j=\frac{2}{N}\sum\limits_{k=0}^N{}^{\prime\prime} f(z_k) T_j(z^T(z_k))
=\frac{2}{N}\sum\limits_{k=0}^N{}^{\prime\prime} f(z_k) T_j(z^T_{k})   .
\label{fC2}
\eeq
Following Ref.\ \cite{Mihaila02}, it is useful for numerical purposes to reformulate 
the expansion by inserting the coefficients $b_j$
into Eq.\ (\ref{fC1}), which yields
\beq
f(z)=
\sum\limits_{k=0}^N{}^{\prime\prime} f(z_k) 
\Bigg[
\frac{2}{N}
\sum\limits_{j=0}^N{}^{\prime\prime} 
T_j(z^T_k)
T_j(z^T(z))
\Bigg]        ,
\label{fC3}
\eeq
and in which the expansion coefficients are just the values
of the original function $f(z)$ at the grid points. In this form, the numerical treatment 
of the integral equations can be done in direct space, there is no need for 
transforming to and working in the space of the expansion coefficients.

For the three dimensional case, 
we first form a tensor product basis, with basis functions 
\beq
T_i\left(x^T(x)\right)\,T_j\left(p^T_x(p_x)\right)\,T_n\left(t^T(t)\right) .
\eeq
Analogously then to the one dimensional case in Eq.\ (\ref{fC1}),
$Q(x,p_x,t)$ can be expanded as 
\beq
Q(x,p_x,t) = 
\sum\limits_{i=0}^{N_x}{}^{\prime\prime} 
\sum\limits_{j=0}^{N_p}{}^{\prime\prime} 
\sum\limits_{n=0}^{N_t}{}^{\prime\prime} 
\hat{Q}_{ijn}\,T_i\left(x^T(x)\right)\,
               T_j\left(p^T_x(p_x)\right)\,
               T_n\left(t^T(t)\right)     ,
\label{Qexpexp}
\eeq
with the expansion coefficients $\hat{Q}_{ijn}$ given by (cf.\ Eq.\ (\ref{fC2}))
%\bea
\beq
\hat{Q}_{ijn} =%&=& 
\Bigg[\frac{2}{N_t}\sum\limits_{m=0}^{N_t}{}^{\prime\prime} 
\Bigg[\frac{2}{N_p}\sum\limits_{l=0}^{N_p}{}^{\prime\prime} 
\Bigg[\frac{2}{N_x}\sum\limits_{k=0}^{N_x}{}^{\prime\prime} 
Q(x_k,p_{x,l},t_m) \,
%\nonumber\\
%&&\ \ \ \ \ \ \ \times 
T_i(x^T_k)\Bigg] 
T_j(p^T_{x,l})\Bigg]
T_n(t^T_m)\Bigg]  .
\eeq
As for Eq.\ (\ref{fC3}) in one dimensions, we insert the expansion coefficients
into Eq.\ (\ref{Qexpexp}), so that 
the expansion is expressed in terms of the values of $Q(x,p_x,t)$
at the grid points $x_k,p_{x,l},t_m$. After rearranging and regrouping the sums, 
the expansion takes the final form 
\bea
Q(x,p_x,t) &=& 
\sum\limits_{m=0}^{N_t}{}^{\prime\prime} 
\sum\limits_{l=0}^{N_p}{}^{\prime\prime} 
\sum\limits_{k=0}^{N_x}{}^{\prime\prime} 
Q(x_k,p_{x,l},t_m) 
\Bigg[
\frac{2}{N_x}
\sum\limits_{i=0}^{N_x}{}^{\prime\prime} 
T_i(x^T_k) 
T_i\left(x^T(x)\right)
\Bigg]
\Bigg[
\frac{2}{N_p}
\sum\limits_{j=0}^{N_p}{}^{\prime\prime} 
T_j(p^T_{x,l})
T_j\left(p^T_x(p_x)\right)
\Bigg]
\nonumber\\
&&\ \ \ \ \ \ \ \times 
\Bigg[
\frac{2}{N_t}
\sum\limits_{n=0}^{N_t}{}^{\prime\prime} 
T_n(t^T_m)
T_n\left(t^T(t)\right)
\Bigg]              .
\label{expa}
\eea

Turning now to the integral equation for $Q(x,p_x,t)$, Eq.\ (\ref{rwQ1Db}), 
we insert the expansion of Eq.\ (\ref{expa}) for $Q(x^\prime,p_x^\prime,t^\prime)$
under the integral in Eq.\ (\ref{rwQ1Db}), which yields 
\begin{eqnarray}
Q(x,p_x,t)
&=&
\int d p_x^\prime 
\int\limits_{\max[x-\vert v_x\vert t,\,L]}^{\min[x+\vert v_x\vert t,\,L]} d x^\prime 
 \nonumber \\
&&\ \ \times
\sum\limits_{m=0}^{N_t}{}^{\prime\prime} 
\sum\limits_{l=0}^{N_p}{}^{\prime\prime} 
\sum\limits_{k=0}^{N_x}{}^{\prime\prime} 
Q(x_k,p_{x,l},t_m) 
\Bigg[
\frac{2}{N_x}
\sum\limits_{i=0}^{N_x}{}^{\prime\prime} 
T_i(x^T_k) 
T_i\left(x^T(x^\prime )\right)
\Bigg]
\Bigg[
\frac{2}{N_p}
\sum\limits_{j=0}^{N_p}{}^{\prime\prime} 
T_j(p^T_{x,l})
T_j\left(p^T_x(p_x^\prime)\right)
\Bigg]
\nonumber\\
&&\ \ \ \ \ \ \ \ \ \ \ \ \ \ \ \ \ \ \ \ \ \ \ \ \ \ \ \ \ \ \ \ \ \ \ \ \ \ 
\ \ \ \  
\times\Bigg[
\frac{2}{N_t}
\sum\limits_{n=0}^{N_t}{}^{\prime\prime} 
%\nonumber\\
%&&\ \ \ \ \ \ \ \times 
T_n(t^T_m)
T_n\left(t^T(t-\vert x - x^\prime\vert/\vert v_x\vert)\right)
\Bigg]
\nonumber \\
&&\ \ \times q_{\Delta x}(x-x^\prime ) 
          \, q_{\Delta p_x}(p_x-p_x^\prime )  \nonumber \\
&+& 
\delta(t)P(x,p_x,0)  
+ S(x,p_x,t)      .
\label{interm2}
\end{eqnarray}
In the same way, Eq.\ (\ref{rwP1Dd}) for $P(x,p_x,t)$ is treated,
we insert Eq.\ (\ref{expa}) for $Q(x^\prime,p_x^\prime,t^\prime)$ under the integral,
which leads to 
\begin{eqnarray}
P(x,p_x,t)
&=&
\frac{1}{\vert v_x \vert}
\int d p_x^\prime 
\int\limits_{\max[x-\vert v_x\vert t,\,L]}^{\min[x+\vert v_x\vert t,\,L]} d x^\prime 
 \nonumber \\
&&\ \ \times
\sum\limits_{m=0}^{N_t}{}^{\prime\prime} 
\sum\limits_{l=0}^{N_p}{}^{\prime\prime} 
\sum\limits_{k=0}^{N_x}{}^{\prime\prime} 
Q(x_k,p_{x,l},t_m) 
\Bigg[
\frac{2}{N_x}
\sum\limits_{i=0}^{N_x}{}^{\prime\prime} 
T_i(x^T_k) 
T_i\left(x^T(x^\prime )\right)
\Bigg]
\Bigg[
\frac{2}{N_p}
\sum\limits_{j=0}^{N_p}{}^{\prime\prime} 
T_j(p^T_{x,l})
T_j\left(p^T_x(p_x^\prime)\right)
\Bigg]
\nonumber\\
&&\ \ \ \ \ \ \ \ \ \ \ \ \ \ \ \ \ \ \ \ \ \ \ \ \ \ \ \ \ \ \ \ \ \ \ \ \ \ 
\ \ \ \  
\times\Bigg[
\frac{2}{N_t}
\sum\limits_{n=0}^{N_t}{}^{\prime\prime} 
%\nonumber\\
%&&\ \ \ \ \ \ \ \times 
T_n(t^T_m)
T_n\left(t^T(t-\vert x - x^\prime\vert/\vert v_x\vert)\right)
\Bigg]
\nonumber \\
&&\ \ \times \Psi_{\Delta x}(\vert x-x^\prime\vert ) 
          \, q_{\Delta p_x}(p_x-p_x^\prime )  .
\label{interm3}
\end{eqnarray}

%\subsubsection{Collocation}
The next step consists in collocating the equations, i.e.\ we consider the equations 
only at the grid points and replace the free variables $(x,p_{x},t)$ by  
their values $(x_a,p_{x,b},t_c)$ at the grid-points, with the free indices
$a,b,c$ running over the same range as the indices $k,l,m$. 
In this way, Eq.\ (\ref{interm2}) for $Q(x,p_x,t)$ takes the form 
\begin{eqnarray}
Q(x_a,p_{x,b},t_c) 
&=&
\int\limits_{\max[x-\vert v_x\vert t,\,L]}^{\min[x+\vert v_x\vert t,\,L]} d x^\prime 
\int d p_x^\prime  \nonumber \\
&&\ \ \times
\sum\limits_{m=0}^{N_t}{}^{\prime\prime} 
\sum\limits_{l=0}^{N_p}{}^{\prime\prime} 
\sum\limits_{k=0}^{N_x}{}^{\prime\prime} 
Q(x_k,p_{x,l},t_m) 
\Bigg[
\frac{2}{N_x}
\sum\limits_{i=0}^{N_x}{}^{\prime\prime} 
T_i(x^T_k) 
T_i\left(x^T(x^\prime)\right)
\Bigg]
\Bigg[
\frac{2}{N_p}
\sum\limits_{j=0}^{N_p}{}^{\prime\prime} 
T_j(p^T_{x,l})
T_j\left(p^T_x(p_{x}^\prime)\right)
\Bigg]
\nonumber\\
&&\ \ \ \ \ \ \ \ \ \ \ \ \ \ \ \ \ \ \ \ \ \ \ \ \ \ \ \ \ \ \ \ \ \ \ \ \ \ 
\ \ \ \  
\times\Bigg[
\frac{2}{N_t}
\sum\limits_{n=0}^{N_t}{}^{\prime\prime} 
%\nonumber\\
%&&\ \ \ \ \ \ \ \times 
T_n(t^T_m)
T_n\left(t^T(t_c-\vert x_a - x^\prime\vert/\vert v_{x,b}\vert)\right)
\Bigg]
\nonumber \\
&&\ \ \times q_{\Delta x}(x_a- x^\prime) 
          \, q_{\Delta p_x}(p_{x,b}- p_x^\prime) \nonumber \\
&+& 
\delta(t_c) P(x_a,p_{x,b},0)  
+ S(x_a,p_{x,b},t_c) .
\label{interm1}
\end{eqnarray}
In the last step, we remove the double primes on the sums, introducing the explicit factors
\beq
C_i: = \left\{ \begin{array}{ll} 
                  \frac{1}{2}, & i=0,N \\
                   1,          & i=1,2,...,N-1 \\
              \end{array}\right.  ,
\eeq
and we rearrange the sums and integrals in Eq.\ (\ref{interm1}) such that the  
structure of their mutual dependencies becomes more obvious, 
\begin{eqnarray}
Q(x_a,p_{x,b},t_c)
&=&
\sum\limits_{m=0}^{N_t}{} 
\sum\limits_{l=0}^{N_p}{} 
\sum\limits_{k=0}^{N_x}{} 
Q(x_k,p_{x,l},t_m) 
\,C_k \frac{2}{N_x} \sum\limits_{i=0}^{N_x}{} C_i T_i(x^T_k) 
\,C_l \frac{2}{N_p} \sum\limits_{j=0}^{N_p}{} 
\,C_j T_j(p^T_{x,l}) C_m \frac{2}{N_t} \sum\limits_{n=0}^{N_t}{} C_n
%\nonumber\\
%&&\ \ \ \ \ \ \ \times 
T_n(t^T_m)
\nonumber\\
&&\ \ \ \ \ \ \ \times
\int\!\! d p_x^\prime \,
T_j\left(p^T_x(p_{x}^\prime)\right) \,q_{\Delta p_x}(p_{x,b}- p_x^\prime)
\nonumber\\
&&\ \ \ \ \ \ \ \times
\int\limits_{\max[x-\vert v_x\vert t,\,L]}^{\min[x+\vert v_x\vert t,\,L]}\!\! d x^\prime\, 
T_i\left(x^T(x^\prime)\right) 
\,T_n\left(t^T(t_c-\vert x_a - x^\prime\vert/\vert v_{x,b}\vert)\right) 
\,q_{\Delta x}(x_a-x^\prime)
\nonumber \\
&+& 
\delta(t_c) \, P(x_a,p_{x,b},0)  
+ 
S(x_a,p_{x,b},t_c)    .
\label{finalQ}
\end{eqnarray}
If the distributions of increments $q_{\Delta x}$ and $q_{\Delta x}$ do not contain 
any information on $P(x,p_x,t)$ and $Q(x,p_x,t)$, as in the case of the mixed model, then 
the integrals are over given functions and can be done with any appropriate numerical
method. Eq.\ (\ref{finalQ})
is then a linear equation for the $(N_x+1)\times (N_p+1)\times (N_t+1)$ 
values of $Q(x_a,p_{x,b},t_c)$ at the grid points, and it can be solved with
any numerical method for linear systems of equations. 

The propagator $P(x,p_x,t)$ is determined by collocating Eq.\ (\ref{interm3}) at the grid points
[$(x,p_{x},t) \to (x_a,p_{x,b},t_c)$, in the same way as for $Q(x,p_x,t)$ in Eq.\ (\ref{interm1})], 
which, after rearranging the
sums and integrals, yields
\begin{eqnarray}
P(x_a,p_{x,b},t_c)
&=&
\frac{1}{\vert v_{x,b}\vert }
\sum\limits_{m=0}^{N_t}{} 
\sum\limits_{l=0}^{N_p}{} 
\sum\limits_{k=0}^{N_x}{} 
Q(x_k,p_{x,l},t_m) 
\,C_k \frac{2}{N_x} \sum\limits_{i=0}^{N_x}{} C_i T_i(x^T_k) 
\,C_l \frac{2}{N_p} \sum\limits_{j=0}^{N_p}{} 
\,C_j T_j(p^T_{x,l}) C_m \frac{2}{N_t} \sum\limits_{n=0}^{N_t}{} C_n
%\nonumber\\
%&&\ \ \ \ \ \ \ \times 
T_n(t^T_m)
\nonumber\\
&&\ \ \ \ \ \ \ \times
\int\!\! d p_x^\prime \,
T_j\left(p^T_x(p_{x}^\prime)\right) \,q_{\Delta p_x}(p_{x,b}- p_x^\prime)
\nonumber\\
&&\ \ \ \ \ \ \ \times
\int\limits_{\max[x-\vert v_x\vert t,\,L]}^{\min[x+\vert v_x\vert t,\,L]}\!\! d x^\prime\, 
T_i\left(x^T(x^\prime)\right) 
\,T_n\left(t^T(t_c-\vert x_a - x^\prime\vert/\vert v_{x,b}\vert)\right) 
\,\Psi_{\Delta x}(\vert x_a-x^\prime \vert)   .
\label{finalP}
\end{eqnarray}
Again, if the pdf of increments are independent of $P(x,p_x,t)$, then the integrals 
are over given functions, and Eq.\ (\ref{finalP}) can be interpreted as a simple 
matrix multiplication that yields the values of $P(x,p_x,t)$ at the grid points
from the value of $Q(x,p_x,t)$ at the grid-points.

In case of the critical gradient model, the spatial pdf of increments $q_{\Delta x}$
depends non-linearly on the density gradient, which is a function of $P(x,p_x,t)$, so that 
both Eqs.\ (\ref{finalQ}) and (\ref{finalP}) turn into a set of non-linear algebraic
equations, for which an appropriate numerical method would have to be developed.

\end{widetext}

{}


\begin{thebibliography}{}


\bibitem{Bak87}
Bak, P., Tang, C., Wiesenfeld, K., Phys.\ Rev.\ Lett.\ {\bf 59}, 381 (1987)

\bibitem{Baker01}
Baker, D.R., Greenfield, C.M., Burnell, K.H., et al, Phys.\ of Plasmas {\bf 8}, 4128 (2001).

\bibitem{Balescu95}
Balescu, R., Phys.\ Rev. E {\bf 51}, 4807 (1995)

%\bibitem{Balescu98}
%Balescu, R., Vlad, M., Spineanu, F., Phys.\ Rev. E {\bf 58}, 951 (1998)

%\bibitem{Carr02}
%Carreras, B.A., Lynch, V.E., Newman, D.E., Sanchez, R., Phys.\ Rev.\ E 
%{\bf 66}, 011302 (2002)

%\bibitem{Carr02b}
%Carreras, B.A., Lynch, V.E., Newman, D.E., Zaslavsky, G.M., Phys.\ Rev.\ E 
%{\bf 60}, 4770 (2002)

%\bibitem{Carr06}
%Carreras, B.A., Lynch, V.E., van Milligen, B.Ph., Sanchez, R., Phys.\ of 
%Plasmas {\bf 13}, 062301 (2006)

%\bibitem{Carr01}
%Carreras, B.A., Lynch, V.E., Zaslavsky, G.M., Phys. of Plasma {\bf 8},
%5096 (2001)

%\bibitem{Diamond95}
%Diamond, P.H., Hahn, T.S., Phys.\ Plasmas {\bf 2}, 3640 (1995)

%\bibitem{Gavnholt05}
%Gavnholt, J., Juul Rasmussen, J., Garcia, O.E., Naulin, V., Nielsen, A.H.,
%Physics of Plasmas {\bf 12}, 084501 (2005).

%\bibitem{Gentle97}
%Gentle, K.W., Bravenec, R.V., Cima, G., et al., Phys. of Plasmas {\bf 4}, 3599 (1997)

\bibitem{Hoang01}
Hoang, G.T., Bourdelle, C., Garbet, X., et al., Phys.\  Rev.\ Lett.\ {\bf 87},
125001-1 (2001)

%\bibitem{Hoang03}
%Hoang, G.T., Bourdelle, C., P\'egouri\'e, B., et al., Phys.\  Rev.\ 
%Lett.\ {\bf 90}, 155002 (2003).

\bibitem{Isliker03}
Isliker, H., Vlahos, L., Phys. Rev. E {\bf 67},026413 (2003).

\bibitem{Garbet04}
Garbet, X., Mantica, P., Angioni, C., et al., Plasma Phys.\ Control.\ Fusion
{\bf 46}, B557 (2004).

\bibitem{Garbet04b}
Garbet, X., Mantica, P., Ryter, F., et al., Plasma Phys.\ Control.\ Fusion
{\bf 46}, 1351 (2004).

%\bibitem{Hwa92}
%Hwa, T., Kardar, M., Phys.\ Rev.\ A {\bf 45}, 7002 (1992).

\bibitem{Imbeau01}
Imbeau, F., Ryter, F., Garbet, X., Plasma Phys.\ Control.\ Fusion {\bf 43}, 1503 (2001).

\bibitem{Jackson62}
Jackson, J.D., Classical Electrodynamics, John Wiley \& Sons, New York, 1962

%\bibitem{Jha03}
%Jha, R., Kaw, P.K., Kulkami, D.R., Panikh, J.C., and ADITYA Team, 
%Phys.\ of Plasmas {\bf 10}, 699 (2003) 

%\bibitem{Kissick98}
%Kissick, M.W., Callen, J.D., Fredrickson, E.D., Nuclear Fusion {\bf 38}, 821 (1998).

\bibitem{Lemoine05}
Lemoine, N., Gresillon, D.M., Phys.\ of Plasmas {\bf 12}, 092301 (2005).

\bibitem{Lopez95}
Lopes Cardoso, N.J., Plasma Phys.\ Control. Fusion {\bf 37}, 799 (1995).

\bibitem{Luce92}
Luce, T.C., Petty, C.C., der Haas, J.C.M., Phys. Rev. Lett. {\bf 68}, 52, (1992)

\bibitem{Mihaila02}
Mihaila, B., Mihaila, I., J.\ Phys.\ A {\bf 35}, 731 (2002),

%\bibitem{vanMil02}
%van Milligen, B.\ Ph., de la Luna, E., Tabar\'es, F.L., et al., 
%Nuclear Fusion {\bf 42}, 787 (2002)

\bibitem{vanMil04}
van Milligen, B.\ Ph., S\'anchez, R., Carreras, B.A., Phys. of Plasmas 
{\bf 11}, 2272 (2004)

\bibitem{vanMil04b}
van Milligen, B.\ Ph., Carreras, B.A., S\'anchez, R., Physics of Plasmas
{\bf 11}, 3787 (2004).

%\bibitem{vanMil05}
%van Milligen, B.\ Ph., Carreras, B.A., S\'anchez, R., Proc. of the 
%EPS conference, Taragona, Spain (2005).

\bibitem{Metzler00}
Metzler, R., Klafter, J., Physics Reports {\bf 339}, 1 (2000).

\bibitem{Montroll65}
E.W.\ Montroll, G.H.\ Weiss, J.\ Math.\ Phys.\ {\bf 6}, 167 (1965).

%\bibitem{Newman96}
%Newman, D.E., et al., Phys.\ Plasmas {\bf 3}, 1858 (1996)

%\bibitem{SOC}
%P.A.\ Politzer, Phys.\ Rev.\ Lett.\ 84, 1192 (2000);
%S.C.\ Chapman, R.O.\ Dendy, B.\ Hnat, Phys.\ Rev.\ Lett.\ 86, 2814 (2001);
%B.A.\ Carreras, D.\ Newman, V.E.\ Lynch, P.H.\ Diamond,
%Phys.\ of Plasmas {\bf 3}, 2903 (1996).

%\bibitem{Carbone}
%E.\ Spada et al., Phys.\ Rev.\ Lett.\ {\bf 86}, 3032 (2001);
%%Spada, E., Carbone, V., Cavazzana, R., Fattorini, L., Regnoli, G.,
%%Vianello, N., Antoni, V., Martines, E., Serianni, G., Spolaore, M.,
%%Tramontin, L., Phys.\ Rev.\ Lett.\ 86, 3032 (2001)
%V.\ Antoni et al., Phys.\ Rev.\ Lett.\ {\bf 87}, 5001 (2001).
%%Antoni, V., Carbone, V., Cavazzana, R., Regnoli, G., Vianello, N.,
%%Spada, E., Fattorini, L., Martines, E., Serianni, G., Spolaore, M.,
%%Tramontin, L., Veltri, P., Phys.\ Rev.\ Lett.\ 87, 5001 (2001)

%\bibitem{confined}
%R.\ Balescu, Phys.\ Rev. E 51, 4807 (1995);
%E.\ Barkai, J.\ Klafter, in \textit{Lecture Notes in Phys.\ 511}, edited by
%S.\ Benkadda, G.M.\ Zaslavsky, (Springer-Verlag, Berlin, 1998), p.\ 373;
%G.\ Zimbardo, A.\ Greco, P.\ Veltri, Phys.\ of Plasmas {\bf 7}, 1071 (2000).

\bibitem{Petty94}
Petty, C.C., Luce, T.C., Nuclear Fusion {\bf 34}, 121 (1994)

\bibitem{Peyret02}
Peyret, R., Spectral Methods for Incompressible Viscous Flow
(Applied mathematical sciences 148), Berlin (Springer-Verlag, Berlin), 2002.

%\bibitem{Rhodes99}
%Rhodes, T.L., Moyer, R.A., Groebner, R., Doyle, E.J., Lehmer, R.,
%Peebles, W.A., Rettig, C.L., Physics Letters A {\bf 253}, 181 (1999).

\bibitem{Ryter01}
Ryter, F., Angioni, C., Beurskens, M., et al., Plasma Phys.\ and Control.\ 
Fusion {\bf 43}, A323 (2001).

\bibitem{Ryter03}
Ryter, F., Tardoni, G., De Luca, F., et al. Nuclear 
Fusion {\bf 43}, 1396 (2003).

%\bibitem{Sanch02}
%S\'anchez, R., Newman, D.E., Carreras, B.A, Phys.\ Rev.\ Lett.\ {\bf 88},
%068302 (2002)

%\bibitem{Sanch05}
%S\'anchez, R., van Milligen, B.Ph., Carreras, B.A, Phys.\ of Plasmas {\bf 12},
%056105-1 (2005)

%\bibitem{Sattin05}
%Sattin, F., Paccagnella, R., D'Angelo, F., Physica A {\bf 358}, 273 (2005).

%\bibitem{Sattin06}
%Sattin, F., Baiesi, M., Phys.\ Rev.\ Lett.\ {\bf 96}, 105005 (2006).

%\bibitem{Shlesinger89}
%Shlesinger, M.F., Klafter, J., Journal of Physical Chemistry {\bf 93}, 7023
%(1989).

\bibitem{Shlesinger87}
Shlesinger, M.F., West, B.,  Klafter, J., Phys. Rev. Lett. {\bf 58}, 1100 (1987).

%\bibitem{Stroth98}
%Stroth, U., Plasma Phys.\ Control.\ Fusion {\bf 40}, 9 (1998)

%\bibitem{Tamura05}
%Tamura, N., Inagaki, S., Ida, K.,  et al., Phys. of Plasmas {\bf 12}, 110705 (2005)

\bibitem{Vlahos04}
Vlahos, L., Isliker, H., Lepreti, F., Astrophys.\ Journ. {\bf 608}, 540 (2004).

%\bibitem{Zas00}
%Zaslavsky, G.M., Edelman, M., Weitzner, H., et al., Phys. of Plasmas {\bf 7},
%3691 (2000).

\bibitem{Zumofen93}
Zumofen, G., Klafter, J., Phys.\ Rev.\ E {\bf 47}, 851 (1993). 

\end{thebibliography}
\end{document}